\documentclass[hyper]{JHEP3}

\input{epsf}
\usepackage{epsfig}
\usepackage{amssymb}
\usepackage{amsfonts}
\usepackage{amsbsy}
\usepackage[all]{xy}
\usepackage{amsmath}

\usepackage{amssymb,amscd}
\usepackage{mathrsfs}
\usepackage{amsmath,amsthm}
\usepackage{xspace}

\def\re{\mbox{Re }}
\def\im{\mbox{Im }}
\def\mod{{\rm mod}}

\def\IC{\mathbb{C}}

\def\IZ{{\mathbb{Z}}}
\def\IR{{\mathbb{R}}}
\def\IP{\mathbb{P}}
\def\IQ{\mathbb{Q}}

\def\CI {{\cal I}}
\def\CM {{\cal M}}
\def\CN{{\cal N}}
\def\CK {{\cal K}}
\def\CN {{\cal N}}

\def\CF {{\cal F}}

\def\CP {{\cal P }}
\def\CL {{\cal L}}
\def\CV {{\cal V}}
\def\CW {{\cal W}}
\def\CX {{\cal X}}

\def\CO {{\cal O}}
\def\CZ {{\cal Z}}

\def\CG {{\cal G}}
\def\CH {{\cal H}}
\def\CB {{\cal B}}

\def\CA{{\cal A}}
\def\CK{{\cal K}}

\def\CU{{\cal U}}
\def\CZ{{\cal Z}}
\def\CT{{\cal T}}

\def\half{\frac{1}{2}}

\renewcommand{\Im}{{\rm Im }}

\def\one{{\hbox{ 1\kern-.8mm l}}}

\def\sgn{{\rm sgn\,}}
\def\vol{{\rm vol\,}}
\def\p{\partial}

\def\be{\bar{e}}

\def\half{\frac{1}{2}}

\def\hk{hyperk\"ahler\xspace}
\def\qk{quaternionic-K\"ahler\xspace}
\def\kahler{K\"ahler\xspace}
\newcommand{\abs}[1]{\lvert#1\rvert}
\newcommand{\norm}[1]{\lVert#1\rVert}
\def\pz{{(\zeta)}}
\newcommand{\ti}[1]{\textit{#1}}

\def\p{\partial}

\def\Tr{{\rm Tr}}
\def\Res{{\rm Res}}

\def\be{\begin{equation}
}

\def\ee{\end{equation}}

\newcommand{\inprod}[1]{\langle#1\rangle}

\renewcommand\sf{{\mathrm{sf}}}
\newcommand\inst{{\mathrm{inst}}}
\newcommand{\plog}[1]{#1 \partial_{#1}}
\newcommand\elec{{e}}
\newcommand\magn{{m}}
\newcommand\degen{{\Omega}}
\newcommand\hsymp{{\varpi}}
\newcommand\clat{{\Gamma}}
\newcommand\vm{{\CB}}
\newcommand\eps{\epsilon}

\newcommand{\cwarrow}{\text{\Large$\curvearrowright$}}
\newcommand{\ccwarrow}{\text{\Large$\curvearrowleft$}}

\title{Four-dimensional wall-crossing via three-dimensional field theory}

\author{Davide Gaiotto$^1$, Gregory W. Moore$^2$, Andrew Neitzke$^1$\\
$^1$ School of Natural Sciences, Institute for Advanced Study, \\
Princeton, NJ 08540, USA\\
 $^2$ NHETC and Department of Physics and Astronomy,
Rutgers University,\\
Piscataway, NJ 08855--0849, USA\\
\\
{\tt dgaiotto@ias.edu, gmoore@physics.rutgers.edu, neitzke@ias.edu} }

\abstract{We give a physical explanation of the Kontsevich-Soibelman wall-crossing formula
for the BPS spectrum in Seiberg-Witten theories.  In the process we give an exact description
of the BPS instanton corrections to the \hk metric of the moduli space of the theory on $\IR^3 \times S^1$.
The wall-crossing formula reduces to the statement that this metric is continuous.  Our construction
of the metric uses a four-dimensional analogue of the two-dimensional $tt^*$ equations.}

\begin{document}

\bibliographystyle{utphys}


\section{Introduction and Summary}
\label{sec:Introduction}

The main subject of this paper is a wall-crossing formula (WCF) for
the degeneracies of BPS states in quantum field theories with $d=4$,
$\CN=2$ supersymmetry. Our conventions and a summary of relevant
definitions can be found in Section \ref{sec:preliminaries}. The
space $\CH_{\gamma,BPS}$ of BPS states of charge $\gamma$ is the
space of states in the one-particle Hilbert space, of
electromagnetic charge $\gamma$, saturating the BPS bound $M \geq
\vert Z_\gamma(u)\vert $. Here $u$ denotes a point in the
vector multiplet moduli space $\vm$, that is, in the
Coulomb branch of the moduli space of vacua.

The only available index for $d=4$, $\CN=2$ supersymmetry is the
second helicity supertrace:
\be \degen(\gamma;u) := -\half
{\Tr}_{\CH_{BPS,\gamma}}(-1)^{2J_3}(2J_3)^2, \ee
where $J_3$ is any generator of the rotation subgroup of the massive
little group. It has been known for a long time that such indices
are generally not independent of $u$ but are only piecewise constant
\cite{Cecotti:1992qh}. Indeed, $\degen(\gamma;u)$ can jump across
walls of marginal stability, where $\gamma = \gamma_1+\gamma_2$ and
$\arg Z_{\gamma_1}(u) = \arg Z_{\gamma_2}(u)$.  This fact played an
important role in the development of Seiberg-Witten theory
 \cite{Seiberg:1994rs,Seiberg:1994aj}.

In recent years a more systematic
understanding of the $u$-dependence of the index has begun to
emerge.
Formulae for the change $\Delta\degen$ across walls of marginal
stability were given in \cite{Denef:2007vg} when at least one of the
constituents in the decay $\gamma \to \gamma_1+\gamma_2$ is
primitive. These primitive and semiprimitive wall-crossing formulae
were derived from physical pictures based on multicentered
solutions of supergravity \cite{Denef:2000nb,Denef:2000ar}. However,
when both constituents have nonprimitive charges, the methods of
\cite{Denef:2007vg} are difficult to employ.

Kontsevich and Soibelman \cite{ks1} have proposed a
remarkable wall-crossing formula for the $\Delta \degen$ which
applies to all possible decays. We review their formula, which we
sometimes refer to as the KS formula, in Section \ref{sec:ReviewKS}.

On the one hand, Kontsevich and Soibelman's ``Donaldson-Thomas
invariants'' $\hat\degen(\gamma;u)$ are not obviously the same as
the $\degen(\gamma;u)$ of interest in physics, and the techniques
they use to arrive at their formula seem somewhat removed from
standard physical considerations. On the other hand, their WCF
involves striking new concepts compared to the formulation of the
semiprimitive wall-crossing formulae of \cite{Denef:2007vg}. In
particular, the WCF is expressed in terms of a certain product of
symplectomorphisms of a torus (see (\ref{eq:ks-product})
below) which depends on the $\hat\degen(\gamma;u)$, and hence
\emph{a priori} depends on $u$. The statement of the WCF is that
this product is, in fact, independent of $u$. That in turn
determines the $u$-dependence of $\hat\degen(\gamma;u)$. This
development raises the question of the physical derivation and
interpretation of the KS formula and holds out the promise that some
essential new physical ideas are involved. This will indeed prove to
be the case.

In this paper we give a physical
interpretation and proof of the KS formula in the case of $d=4$,
$\CN=2$ \ti{field theories}.  The generalization to supergravity is
an interesting and important problem for future work.

Here is a sketch of the main ideas and the basic strategy. We
consider the gauge theory on the space $\IR^3 \times S^1$ where
$S^1$ has radius $R$.  At low energies this theory is described by a
$d=3$ sigma model with \hk target space $(\CM,g)$. This sigma model
receives corrections from BPS instantons, in which the world-line of
a BPS particle of the $d=4$ theory is wrapped around $S^1$.
Expanding the metric $g$ at large $R$, one can therefore read off
the degeneracies $\degen(\gamma;u)$ of the BPS particles. This
immediately raises a puzzle:  we know that the
$\degen(\gamma;u)$ are discontinuous, but $g$ should be
continuous!  The continuity of the metric is based on the physical
principle (which was crucial in
\cite{Seiberg:1994rs,Seiberg:1994aj}) that the only singularities in
the low energy effective field theory Lagrangian arise from the
appearance at special moduli of massless particles (which should not
have been integrated out in the effective theory).

Physically the resolution of this puzzle is similar to one recently
discussed in \cite{GarciaEtxebarria:2007zv}.  The exact metric $g$
is indeed smooth, but it receives corrections from multi-particle as
well as single-particle states.  The disappearance of a
$1$-instanton contribution when a particle decays is compensated by
a discontinuity in the multi-instanton contribution from its decay
products. Similarly, disappearing $n$-instanton contributions are
compensated by discontinuities in the $m$-instanton contributions for $m > n$. To put
this more precisely, the $n$-instanton corrections have the form
$\sum \prod^n_{i=1} \degen(\gamma_i;u) F^{(n)}_{(\gamma_i)}(u)$,
where the sum runs over all $n$-tuples $\{\gamma_i\}$ of charges,
and $F^{(n)}$ are essentially universal functions of $R$ and the
$Z_{\gamma_i}$.  Upon crossing the wall each $\degen(\gamma;u)$ has
a discontinuity proportional to a sum of products of
$\degen(\gamma_j;u)$ with $\sum \gamma_j = \gamma$.  At the same
time, the functions $F^{(n)}$ have discontinuities proportional to
the functions  $F^{(n')}$ with $n'<n$.  We will see that the
Kontsevich-Soibelman wall-crossing formula expresses the consistency
of this tower of cancellations.

The main technical hurdle in understanding the WCF
is thus to give an efficient description of the corrections to $g$
coming from the BPS instantons.  A \hk metric is a complicated
object and it is hard to make progress by studying, say, the
corrections to its components; nor is there generally a simple additive object
like the \kahler potential available.  To overcome this problem we
borrow some ideas from twistor theory. Recall that a \hk manifold
is complex-symplectic with respect to a whole $\IC\IP^1$ worth of complex
structures. The basic idea is that studying $g$ is equivalent to
studying the holomorphic Darboux coordinates on $\CM$, provided that we
consider \ti{all} of these complex structures at once.

In the main body of this paper, we assume that the
Kontsevich-Soibelman wall-crossing formula holds for
$\degen(\gamma;u)$.  Under this assumption we construct the
metric on $\CM$, by giving a canonical set of
functions $\CX_\gamma(u, \theta;\zeta)$ on $\CM \times \IC^\times$, indexed by an electromagnetic charge
$\gamma$.  Here $(u, \theta)$ specifies a point of $\CM$ and the parameter $\zeta$
labels the complex structures on $\CM$.
Each $\CX_\gamma$ is \ti{piecewise} holomorphic in $\zeta$;
the effect of the BPS instantons is to create
discontinuities in the $\CX_\gamma(u, \theta;\zeta)$, along rays $\ell$ in the
$\zeta$-plane.  These discontinuities are identified with the
symplectomorphisms introduced by Kontsevich and Soibelman.  In this
approach the continuity of the metric is a consequence of the
WCF.  In the final section, we run the argument in
reverse:  using general principles of supersymmetric gauge theory,
we deduce properties of the metric $g$ which are sufficient to
prove the WCF.

\subsection*{Summary}

We begin in Section \ref{sec:preliminaries} with a review of
the Seiberg-Witten solution of $d=4$, $\CN=2$ gauge theories and the Kontsevich-Soibelman
wall-crossing formula.  We then
discuss the formulation of the theory on $\IR^3 \times S^1$.  It is a sigma model into a
manifold $\CM$, which is topologically the Seiberg-Witten
torus fibration over the $d=4$ moduli space $\vm$, equipped with a \hk metric $g$.
This metric depends on the radius $R$ of $S^1$.
As $R \to \infty$ it approaches a simple form, which can be obtained
by naive dimensional reduction of the $d=4$ theory; we call this simple metric $g^\sf$ (for ``semiflat'').

In Section \ref{sec:twistor} we explain our ``twistorial''
construction of \hk metrics:  given a collection of functions $\CX_\gamma(u,\theta;\zeta)$ on
$\CM$, varying holomorphically with $\zeta\in \IC^\times$ and obeying certain additional conditions,
there is a \hk metric for which $\CX_\gamma(u,\theta;\zeta)$ are holomorphic Darboux coordinates.  In particular,
we give the functions $\CX^\sf_\gamma(u,\theta;\zeta)$ corresponding to the semiflat metric $g^\sf$.

With this background in place we are ready to consider the instanton
corrections.  We begin this study in Section \ref{sec:periodicnut} with
the simple case of a $U(1)$ gauge theory coupled to a
single matter hypermultiplet of electric charge $q$.  In this theory
the corrected metric $g$ is known exactly \cite{Ooguri:1996me,Seiberg:1996ns}.
We explain how to obtain this corrected metric by including instanton corrections
which modify the functions $\CX^\sf_\gamma(u,\theta;\zeta)$ to new ones $\CX_\gamma(u,\theta;\zeta)$.
In this construction we already see the building
blocks of the Kontsevich-Soibelman formula appear:  our $\CX_\gamma(u,\theta;\zeta)$
naturally come out with discontinuities in the $\zeta$-plane, which are precisely
the elementary Kontsevich-Soibelman symplectomorphisms corresponding to the
electric charges $\pm q$.

We then turn in Section \ref{sec:constructing-moduli} to the more interesting case where
we have multiple kinds of BPS instanton corrections,
coming from mutually non-local BPS particles in $d=4$.  In this case we find a natural ansatz for the
$\CX_\gamma(u,\theta;\zeta)$:  essentially we just require that each BPS particle independently contributes a discontinuity
like the one we found for a single particle.  This discontinuity is most naturally located along a
ray in the $\zeta$-plane determined by the phase of the central charge of the BPS particle.
The Kontsevich-Soibelman factors for mutually non-local particles
do not commute; but this generically presents no problem since these particles have non-aligned
central charges, and hence their discontinuities appear on distinct rays in the $\zeta$-plane.
The separation between rays disappears exactly at the walls of marginal stability; here the discontinuities pile up
into products of Kontsevich-Soibelman factors.  The WCF is the
statement that this product is the same as we approach the wall from either side.  This requirement
is essential for us:  it implies that the metric we construct from the $\CX_\gamma(\zeta)$ is continuous.

More precisely, to determine the $\CX_\gamma(u,\theta;\zeta)$ we specify both their
discontinuities in the $\zeta$-plane and also their asymptotics as $\zeta \to 0, \infty$.
In other words, we formulate an infinite-dimensional
``Riemann-Hilbert problem'' whose solution is the $\CX_\gamma(u,\theta;\zeta)$. We
do not construct its solution exactly; rather we follow a
strategy closely analogous to that employed by Cecotti and Vafa,
who encountered a similar (but finite-dimensional)
Riemann-Hilbert problem in the study of $d=2$ theories with
$\CN=(2,2)$ supersymmetry \cite{Cecotti:1993rm}. A variation of
their arguments allows us to show that the solution to our problem exists,
at least for sufficiently large $R$. Indeed, in the large
$R$ limit the desired $\CX_\gamma$ can be obtained by successive approximations,
where the zeroth approximation is just $\CX^\sf_\gamma$,
and the $n$-th approximation incorporates multi-instanton
effects up to $n$ instantons.

Having constructed the functions $\CX_\gamma(u,\theta;\zeta)$ and hence
the metric $g$, we check that $g$ has various properties
which are expected on general field theory grounds; it passes all of these tests
and we therefore argue that it should be the correct physical metric
on $\CM$, generalizing a similar argument
in \cite{Seiberg:1996nz}.

As we mentioned above, our construction of the $\CX_\gamma(u,\theta;\zeta)$ bears a
striking similarity to constructions which appeared in the $d=2$
case \cite{Cecotti:1993rm}.  In that case a wall-crossing formula
for the degeneracies of BPS domain walls was proven using the flat
``$tt^*$ connection'' in the bundle of vacua of the $d=2$ theory.
Two components of this connection give differential equations expressing the R-symmetry
and scale invariance of the $d=2$ theory.
Our construction can similarly be phrased in terms of a flat connection $\CA$ over
$\vm \times \IC\IP^1 \times \IR_+$, in the infinite-dimensional bundle of real-analytic functions
on the torus fibers $\CM_u$ of $\CM$.
The Riemann-Hilbert construction guarantees the existence of this $\CA$.
Each $\CX_\gamma$ defines a flat section.
In particular, this flatness gives a pair of differential equations for the $\zeta$ and $R$
dependence of $\CX_\gamma$, which have their physical origin in the anomalous R-symmetry and scale invariance of
the $d=4$ theory.

As we describe in Section \ref{sec:proof}, this $tt^*$-like flat connection can be discovered
using only general principles of supersymmetric gauge theory.
Moreover, its mere \ti{existence}
is strong enough to justify our ansatz for the metric \ti{a priori}.  In particular, the
wall-crossing formula, which appeared as a consistency requirement working within that ansatz, can be
understood as the existence of an ``isomonodromic deformation'' constructed from $\CA$.
This gives a physical proof of the wall-crossing formula.

For convenience, in most of this paper we use a simple form of the wall-crossing formula which
does not include information about flavor symmetries, and correspondingly we set all
flavor masses to zero.  In Section \ref{sec:masses} we explain how to restore the flavor charge
and mass information.

We include several appendices with additional details. In Appendix
\ref{app:ks-product} we explain a direct verification that the
wall-crossing formula gives the correct BPS degeneracies in the case
of the pure $SU(2)$ theory.  In Appendix \ref{app:holomorphy} we
describe the Cauchy-Riemann equations on $\CM$, in a way that makes
contact with our construction of the \hk metric and with the $tt^*$
equations of \cite{Cecotti:1991me}.  In Appendix
\ref{app:integral-asymptotics} we give the asymptotic analysis
necessary for extracting the large-$R$ corrections to the metric
from our Riemann-Hilbert problem.  In Appendix
\ref{app:diff-asymptotics} we discuss some details of how to extract
the differential equations from the solution of the Riemann-Hilbert
problem. Finally, Appendix \ref{app:TBA} explains a curious relation
of one of our main results, equation
\eqref{eq:X-integral-mult-explicit}, with the Thermodynamic Bethe
Ansatz. There is much more to be said about this connection, but we
leave that for another occasion.

Several subsubsections of the paper are devoted to global issues
which are related to a subtle but important sign in the KS
formula.  On a first reading it would be reasonable to skip this
discussion.  Readers who choose this course should allow themselves to confuse
$T$ and $\tilde{T}$, as well as $\CM$ and $\widetilde{\CM}$, in
the main text.

\subsection*{Discussion}

Let us remark on a few particularly interesting points.

\begin{itemize}
\item Physically, our construction of the metric on $\CM$ amounts to a rule for ``integrating out''
mutually non-local particles in $d=4$.  This problem \ti{a priori}
appears to be difficult because one cannot find any duality frame in
which all of the particles are electrically charged, so it is
difficult to write a Lagrangian which includes all of the relevant
fields.  Here we have circumvented that difficulty.

\item Our construction of the metric uses its twistorial description.  The most natural physical context
in which the twistor space occurs is projective superspace \cite{Karlhede:1984vr,Lindstrom:1987ks,Lindstrom:2008gs},
 in which the parameter $\zeta$ is
a bosonic superspace coordinate.  The fact that the corrections to $g$ come only from BPS
instantons, and that they are localized at specific rays in the $\zeta$-plane, should have a natural
explanation in the projective superspace language.

\item One of the inspirations for the Kontsevich-Soibelman WCF
was their earlier work \cite{MR2181810}, in which they gave a construction of the sheaf of holomorphic
functions on a K3 surface, by ``correcting'' the sheaf of functions on the semiflat K3.  The corrections
were formulated in terms of products of symplectomorphisms similar to those which appear in the wall-crossing
formula.  This construction is closely related to ours, with K3 replaced by $\CM$.
The key new ingredient in our work is to consider all complex structures at once,
thus introducing the parameter $\zeta \in \IC\IP^1$; having done so, we can formulate the
crucial Riemann-Hilbert problem.  This idea might also be useful in the original K3 context.

\item The multi-instanton expansion of $g$ is given as a sum of basic building blocks weighted by products
of the BPS degeneracies $\degen(\gamma;u)$.  These basic building
blocks have intricate discontinuities at the walls of marginal
stability, which conspire with the jumps of $\degen(\gamma;u)$ to
make $g$ continuous in $u$.  All this is reminiscent of recent work
of Joyce on wall-crossing \cite{Joyce:2006pf}. Moreover, Joyce's
work was interpreted by Bridgeland and Toledano Laredo in
\cite{stab-stokes} in terms of isomonodromic deformation of a
connection on $\IC\IP^1$, which somewhat resembles the one we
consider here, but has a slightly different form: it has an
irregular singularity only at $t = 0$, while ours has them both at
$\zeta = 0$ and $\zeta = \infty$.  There is an interesting scaling
limit of our connection, $R \to 0$ and $\zeta \to 0$ with $\zeta / R
= t$ fixed, which brings it into the form of the one in
\cite{stab-stokes} (albeit with a different structure group).  This
limit retains the information about the BPS degeneracies and their
wall-crossing.  It would be interesting to see whether there is any
sense in which it relates our connection to the one in
\cite{stab-stokes}.

\item In our discussion we studied structures defined over the vector multiplet moduli space $\vm$.
However, both Kontsevich-Soibelman and Joyce formulate their invariants over a larger space,
the space of ``Bridgeland stability conditions'' \cite{MR2373143}.  We do not understand
the meaning of our constructions when extended to this larger space.

\item The wall-crossing formula as formulated by Kontsevich-Soibelman makes sense not only for $\CN=2$ field
theories but also for supergravity, and indeed this was the main
focus of \cite{Denef:2007vg}.  The moduli space $\CM$ of the theory
on $\IR^3 \times S^1$ is then a \qk manifold rather than \hk.
 Nevertheless, most of our considerations seem to make
sense in that context, with appropriate modifications. For example,
Hitchin's theorem is replaced by LeBrun's theorem characterizing the
twistor space of a \qk manifold in terms of holomorphic contact
structures. In particular, there is still a natural notion of a
``holomorphic'' function $\CX_\gamma(x, \zeta)$ (namely, a
holomorphic function on the twistor space of $\CM$), and the \qk
analogue of $\CX^\sf_\gamma$ has been worked out in
\cite{Neitzke:2007ke}.  We expect that the instanton-corrected
metric $g$ on $\CM$ can be obtained by a method parallel to the one
employed in this paper: formulate a Riemann-Hilbert problem for
$\CX_\gamma$, using $\CX^\sf_\gamma$ to fix the asymptotics, and the
Kontsevich-Soibelman factors to fix the discontinuities. One
important difficulty to overcome is that in gravity the degeneracies
$\degen(\gamma;u)$ grow very quickly with $\gamma$; this makes the
convergence of the iterative solution for $\CX_\gamma$ less obvious
in this case. As in the \hk case, the WCF should arise as a
consistency condition ensuring that $g$ is smooth.

\item The analogy between the \hk geometry of the fibration $\CM \to \vm$
and the $tt^*$ geometry of \cite{Cecotti:1991me,Cecotti:1992qh,Cecotti:1993rm} is striking:
the two structures are very similar although one has to do with field theories in $d=4$, the other in $d=2$.
Is there a direct relation between the two?  One possibility is to relate them just by
compactification, e.g.\ on $S^2$.  Different values of the $U(1)$ fluxes on $S^2$ would then correspond to
different vacua of the $d=2$ theory, and BPS states of the $d=4$ theory could be identified with domain walls
interpolating between these vacua in $d=2$.\footnote{This picture has been advocated to us by Cumrun Vafa.}
Related ideas have appeared in the literature before ---
in particular see \cite{Losev:1997tp,Losev:1999nt,Ooguri:2005vr}.
See also \cite{Dorey:1999zk,Ritz:2006rt} for a slightly different link between BPS spectra in $d=2$
and $d=4$.

\item Infinitesimal deformations of a class of \hk manifolds which
include the semiflat geometry have been recently studied in
\cite{Alexandrov:2008ds}. It would be interesting to describe the
leading correction to the semiflat geometry in their language. Our equation
\eqref{eq:FullX2} resembles their equation (3.38), with an appropriate choice of $H$ and
contours of integration.

\end{itemize}

\section{Preliminaries} \label{sec:preliminaries}

\subsection{$d=4$, $\CN=2$ gauge theory}

We consider a gauge theory in $d=4$ with $\CN=2$ supersymmetry,
gauge group $G$ of rank $r$, and a characteristic (complex) mass
scale $\Lambda$. Seiberg-Witten theory (initiated in
\cite{Seiberg:1994rs,Seiberg:1994aj}, and reviewed more generally in
e.g. \cite{Lerche:1997sm,Freed:1997dp,ias-volumes}) gives a
rather complete description of the behavior of such a gauge theory
on its Coulomb branch at energies $\mu \ll \Lambda$, as follows.

The Coulomb branch is a complex manifold $\vm$ of complex
dimension $r$, parameterized by the vacuum expectation values of the
vector multiplet scalars.  We denote a generic coordinate system on
$\vm$ as $(u^1, \dots, u^r)$.  At each point $u \in \vm$ the gauge
group is broken to a maximal torus $U(1)^r$.  There is a lattice $\clat_u
\simeq \IZ^{2r}$ of electric and magnetic charges, equipped with an
integral-valued symplectic pairing $\inprod{,}$.  This lattice is
the fiber of a local system $\clat$ over $\vm$. That is, there is a
fibration of lattices with fiber $\clat_u$ over $u \in \vm$, with nontrivial monodromy around
the singular loci in $\vm$, of complex codimension 1, where some
BPS particles become massless.
We sometimes write ``$\gamma \in \clat$'' informally, meaning
that $\gamma$ is a local section of $\clat$.

There is a vector $Z(u) \in \clat_u^* \otimes_\IZ \IC$ of ``periods,'' which varies
holomorphically with $u$.  For any $\gamma \in \clat$ we define the
central charge $Z_\gamma(u)$ by
\begin{equation}
Z_\gamma(u) = Z(u) \cdot \gamma.
\end{equation}
$Z(u)$ plays a fundamental role in the description both of the massless and the massive sectors.

We begin with the massless part.
Locally on $\vm$ one can choose a splitting $\clat = \clat^m \oplus \clat^e$ into
Lagrangian sublattices of ``magnetic'' and ``electric'' charges respectively.
$\clat^m$ and $\clat^e$ are then dual to one another using
the pairing on $\clat$.  Such a splitting is called an electric-magnetic duality frame.
Concretely we may choose a basis
$\{\alpha_1, \dots, \alpha_r\}$ for $\clat^m$ and $\{\beta^1, \dots, \beta^r\}$ for $\clat^e$
such that
\begin{equation}
\inprod{\alpha_I, \alpha_J} = 0, \quad \inprod{\beta^I, \beta^J} = 0, \quad \inprod{\alpha_I, \beta^J} = \delta_I^J
\end{equation}
with $I, J = 1, \dots, r$.
After choosing such a frame, we obtain a system of ``special
coordinates'' $a^I$ on $\vm$, which are nothing but the electric
central charges, i.e.
\begin{equation} \label{eq:special-coords}
Z_{\beta^I} = a^I.
\end{equation}
The magnetic central charges are then holomorphic functions of the $a^I$.
They are determined in terms of a single function
$\CF(a^I)$ (depending on the chosen frame), the $\CN=2$ prepotential:
\begin{equation} \label{eq:prepot}
Z_{\alpha_I} = \frac{\partial \CF}{\partial a^I}.
\end{equation}
This implies in particular that $Z$ is not arbitrary:  from the symmetry of mixed partial derivatives
one obtains
\begin{equation} \label{eq:lag-constraint}
\inprod{dZ, dZ} = 0.
\end{equation}
On the left side of \eqref{eq:lag-constraint}
we are using the antisymmetric pairing $\inprod{,}$ and also the antisymmetric wedge
product of 1-forms on $\vm$; the combined pairing is symmetric, so this condition is not vacuous.
Indeed, \eqref{eq:lag-constraint} says that around $u$, $\CB$ can be locally identified with a complex Lagrangian
submanifold of $\Gamma^*_u \otimes_\IZ \IC$.

The prepotential completely determines the two-derivative effective Lagrangian, written
in terms of the electric vector multiplets.  To write this Lagrangian we introduce the symmetric matrix $\tau$
defined by
\begin{equation}
\tau_{IJ}(a) = \frac{\partial^2 \CF}{\partial a^I \partial a^J},
\end{equation}
and then adopt a notation that suppresses the gauge index,
e.g. $\tau \abs{da}^2$ for $\tau_{IJ} da^{I} \wedge \star d\bar{a}^J$.
Then the bosonic part of the Lagrangian is
\begin{equation} \label{eq:sw-lagrangian}
\CL^{(4)} = \frac{\im \tau}{4\pi} \left(- \abs{da}^2 - F^2 \right) + \frac{\re \tau}{4\pi} F \wedge F.
\end{equation}

The central charges $Z_\gamma$ are also of fundamental importance for the massive
spectrum.  Indeed, the mass of any 1-particle state with charge $\gamma$ obeys
\begin{equation} \label{eq:bps-bound}
M \ge \abs{Z_\gamma}
\end{equation}
with equality if and only if the state is BPS.  BPS states belong to massive short multiplets of the super
Poincare symmetry; under the little group $SU(2)$ the states at rest in such a multiplet transform as
\begin{equation}
[j] \otimes \left(\left[1/2\right] + 2[0]\right).
\end{equation}
Choosing $j=0$ gives the massive hypermultiplet, while $j = \half$ is the massive vector multiplet.

There is a standard index which ``counts'' the short multiplets, namely the second helicity supertrace
$\degen(\gamma;u)$.  This supertrace
receives the contribution $+1$ for each massive hypermultiplet of charge $\gamma$ in the spectrum
of the theory at $u \in \vm$, and similarly $-2$ for each massive vector multiplet.
$\degen(\gamma;u)$ is invariant under any deformation of the theory in which
the 1-particle states do not mix with the continuum of multiparticle states.
From \eqref{eq:bps-bound} it follows that such mixing is very restricted; a BPS particle can decay
only into other BPS particles, and then only if their central charges all have the same phase.
Hence $\degen(\gamma;u)$ is locally constant in $u$, away from the \ti{walls of marginal stability}
in $\vm$.  These walls of marginal stability are of real codimension $1$ and
are defined, for a pair of linearly independent charges $\gamma, \gamma'$, to be the locus of $u\in \vm$ where
$Z_\gamma$ and $Z_{\gamma'}$ are nonzero and have the same phase.

Understanding the jumping behavior of $\degen(\gamma;u)$ as $u$ crosses a wall of marginal stability is
one of the main motivations of this paper.  We turn to it next.

\subsection{The Kontsevich-Soibelman wall-crossing formula}
\label{sec:ReviewKS}

In this section we review the Kontsevich-Soibelman wall-crossing
formula.  As originally proposed in \cite{ks1} this formula
determines the jumping behavior of ``generalized Donaldson-Thomas
invariants'' $\hat\degen(\gamma;u)$.
As we will see below, if we identify the Donaldson-Thomas invariants
with the helicity supertraces, $\hat\degen(\gamma;u) = \degen(\gamma;u)$,
then the wall-crossing formula gives the
physically expected answer in several nontrivial examples: in
particular, it reproduces the ``primitive wall-crossing formula'' of
\cite{Denef:2007vg}, as well as the wall-crossing behavior of the
BPS spectrum of Seiberg-Witten theory with gauge group $SU(2)$.

A technical point:  for the KS formula to make sense,
the $\degen(\gamma;u)$ are not allowed to be completely arbitrary.  Introducing a
positive definite norm on $\clat$, one must require that there exists some $K > 0$ such that
\begin{equation}
\frac{\abs{Z_\gamma}}{\norm{\gamma}} > K
\end{equation}
for all $\gamma$ such that $\hat\degen(\gamma;u) \neq 0$.  Throughout this paper
we will assume that this property, called the ``Support Property,'' holds.

\subsubsection*{The Kontsevich-Soibelman algebra}

The wall-crossing formula is given in terms of a Lie
algebra defined by generators $e_\gamma$, with $\gamma \in \clat$,
and a basic commutation relation
\begin{equation} \label{eq:ks-commutator}
[e_{\gamma_1}, e_{\gamma_2}] = (-1)^{\langle
\gamma_1,\gamma_2\rangle} \langle \gamma_1,\gamma_2\rangle
e_{\gamma_1+\gamma_2}.
\end{equation}
In this paper it will be important to realize this abstract Lie
algebra as an algebra of complex symplectomorphisms of a
complexified torus.
Modulo a subtlety which will appear at the end of this section, this torus is
the fiber $\tilde{T}_u$ of the local system $\tilde{T} := \clat^* \otimes_{\IZ} \IC^\times$.

Any $\gamma \in \clat$ gives a corresponding function $X_\gamma$ on $\tilde{T}_u$,
with $X_\gamma X_{\gamma'} = X_{\gamma + \gamma'}$.
Upon choosing a basis $\{\gamma^1, \dots, \gamma^{2r}\}$ for $\clat$,
we can choose $X^i := X_{\gamma^i}$ as coordinates for $\tilde{T}_u$.
The symplectic pairing on $\clat^*$ gives a holomorphic symplectic form $\hsymp^{\tilde{T}}$ on $\tilde{T}_u$:
if $\epsilon^{ij} = \inprod{\gamma^i, \gamma^j}$, and $\epsilon_{ij}$ is its inverse,
\begin{equation}
\hsymp^{\tilde{T}} = \frac{1}{2}\epsilon_{ij} \frac{dX^i}{X^i} \wedge \frac{dX^j}{X^j}.
\end{equation}

We would like to identify $e_\gamma$ with the infinitesimal
symplectomorphism of $\tilde{T}_u$ generated by the Hamiltonian $X_\gamma$.
This almost gives the algebra \eqref{eq:ks-commutator}, but misses
the extra sign $(-1)^{\langle \gamma_1,\gamma_2\rangle}$.
This sign will be important below in comparing to wall-crossing
formulas known from physics; in that context it is related to
the fact that the fermion number of a bound state of two
particles of charges $\gamma_1,\gamma_2$ is shifted by
$\langle \gamma_1,\gamma_2\rangle$.

Over a local patch of $\vm$,
we can absorb this sign by introducing a ``quadratic refinement'' of the
$\IZ_2$-valued quadratic form $(-1)^{\langle \gamma_1,\gamma_2\rangle}$:  this means a $\sigma: \clat \to \IZ_2$
obeying
\begin{equation}
\sigma(\gamma_1)\sigma(\gamma_2) = (-1)^{\langle \gamma_1,\gamma_2\rangle} \sigma(\gamma_1 + \gamma_2).
\end{equation}
One way to get such a $\sigma$ is to choose a local electric-magnetic duality
frame $\clat \cong \clat^e \oplus \clat^m$, write $\gamma = \gamma^e + \gamma^m$, and
set $\sigma(\gamma)=(-1)^{\inprod{\gamma^e,\gamma^m}}$.  Notice that
\begin{equation}
\sigma(\gamma_1)\sigma(\gamma_2) = \sigma(\gamma_1 + \gamma_2)
(-1)^{\inprod{\gamma_1^e,\gamma_2^m} + \inprod{\gamma_2^e,\gamma_1^m}} =
\sigma(\gamma_1 + \gamma_2) (-1)^{\langle \gamma_1,\gamma_2\rangle}
\end{equation}
as needed.
At any rate, having chosen any $\sigma(\gamma)$, we could identify $e_\gamma$ with the symplectomorphism
generated by the Hamiltonian $\sigma(\gamma) X_\gamma$.

Any two refinements $\sigma$, $\sigma'$
obey $\sigma(\gamma) \sigma'(\gamma) = (-1)^{c(\sigma, \sigma') \cdot \gamma}$ for some fixed
$c(\sigma, \sigma') \in \clat^* / 2 \clat^*$.
The Hamiltonians $\sigma(\gamma) X_\gamma$ and $\sigma'(\gamma) X_\gamma$
associated to these two refinements are related by the automorphism of $\tilde{T}_u$ which sends
$X_\gamma \to (-1)^{c(\sigma, \sigma') \cdot \gamma} X_\gamma$ .

\subsubsection*{The wall-crossing formula}

Now we are ready to formulate the wall-crossing formula.
Its basic building block is the group element
\begin{equation}
\CK_\gamma := \exp \sum_{n=1}^{\infty} \frac{1}{n^2} e_{n \gamma}.
\end{equation}
Under our identification, $\CK_\gamma$ becomes a symplectomorphism
acting on $\tilde{T}_u$, given by
\begin{equation} \label{eq:ks-symplect}
\CK_\gamma : X_{\gamma'} \to X_{\gamma'} (1- \sigma(\gamma)
X_\gamma)^{\langle \gamma',\gamma \rangle}.
\end{equation}

Associate to each BPS
particle of charge $\gamma$ a ray in the complex $\zeta$-plane, determined
by the central charge,
\begin{equation}
\ell_\gamma := \{\zeta: Z_\gamma(u) / \zeta \in \IR_- \}.
\end{equation}
 As we
vary $u \in \vm$ these rays rotate in the $\zeta$-plane. The
cyclic ordering of the rays changes only when $u$ reaches a wall of
marginal stability.  At such a wall a set of BPS rays $\ell_\gamma$ come
together, corresponding to a set of charges $\gamma$ for which $Z_\gamma$
become aligned.
At a generic point on the wall of marginal stability, this set
of charges can be parameterized as\footnote{To establish the existence of
these $\gamma_1$, $\gamma_2$ we need to use the Support Property:  otherwise one
can easily imagine situations in which the aligned $Z_\gamma$ accumulate
near the origin.}
$\{n \gamma_1 + m \gamma_2: m, n > 0\}$, for some primitive vectors $\gamma_1, \gamma_2$ with
$Z_{\gamma_1} / Z_{\gamma_2} \in \IR_+$.

Now associate the group element $\CK_\gamma$ to each BPS
particle of charge $\gamma$, and
form the product over states which become aligned at the wall:
\begin{equation} \label{eq:ks-product}
A := \prod^\cwarrow_{\substack{\gamma = n \gamma_1 + m \gamma_2 \\ m>0,\,n>0}}
\CK_\gamma^{\degen(\gamma;u)},
\end{equation}
where the ordering of the factors corresponds to clockwise ordering of the rays $\ell_\gamma$.
We can consider this product for $u$ on either side of the wall.
As $u$ crosses the wall, the order of the factors is
reversed, and the $\degen(\gamma;u)$ jump.  The statement of the
wall-crossing formula is that the whole product $A$ is unchanged.

This condition is strong enough to determine the $\degen(\gamma;u_+)$ from
the $\degen(\gamma;u_-)$, where $u_\pm$ are points infinitesimally displaced
on opposite sides of the wall.  To understand how to do this in practice we first have to deal
with an important subtlety:  since the spectrum of BPS states is
typically infinite, the product \eqref{eq:ks-product} generally
involves infinitely many factors.
Following \cite{ks1}, we can understand it as follows.  The product
only involves the generators $e_{n \gamma_1 + m \gamma_2}$, where
$m,n>0$. The Lie algebra they generate can be consistently truncated
by fixing some integer $L$ and then setting $e_{n \gamma_1 + m
\gamma_2} = 0$ whenever $n+m > L$. (That is, the Lie algebra is
filtered by Lie subalgebras with $n+m>L$, and we can take quotients
by subalgebras with successively larger values of $L$.)  After such
a truncation \eqref{eq:ks-product} involves only finitely many
nontrivial terms; the infinite product can be understood as the
limit of these truncated products as $L \to \infty$.

In a similar spirit consider the power expansion of the
symplectomorphism $A$,
\begin{equation} A : X_{\gamma'} \to  ( 1 + \sum_{m>0,n>0} c^{m,n}_{\gamma'} X_{n
\gamma_1 + m \gamma_2}) X_{\gamma'}, \end{equation} and truncate it
to $n+m \le L$.  We can compute this expansion on one side of the wall of
marginal stability, and then recursively identify the
$\degen(\gamma;u)$ on the other side of the wall. Concretely, first
set $L=1$; then $\degen(\gamma_1;u)$ and $\degen(\gamma_2;u)$ are
fixed by the requirement that they correctly reproduce
$c^{1,0}_{\gamma'}$ and $c^{0,1}_{\gamma'}$.  Next set $L=2$ and
consider the expansion of $A \CK_{\gamma_1}^{-\degen(\gamma_1;u)}
\CK_{\gamma_2}^{-\degen(\gamma_2;u)}$ to extract the next set of
degeneracies. This iteration can be continued in a straightforward
way to determine all of the $\degen(n \gamma_1 + m \gamma_2 ; u)$.
What is far from obvious --- but conjectured in \cite{ks1} --- is that the
$\degen(\gamma ; u)$ computed in this way are integers!

\subsubsection*{Some examples}

In the above interpretation of the Kontsevich-Soibelman formula we
identified their generalized Donaldson-Thomas invariants with the
physically defined $\degen(\gamma;u)$. To motivate this
identification, we now describe a few examples.

As explained above,
at a generic point on a wall of marginal stability the symplectomorphisms
which enter the WCF are generated by a two-dimensional lattice of charges,
$\gamma = (p,q) \in \IZ^2$ with canonical symplectic form
$\inprod{(p,q), (p',q')} = pq' - q p'$.  We write correspondingly
$X_{1,0}=x$, $X_{0,1}=y$.
The symplectomorphisms $\CK_{p,q}$ are then determined by their action on $x$ and $y$,
which is explicitly
\begin{equation}
\CK_{p,q}: (x,y) \to \biggl(  \bigl(1-(-1)^{pq} x^p y^q \bigr)^q x,
\bigl(1-(-1)^{pq} x^p y^q\bigr)^{-p} y \biggr).
\end{equation}

Consider a wall of marginal stability where the central charges for
a single BPS particle of primitive charge $(1,0)$ and a single
particle of primitive charge $(0,1)$ come together. Kontsevich and
Soibelman notice a beautiful ``pentagon identity'':
\begin{equation} \label{eq:product-formula-primitive}
\CK_{1,0} \CK_{0,1} = \CK_{0,1} \CK_{1,1} \CK_{1,0}.
\end{equation}
Hence the WCF predicts that crossing the wall, only one extra
particle will be created, a dyonic bound state of one electrically
charged particle and one magnetically charged particle. Indeed the
``primitive wall-crossing formula'' from supergravity (which is also
valid in field theory) \cite{Denef:2007vg} predicts that this pair
of particles will form a single bound state in a hypermultiplet
representation. It also predicts that a single particle of charge
$(1,0)$ cannot be bound to more than one particle of charge $(0,1)$.
It is quite hard to count more general bound states of several
particles of different type. Their absence is already a non-trivial
prediction of the KS wall-crossing formula.

A further comparison with the primitive wall-crossing formula helps us
understand the role of the sign in the commutation relation
\eqref{eq:ks-commutator} of the $e_\gamma$. Consider the product
$\CK_{\gamma_1} \CK_{\gamma_2}$ and try to rewrite it as a product in
the opposite direction, (i.e. with the slopes of $Z_{\gamma_i}$
increasing instead of decreasing) of the form $\CK_{\gamma_2} \cdots
\CK_{\gamma_1}$.  Suppose $\gamma_1,\gamma_2$ are primitive and
consider the subalgebra generated by $e_{n\gamma_1 + m \gamma_2}$
quotiented by that with $n\geq 2, m\geq 2$. The result is a
Heisenberg algebra. The KS formula in the truncated Heisenberg group
reads
\begin{equation}
\CK_{\gamma_1}^{\degen(\gamma_1;u_+)}\CK_{\gamma_1+\gamma_2}^{\degen(\gamma_1+\gamma_2;u_+)}
\CK_{\gamma_2}^{\degen(\gamma_2;u_+)}=
\CK_{\gamma_2}^{\degen(\gamma_2;u_-)}\CK_{\gamma_1+\gamma_2}^{\degen(\gamma_1+\gamma_2;u_-)}
\CK_{\gamma_1}^{\degen(\gamma_1;u_-)}
\end{equation}
where $u_\pm$ are points infinitesimally displaced on either side of
the wall.  Now, at a generic point on the wall of marginal stability
we have $\degen(\gamma_i; u_+) = \degen(\gamma_i; u_-)$ for $i=1,2$.
Moreover, $\CK_{\gamma_1+\gamma_2}$ is central in the Heisenberg
group, and therefore, computing the group commutator we reproduce
the corollary of the primitive wall-crossing formula:
\begin{equation}
\Delta \degen = (-1)^{\langle \gamma_1,\gamma_2\rangle -1} \langle
\gamma_1,\gamma_2\rangle \degen(\gamma_1;u) \degen(\gamma_2;u).
\end{equation}
A more elaborate version of this argument allows one to extract the
semiprimitive wall-crossing formula of \cite{Denef:2007vg} from the
KS formula.\footnote{This was shown in unpublished work with Wu-yen
Chuang.}

The example in \eqref{eq:product-formula-primitive} is exceptional
in that both sides involve a finite number of terms. More typically
one encounters infinite products.  A second beautiful example
presented by Kontsevich and Soibelman is the following:
\begin{gather}
\CK_{1,0}^2 \CK_{0,1}^2 = \left( \CK_{0,1}^2 \CK_{1,2}^2 \CK_{2,3}^2 \cdots \right)
\CK_{1,1}^{4} \CK_{2,2}^{-2} \left( \cdots \CK_{3,2}^2 \CK_{2,1}^2
\CK_{1,0}^2 \right).
\end{gather}
We give an instructive proof of this identity in Appendix
\ref{app:ks-product}.
By a change of basis we obtain a physically very interesting
formula,
\begin{gather} \label{eq:nf2}
\CK_{1,-1}^2 \CK_{0,1}^2 = \left( \CK_{0,1}^2 \CK_{1,1}^2 \CK_{2,1}^2 \cdots \right) \CK_{1,0}^4 \CK_{2,0}^{-2}
\left( \cdots \CK_{3,-1}^2 \CK_{2,-1}^2 \CK_{1,-1}^2 \right),
\end{gather}
which captures the spectrum of an $SU(2)$ Seiberg-Witten theory with
two massless flavors (more precisely, hypermultiplets transforming
in the vector representation of an $SO(4)=SU(2)_A \times SU(2)_B$
flavor symmetry) as described in \cite{Bilal:1996sk}.\footnote{The
relation of the identity \eqref{eq:nf2}  to  Seiberg-Witten theory
was first suggested by Frederik Denef. The precise relation of
\eqref{eq:nf2} to the $N_f=2$ theory was worked out in collaboration
with Wu-yen Chuang. } On the right side we see the full weak
coupling spectrum: one $W$ boson of charge $(2,0)$ (which
contributes $-2$ to the helicity supertrace), the four
hypermultiplets of charge $(1,0)$, and a set of dyons of
charge $(n,\pm 1)$, with multiplicity $2$.
(In fact these dyons are in doublets of $SU(2)_A$ or $SU(2)_B$,
depending on the parity of $n$.) On the left side we see the strong
coupling spectrum: a single monopole with multiplicity $2$ (a
doublet of $SU(2)_A$) and a single dyon with multiplicity $2$ (a
doublet of $SU(2)_B$.)

The small change of variables $y \to -y^2$ converts the
product formula \eqref{eq:nf2} into
\begin{gather} \label{eq:nf0}
\CK_{2,-1} \CK_{0,1} = \left( \CK_{0,1} \CK_{2,1} \CK_{4,1} \cdots \right)
\CK_{2,0}^{-2} \left( \cdots \CK_{6,-1} \CK_{4,-1} \CK_{2,-1} \right).
\end{gather}
This formula captures the wall-crossing behavior of the pure $SU(2)$
Seiberg-Witten theory.\footnote{The close resemblance between
(\ref{eq:nf2}) and (\ref{eq:nf0}) arises because the Seiberg-Witten
curve for the $N_f=2$ theory with zero masses is a double cover of that for
the $N_f=0$ theory.} The left side includes the two particles present at
strong coupling \cite{Ferrari:1996sv}: a monopole of charge $(0,1)$ and a dyon of charge
$(2,-1)$. The right side includes the infinite spectrum of dyons at
weak coupling, together with the $W$ boson contribution
$\CK_{2,0}^{-2}$.

\subsubsection*{Adding flavor information}

The product \eqref{eq:nf2} describes the BPS spectrum of $SU(2)$
Seiberg-Witten theory with $N_f = 2$, but does not carry information
about the flavor charges of the BPS particles.  We now describe a
conjectural variant of the KS formula which includes the information
about flavor charges. (We will see the physical motivation for this
formula in Section \ref{sec:masses}.) Introduce a new lattice of
flavor charges $\clat^f$, and a new parameter $\log \mu \in
(\clat^f)^* \otimes_\IZ \IC^\times$. Then generalize the $X_\gamma$
to new functions labeled by $(\gamma,\gamma^f) \in \clat \oplus
\clat^f$:\footnote{Strictly speaking, the full local system
$\hat\Gamma$ of charges does not split into $\Gamma \oplus \Gamma^f$
globally; we really have an extension $0 \to \Gamma^f \to \hat\Gamma
\to \Gamma \to 0$.  However, we can always split this extension
locally, and this is sufficient for our purposes.} letting $a$ run
over a basis for $\clat^f$,
\begin{equation}
X_{\gamma,\gamma^f} := \prod_i (X^i)^{\gamma_i} \prod_a
(\mu^a)^{\gamma^f_a} = X_\gamma \prod_a (\mu^a)^{\gamma^f_a}.
\end{equation}
Define refined symplectomorphisms carrying flavor information:
\begin{equation}
\CK_{\gamma,\gamma^f} : X_{\gamma'} \to X_{\gamma'} (1- \sigma(\gamma)
X_{\gamma,\gamma^f})^{\langle \gamma',\gamma \rangle}.
\end{equation}
The central charge now depends on the masses $m^a$,
$Z_{\gamma,\gamma^f}(u) = Z_\gamma(u) + \gamma^f_a m^a$, and
determines new walls of marginal stability.  ($\mu^a$ are functions
of the $m^a$. See section \ref{sec:masses} below.) We introduce a
product analogous to \eqref{eq:ks-product},
\begin{equation} \label{eq:ks-product-masses}
A := \prod^\cwarrow_{\substack{(\gamma,\gamma^f) = n \gamma_1 + m \gamma_2 \\ m>0,\,n>0}}
\CK_{\gamma,\gamma^f}^{\degen(\gamma,\gamma^f;u)},
\end{equation}
The extended WCF states the continuity of $A$ across the walls.  We derive a
refined version of the infinite product \eqref{eq:nf2}, including
the flavor charges, in Appendix \ref{app:ks-product}; combining this with the extended WCF
we obtain the correct wall-crossing for the $SU(2)$ theory with $N_f = 2$.

\subsubsection*{Global issues}

So far in this section we have worked over a local patch in $\vm$, and chosen
a fixed quadratic refinement $\sigma$ in order to identify the Kontsevich-Soibelman
algebra with an algebra of symplectomorphisms of the complexified torus $\tilde{T}_u$,
a fiber of the local system $\tilde{T}$.
It is impossible in general to choose such a refinement globally over $\vm$,
because of the monodromies of the local system $\clat$.
Hence it is not true globally that the Kontsevich-Soibelman algebra is the algebra of
symplectomorphisms acting on $\tilde{T}$.

However, by an appropriate twisting of $\tilde{T}$ we can define a closely related complexified torus
fibration $T$, on which the Kontsevich-Soibelman
algebra does act.  $T$ is defined so that a local choice of quadratic refinement gives an identification
$T \simeq \tilde{T}$, and given two different refinements $\sigma, \sigma'$, the corresponding
identifications differ by the map $X_\gamma \to (-1)^{c(\sigma, \sigma') \cdot \gamma} X_\gamma$
on $\tilde{T}$.
The fiberwise symplectic form $\hsymp^{\tilde T}$ induces a corresponding fiberwise symplectic form $\hsymp^T$ on $T$.

We can construct a twisted fibration $T$ with the above properties
as follows.  Let $R$ denote the local system
over $\vm$ whose local sections are refinements $\sigma$.  $R$ is a torsor for
$\clat^*/2\clat^*$, and $T$ is the associated fibration,
\begin{equation}
T := \left( \tilde{T} \times R \right) / \left( (X_\gamma, \sigma) \sim
((-1)^{c(\sigma, \sigma') \cdot \gamma} X_\gamma, \sigma') \right).
\end{equation}

\subsection{The low energy effective theory on $\IR^3 \times S^1$} \label{sec:reduction}

Our goal is to explain the Kontsevich-Soibelman WCF as a statement about the
gauge theory on $\IR^3 \times S^1$, with $S^1$ of radius $R$.
We study the theory at an energy scale $\mu$ which is low compared to all other scales, i.e., $\mu \ll \Lambda$
and also $\mu \ll 1/R$.  At this energy the theory looks effectively three-dimensional.  In this section
we describe some of its basic properties.

In the limit of large radius, $R \gg 1/\Lambda$, we can determine the three-dimensional dynamics using
the infrared Lagrangian \eqref{eq:sw-lagrangian}.
The dynamical degrees of freedom are just the $x^4$-independent modes of the four-dimensional fields.
These include of course the scalars $a^I$.
In addition, from the gauge sector we get the ``electric'' Wilson lines
\begin{equation}
\theta_\elec^I := \oint_{S^1} A_4^I dx^4,
\end{equation}
as well as another set of periodic scalars $\theta_{\magn,I}$ obtained by
dualizing the $d=3$ gauge fields $A^I_\alpha dx^\alpha$. We will
often think of these as ``magnetic'' Wilson lines,
\begin{equation}
\theta_{\magn,I} := \oint_{S^1} (A_{D,4})_I dx^4.
\end{equation}
We can define $\theta_{\magn,I}$ either by working in a formulation
treating the gauge fields as self-dual, or by working at fixed
magnetic quantum numbers $P^I$ and introducing $\theta_{\magn,I}$ as their
Fourier duals.

All these periodic scalars coordinatize a $2r$-torus $\CM_u$ at any
fixed $u \in \vm$.  Letting $u$ vary we obtain a torus fibration
$\CM$.  The fiber $\CM_u$ degenerates over the singular loci in
$\vm$. The low energy theory on $\IR^3$ is a sigma model with target
space $\CM$.

More precisely, $\theta = (\theta_\elec^I, \theta_{\magn,I})$ is an element in the
fiber of a local system of $2r$-tori
$\widetilde\CM := \clat^*\otimes_{\IZ} (\IR/2\pi \IZ)$.  For any $\gamma \in
\clat$, we get an angular coordinate on $\widetilde\CM$ denoted
$\theta_\gamma := \gamma \cdot \theta$.  $\widetilde\CM$ is not exactly
the same as $\CM$; there is a global twisting which we glossed over above, and
which we discuss at the end of this section.

In sum, the three-dimensional theory is a sigma model into a Riemannian
manifold $\CM$ of real dimension $4r$, which is topologically a $2r$-torus
fibration over $\vm$.
The theory enjoys $\CN=4$ supersymmetry (8
real supercharges), which implies that the metric on $\CM$ is \hk.
This metric is the main object of study in this paper.
It was studied previously in \cite{Seiberg:1996nz}, where in
particular the $R \to 0$ limit for pure $SU(2)$ gauge theory was
identified as the Atiyah-Hitchin manifold. In this paper we are more
interested in the opposite limit $R \to \infty$, because in this
limit one can read off the imprint of the full BPS spectrum of the
theory in $d=4$.  In the next section we begin by considering the
leading behavior in this limit.

\subsubsection*{Global issues}

In the description above we were slightly naive about the precise definition of the Wilson lines.  Our
description is adequate over a local patch in $\vm$, but as we will see in Section \ref{sec:periodicnut},
it cannot be quite correct globally.  Indeed, in order for the metric on $\CM$ to be smooth, we will see that
the monodromies around paths in $\vm$ must
generally be accompanied by shifts of the Wilson lines by $\pi$.  This contradicts our naive description,
since the torus fibration $\widetilde\CM$ comes with a distinguished zero section.

We propose that the correct global picture is as follows:  at any fixed $u \in \vm$, the Wilson
lines live in a torus $\CM_u$ which is isomorphic
to $\widetilde\CM_u$, but not canonically isomorphic.  One obtains an isomorphism $\CM_u \simeq \widetilde\CM_u$
upon choosing a refinement $\sigma$ of the quadratic form
$(-1)^{\inprod{\gamma_1,\gamma_2}}$ on $\clat_u$.\footnote{Such
quadratic refinements frequently appear in the precise formulations of self-dual gauge
theories \cite{Witten:1996hc,Freed:2000ta,Hopkins:2002rd,Belov:2006jd}. It
seems likely that the origin of $\sigma$ here can be
explained in this way.}  Such a refinement generally exists only locally, so the
fibrations $\CM$ and $\widetilde\CM$ are globally different.
Given two local refinements $\sigma$, $\sigma'$ the corresponding
two local isomorphisms $\CM \simeq \widetilde\CM$ differ by the shift $\theta \to \theta + \pi c(\sigma,\sigma')$
acting on $\widetilde\CM$.  Of course, this discussion is closely parallel to the relation between
the torus fibrations $T$ and $\tilde{T}$ which we described at the end of Section \ref{sec:ReviewKS}.

\subsection{The semiflat geometry} \label{sec:semiflat}

The leading behavior of the metric on $\CM$ in the $R \to \infty$ limit is governed by the
$d=3$ effective action obtained by simply   truncating
\eqref{eq:sw-lagrangian} to its $x^4$-independent sector. This gives
\begin{equation}
\CL^{(3)} = (\im \tau) \left(- \frac{R}{2} \abs{da}^2 - \frac{R}{2}
F^{(3)} \wedge \star F^{(3)} -  \frac{1}{8 \pi^2 R} d\theta_\elec^2
\right) +  (\re \tau) \left(\frac{1}{2\pi} d \theta_\elec \wedge
F^{(3)}\right).
\end{equation}
Then dualizing the $d=3$ gauge field $A^I$ to a scalar $\theta_{\magn,I}$ gives after a little rearranging
\begin{equation}
\CL^{(3)}_{dual} = - \frac{R}{2} (\im \tau) \abs{da}^2 - \frac{1}{8 \pi^2 R} (\im \tau)^{-1} \abs{d \theta_\magn - \tau d\theta_\elec}^2.
\end{equation}
This is the Lagrangian for a sigma model into $\CM$, with metric locally given by
\begin{equation} \label{eq:semiflatmetric}
g^\sf = R (\im \tau) \abs{da}^2 + \frac{1}{4 \pi^2 R} (\im \tau)^{-1} \abs{dz}^2,
\end{equation}
where we introduced
\begin{equation}
dz_I = d \theta_{\magn,I} - \tau_{IJ} d \theta_\elec^J.
\end{equation}
(While this notation is very convenient, we should emphasize that
the form ``$dz_I$'' is not closed on the whole $\CM$:  it is only closed when restricted
to each torus fiber $\CM_u$.)

We call $g^\sf$ the ``semiflat'' metric on $\CM$, because in this metric the torus fibers
are flat.  The expression \eqref{eq:semiflatmetric} reflects the fact that $g^\sf$ is \kahler, with respect
to a complex structure on $\CM$ for which $da^I$ and $dz_I$ are a basis for
$\Omega^{1,0}$.  In this complex structure $\CM$ is the Seiberg-Witten fibration by compact \ti{complex}
tori over $\vm$.  (We contrast this with other complex structures on $\CM$ which we
will meet momentarily, in which the tori $\CM_u$ are \ti{not} complex submanifolds.)
The fibers $\CM_u$ all have volume
\begin{equation}
\vol(\CM_u) = \left(\frac{1}{R}\right)^r.
\end{equation}

The expression \eqref{eq:semiflatmetric} is valid only locally, since it uses
a choice of duality frame.  Nevertheless the expressions in different frames glue
together into a smooth metric, everywhere except over the singular loci of $\vm$, where
$g^\sf$ has a singularity.  Such a singularity would be unexpected from the point of
view of effective field theory; we will see that it is resolved by BPS instanton
corrections in the exact quantum-corrected metric $g$.

\section{A twistorial construction of \hk metrics} \label{sec:twistor}

In this section we review some general facts about \hk geometry, and then explain the basic idea
underlying our description of $g$.

\subsection{Holomorphic data from hyperk\"ahler manifolds} \label{sec:hk-general}

We first recall some holomorphic data attached to any \hk manifold.
By definition, a \hk manifold $(\CM, g)$ is
\kahler with respect to a triplet of complex structures $\vec{J}$, obeying the relations
\begin{equation} \label{eq:quaternion-complex-structures}
J_1 J_2 = J_3, \quad J_2 J_3 = J_1, \quad J_3 J_1 = J_2, \quad
J_\alpha^2 = -1.
\end{equation}
Let $\omega_\alpha$ denote the three corresponding \kahler forms.

In fact, any \hk $(\CM,g)$ is \kahler with respect to a more general
complex structure, namely $a^\alpha J_\alpha$ with
$\sum_{\alpha=1}^3 a_\alpha^2 = 1$, with corresponding \kahler form
$a^\alpha \omega_\alpha$. So we have a whole $S^2$ worth of complex
structures. One of the key insights of the twistor approach is that
it is useful to consider this $S^2$ as the Riemann sphere, labeled
by a complex parameter $\zeta$. So we write the general complex
structure and corresponding \kahler form as
\begin{align} \label{eq:Jzeta}
J^\pz &= \frac{i(- \zeta + \bar\zeta)J_1 - (\zeta + \bar\zeta) J_2 + (1 - \abs{\zeta}^2) J_3}{1 + \abs{ \zeta}^2},\\
\omega^\pz &= \frac{i(- \zeta + \bar\zeta)\omega_1 -  (\zeta + \bar\zeta) \omega_2 + (1 - \abs{\zeta}^2) \omega _3}{1 + \abs{\zeta}^2}.
\end{align}
We also organize the \kahler forms into a second combination,
\begin{equation} \label{eq:Omegazeta}
\hsymp(\zeta) = - \frac{i}{2\zeta} \omega_+ + \omega_3 - \frac{i}{2} \zeta \omega_-,
\end{equation}
where we introduced the notation
\begin{equation}
\omega_\pm = \omega_1 \pm i \omega_2.
\end{equation}
The essential property of $\hsymp(\zeta)$ is that for any fixed $\zeta \in
\IC\IP^1$ it is a holomorphic symplectic form on $\CM$ in complex
structure $J^\pz$.   (To
make sense of this statement for $\zeta = 0, \infty$ we have to
rescale $\hsymp(\zeta)$ by $\zeta$, $1/\zeta$ respectively.  Globally one
could say that $\hsymp(\zeta)$ is twisted by the line bundle $\CO(2)$ over $\IC\IP^1$.)

\subsection{Twistorial construction of $g$} \label{sec:twistor-general}

Now we describe the method of determining $g$ from holomorphic data on $\CM$, which will
be used in the rest of this paper.

First we specify our assumptions.
Recall that $\CM$ is topologically a torus fibration over $\vm$.
For any choice of local patch in $\vm$, quadratic refinement, and local section $\gamma$ of the charge lattice
$\clat$, we assume given a locally defined
$\IC^\times$-valued function $\CX_\gamma(u,\theta;\zeta)$ of $(u,\theta) \in \CM$ and $\zeta \in \IC^\times$, with the following properties:
\begin{itemize}

\item The $\CX_\gamma$ are multiplicative,
\begin{equation}
\CX_\gamma \CX_{\gamma'} = \CX_{\gamma + \gamma'}.
\end{equation}

\item The $\CX_\gamma$ obey a reality condition,
\begin{equation} \label{eq:X-real}
\CX_\gamma(\zeta) = \overline{\CX_{-\gamma}(-1 / \bar\zeta)}.
\end{equation}

\item All $\CX_\gamma$ are solutions to a single set of differential equations, of the form
\begin{align}
\frac{\partial}{\partial u^i} \CX &= \left( \frac{1}{\zeta} \CA^{(-1)}_{u^i} + \CA^{(0)}_{u^i} \right)  \CX , \label{eq:cr-1} \\
\frac{\partial}{\partial \bar{u}^{\bar i}} \CX &= \left( \CA^{(0)}_{\bar{u}^{\bar{i}}} + \zeta \CA^{(1)}_{\bar{u}^{\bar{i}}} \right) \CX, \label{eq:cr-2}
\end{align}
where the operators $\CA^{(n)}_{u^i}$, $\CA^{(n)}_{\bar{u}^{\bar{i}}}$ are complex vertical vector fields on
the torus fiber $\CM_u$, with the $\CA^{(-1)}_{u^i}$ linearly independent at every point,
and similarly $\CA^{(1)}_{\bar{u}^{\bar{i}}}$.
(To motivate these equations, note that in Appendix \ref{app:holomorphy} we show
that the Cauchy-Riemann equations on $(\CM, g)$ have this form.)

\item For each fixed $x \in \CM$, $\CX_\gamma(x; \zeta)$ is holomorphic in $\zeta$ on a dense
subset of $\IC^\times$.  (In our application below, $\CX_\gamma(x; \zeta)$ will be holomorphic away
from a countable union of lines.)

\end{itemize}
To state our last three assumptions on the functions $\CX_\gamma$ we first define
\begin{equation} \label{eq:darboux}
\hsymp(\zeta) := \frac{1}{8 \pi^2 R} \eps_{ij} \frac{d \CX_{\gamma^i}}{\CX_{\gamma^i}} \wedge \frac{d \CX_{\gamma^j}}{\CX_{\gamma^j}},
\end{equation}
where by $d$ we mean the fiberwise differential, i.e. we treat $\zeta$ as a fixed parameter.
We assume:
\begin{itemize}

\item $\hsymp(\zeta)$ is globally defined (in particular the $\hsymp(\zeta)$ defined over different local patches of $\vm$ agree with one another) and holomorphic in $\zeta \in \IC^\times$.  (Note that this does not imply that the $\CX_\gamma$ are holomorphic
in $\zeta$; in our application they will be only piecewise holomorphic.)

\item $\hsymp(\zeta)$ is nondegenerate in the appropriate sense for a holomorphic symplectic form, i.e. $\ker \hsymp(\zeta)$ is a
$2r$-dimensional subspace of the $4r$-dimensional $T_\IC \CM$.

\item $\hsymp(\zeta)$ has only a simple pole as $\zeta \to 0$ or $\zeta \to \infty$.

\end{itemize}
In the rest of this section we explain how to define a \hk metric $g$ on $\CM$, such that $\CX_\gamma(\zeta)$
are holomorphic functions in complex structure $J^\pz$, and $\hsymp(\zeta)$ is
the holomorphic symplectic form as in \eqref{eq:Omegazeta}.

We consider the manifold $\CZ := \CM \times \IC\IP^1$.
It has the following properties:

\begin{enumerate}

\item  \ti{$\CZ$ is a complex manifold.}
At any $(x, \zeta)$ the $2r$ equations \eqref{eq:cr-1}, \eqref{eq:cr-2} define
a half-dimensional subspace of $T_\IC \CM$ (if $\zeta = 0$ or $\zeta = \infty$ this is still
true after rescaling one of the equations by a factor $\zeta$).
The direct sum of this subspace
and the one generated by $\partial / \partial \bar{\zeta}$ is a half-dimensional
subspace of $T_\IC \CZ$.  We define $T^{0,1} \CZ$ to be this subspace.
This \ti{a priori} defines only an almost complex structure on $\CZ$.  However,
the existence of the functions $\CX_\gamma$ guarantees that this almost
complex structure is actually integrable.  (Of course, the $\CX_\gamma$ are not everywhere
holomorphic in $\zeta$; but they are holomorphic on a dense set, which is enough to
guarantee the vanishing of the Nijenhuis tensor.  It follows in particular that
there exist complex coordinates on $\CZ$ even around $\zeta = 0$ or $\zeta = \infty$.)

\item  \ti{$\CZ$ is a holomorphic fibration over $\IC\IP^1$.}
The projection is simply $p(x, \zeta) = \zeta$.

\item  \ti{There is a holomorphic section of $\Omega^2
_{\CZ/\IC \IP^1} \otimes \CO(2)$, giving a holomorphic
symplectic form on each fiber $p^{-1}(\zeta)$.}  This is
the globally defined $\hsymp(\zeta)$.

\item  \ti{There is a family of holomorphic sections $s: \IC\IP^1 \to \CZ$,
each with normal bundle $N \simeq \CO(1)^{\oplus 2r}$.}
Indeed, for each $x \in \CM$, we can define a section $s_x: \IC \IP^1 \to \CZ$ by $s_x(\zeta) = (x, \zeta)$.
To see that this is a holomorphic section, note first
that it is holomorphic at least away from $\zeta = 0,\infty$, just
because the local complex coordinates $\CX_{\gamma^i}(x, \zeta)$ of $\CZ$ are
holomorphic in $\zeta$ at fixed $x$; but it extends continuously to
$\zeta = 0,\infty$, so it must be holomorphic there as well by the
Riemann removable singularity theorem.  To show that the normal bundle
$N(s_x) \simeq \CO(1)^{\oplus 2r}$, first note that there is a 1-1 correspondence between
holomorphic sections of $N^*(s_x)$ and holomorphic functions on the first infinitesimal neighborhood of $s_x$
which vanish on $s_x$.  But such functions are determined by their first-order Taylor expansion around $x$,
i.e. they correspond to holomorphic sections of the trivial bundle $p^*((T^*_\IC)_x \CM)$ which
annihilate the subbundle $B \subset p^*((T_\IC)_x \CM)$ defined by the equations \eqref{eq:cr-1}, \eqref{eq:cr-2}.
Dualizing, we have $N(s_x) \simeq p^*((T_\IC)_x \CM) / B$.  On the other hand \eqref{eq:cr-1}, \eqref{eq:cr-2}
give $2r$ trivializing sections of $B \otimes \CO(1)$.  So we conclude that $N(s_x) \otimes \CO(-1)$ is trivial.

\item  \ti{There is an antiholomorphic involution $\sigma: \CZ \to \CZ$, which covers the antipodal
map on $\IC \IP^1$, and preserves $\hsymp$ in the sense that
$\sigma^* \hsymp = \overline{ \hsymp}$.}  This involution is just $\sigma(x, \zeta) = (x, - 1 / \bar\zeta)$.
Using the reality condition \eqref{eq:X-real} we can check that it is antiholomorphic and preserves $\hsymp$.
\end{enumerate}

These are the characteristic properties of the \ti{twistor space} of a \hk manifold as
described in \cite{Hitchin:1986ea,MR1206066}.  In particular, using the recipe of \cite{Hitchin:1986ea,MR1206066},
one can reconstruct a \hk metric $g$ on $\CM$ from $\CZ$.  We can describe $g$ concretely:
note that from $\hsymp(\zeta)^{r+1} = 0$ it follows that $\omega_+^r \wedge \omega_3 = 0$, which implies
that the real 2-form $\omega_3$ is of type $(1,1)$ in complex structure $J_3$.  Therefore we can use $J_3$
and $\omega_3$ to build a \kahler metric $g$ on $\CM$.  This $g$ coincides with the \hk metric guaranteed
by the twistor construction.  In the following sections we will use this approach.

\subsection{Twistorial construction of the semiflat geometry} \label{sec:semiflat-twistor}

The foregoing description of \hk metrics is particularly convenient in the case of the semiflat metric $g^\sf$
which we introduced in Section \ref{sec:semiflat}.
As above, we work over a local patch in $\vm$, and make a local choice of quadratic refinement.
Then for any $\gamma \in \clat$ we write the locally defined function\footnote{This
formula was first obtained in joint work with Boris Pioline, and is essentially the rigid limit of
a formula in \cite{Neitzke:2007ke} for the quaternionic-K\"ahler case.  It provided an important clue to
discovering the constructions described in this paper.}
\begin{equation} \label{eq:X-sf}
\CX_\gamma^\sf(\zeta) := \exp\biggl[ \pi  R \zeta^{-1}  Z_\gamma + i
\theta_\gamma + \pi R \zeta \bar Z_\gamma \biggr].
\end{equation}
These functions obey ``Cauchy-Riemann equations'' of the form \eqref{eq:cr-1}, \eqref{eq:cr-2},
where
\begin{gather}
\CA^{(-1)}_{u^i} = -i \pi R \frac{\partial Z}{\partial u^i} \cdot \frac{\partial}{\partial \theta}, \quad \CA^{(1)}_{\bar{u}^{\bar{i}}} = -i \pi R \frac{\partial \bar{Z}}{\partial \bar{u}^{\bar{i}}} \cdot \frac{\partial}{\partial \theta}, \\
\CA^{(0)}_{u^i} = 0, \quad \CA^{(0)}_{\bar{u}^{\bar{i}}} = 0
\end{gather}
and $Z$ stands for the vector of periods.  Then
$\hsymp^\sf(\zeta)$ is
\begin{align} \label{eq:Omega-sf}
\hsymp^\sf(\zeta) & := \frac{1}{8 \pi^2 R} \eps_{ij} \frac{d \CX^\sf_{\gamma^i}}{\CX^\sf_{\gamma^i}} \wedge \frac{d \CX^\sf_{\gamma^j}}{\CX^\sf_{\gamma^j}} \\
&= \frac{1}{4\pi} \left[ \frac{i}{\zeta} \left\langle d Z, d \theta \right\rangle + \left( \pi R \langle d Z, d \bar{Z} \rangle - \frac{1}{2 \pi R} \langle d \theta , d \theta \rangle \right) + i \zeta \langle d \bar{Z}, d \theta \rangle \right]. \label{eq:Omega-sf-2}
\end{align}
(Note that the vanishing condition \eqref{eq:lag-constraint} ensures that $\hsymp^\sf(\zeta)$ has no terms
of order $\zeta^{-2}$ or $\zeta^2$.)
$\hsymp^\sf(\zeta)$ and $\CX^\sf_\gamma(\zeta)$ obey the necessary conditions for the construction we described in
Section \ref{sec:twistor-general}, so they are the holomorphic symplectic form and complex
coordinates for some \hk metric on $\CM$.  As we now check, this metric is simply $g^\sf$ as desired.

First note that comparing the leading terms in \eqref{eq:Omegazeta} and \eqref{eq:Omega-sf-2} gives
\begin{equation} \label{eq:omegaplus-sf}
\omega_+^\sf = - \frac{1}{2\pi} \inprod{d Z, d \theta}.
\end{equation}
From $\omega_+^\sf$ we can determine complex structure $J^\sf_3$:  indeed, after choosing an
electric-magnetic duality frame, we can rewrite \eqref{eq:omegaplus-sf} as
\begin{equation}
\omega_+^\sf = \frac{1}{2\pi} da^I \wedge dz_I.
\end{equation}
This makes manifest that $\CM$ in complex
structure $J^\sf_3$ is just the Seiberg-Witten fibration by complex tori.  This is the complex structure
we already described in Section \ref{sec:semiflat}.

Similarly, comparing the $\zeta$-independent terms in \eqref{eq:Omegazeta} and \eqref{eq:Omega-sf-2} gives
\begin{equation} \label{eq:omega3-sf}
\omega_3^\sf = \frac{R}{4} \langle d Z , d \bar{Z} \rangle - \frac{1}{8 \pi^2 R} \langle d\theta, d \theta \rangle,
\end{equation}
which we can rewrite as
\begin{equation}
\omega_3^\sf = \frac{i}{2} \left( R (\im \tau)_{IJ} d a^I \wedge d \bar{a}^J + \frac{1}{4 \pi^2 R} ((\im \tau)^{-1})^{IJ} dz_I \wedge d\bar{z}_J \right).
\end{equation}
Comparing this with \eqref{eq:semiflatmetric} we see that $g^\sf$ is indeed
\kahler for complex structure $J^\sf_3$ and \kahler form $\omega_3^\sf$,
and hence it is the \hk metric guaranteed by the twistor construction starting from $\hsymp^\sf(\zeta)$.

In this section we have seen that the semiflat metric on $\CM$ and its \hk structure
can be constructed from the functions $\CX_\gamma^\sf$ defined in
\eqref{eq:X-sf}.  These functions are of fundamental importance for what follows.

\section{Mutually local corrections} \label{sec:periodicnut}

If we considered only the naive dimensional reduction of the massless sector,
then the semiflat metric $g^\sf$ would be
the end of the story.  However, the theory in $d=4$ also contains massive BPS particles.
The metric receives corrections from ``instanton'' configurations in
which one or more of these massive particles go around $S^1$.  These
corrections will be weighted by a factor of at least $e^{-2 \pi R \abs{Z}}$, because of the bound
$M \ge \abs{Z}$ on the energy of states in the $d=4$ theory.

In this section we study these corrections in examples in which all of the BPS particles are mutually local.
This is much more tractable than the general situation, because we can choose a duality
frame in which these particles are all
electrically charged, and hence we can work completely within an effective Lagrangian description.

For most of the section we specialize further to the free $U(1)$
gauge theory coupled to a single charged hypermultiplet.  In
addition to being the simplest example, this theory is physically
relevant because it describes the physics near a generic singularity
in $\vm$, where one BPS particle becomes much lighter than the
others.

\subsection{The exact single-particle metric}

We consider a $U(1)$ gauge theory on $\IR^3 \times S^1$, coupled to
a single hypermultiplet of charge $q>0$ (along with its CPT
conjugate of charge $-q$). The metric we will describe has been
considered previously in \cite{Ooguri:1996me,Seiberg:1996ns}.

The moduli space $\vm$ of the $d=4$
theory is coordinatized by the vector multiplet scalar $a \in \IC$.
More precisely, $\vm$ is only
an open patch in $\IC$, because the $d=4$ theory is not asymptotically free:  there is a
cutoff at $\abs{a} \sim \abs{\Lambda}$.

As we explained in Section \ref{sec:reduction}, the moduli space $\CM$ of the $d=3$ theory is
a 2-torus fibration over $\vm$.
The torus fibers $\CM_a$ are coordinatized (temporarily ignoring the subtlety
about quadratic refinements) by the electric Wilson line $\theta_\elec$ and the magnetic Wilson line
$\theta_\magn$, both with periodicity $2 \pi$.

The semiflat metric $g^\sf$ has an action of $U(1)^2$ by
isometries, because shifts of $\theta_\elec$ and $\theta_\magn$ are exact symmetries.
The electrically charged hypermultiplet couples to $\theta_\elec$, and hence breaks the isometry which
shifts it.  However, there are no magnetically charged BPS states in the theory, so shifts of $\theta_\magn$ are still
exact isometries.  The corrected metric $g$ is therefore of Gibbons-Hawking form.

For comparison with the Gibbons-Hawking ansatz we
introduce a vector $\vec{x}$ by
\begin{equation}
a = x^1 + i x^2, \quad \theta_\elec = 2 \pi R x^3.
\end{equation}
$\theta_\magn$ is a local coordinate on a $U(1)$ bundle over the open subset of $\IR^2 \times S^1$ parameterized by $\vec{x}$.  The metric is
\begin{equation}
g = V(\vec{x})^{-1} \left(\frac{d \theta_\magn}{2\pi} + A(\vec{x})\right)^2 + V(\vec{x}) d\vec{x}^2,
\end{equation}
where $V$ is a positive harmonic function, to be calculated below, and $A$ is a $U(1)$ connection with curvature
\begin{equation} \label{eq:FdV}
F = \star dV.
\end{equation}
This is a slight generalization of the standard Gibbons-Hawking ansatz, in which one
takes $\vec{x}$ to lie in (an open subset of) $\IR^3$.
(We can first work over a suitable subset of $\IR^3$ and then divide by
a $\IZ$-action on the total space
which shifts $x^3$.)
In the standard ansatz all $A$ obeying \eqref{eq:FdV}
are gauge equivalent and so define the same metric.  In our case this is not quite true:  there is one
additional gauge invariant degree of freedom associated to the holonomy around $S^1$.  This choice is related
to the choice of a $\theta$ angle in $d=4$.

$V(\vec{x})$ in our case can be
calculated by integrating out the charged hypermultiplet at one
loop.  Reference \cite{Seiberg:1996ns} asserts a nonrenormalization
theorem which implies that the computation is exact. The resulting
$V$   is a harmonic function with $q$ singularities in $\IR^2 \times
S^1$. The periodicity in $\theta_\elec$ arises because one sums over
the Kaluza-Klein momenta of the charged hypermultiplet on $S^1$:
\begin{equation} \label{eq:V-sum}
V = \frac{q^2 R}{4 \pi} \sum_{n=-\infty}^\infty \left( \frac{1}{\sqrt{q^2 R^2 \abs{a}^2 + (q \frac{\theta_\elec}{2 \pi} + n)^2}} - \kappa_n \right)
\end{equation}
Here $\kappa_n$ is a regularization constant introduced to make the sum converge.
Poisson resummation of \eqref{eq:V-sum} shows that
\begin{equation}
V = V^\sf + V^\inst,
\end{equation}
with
\begin{align}
V^\sf &= -\frac{q^2 R}{4\pi} \left(\log  \frac{a}{\Lambda} + \log \frac{\bar{a}}{\bar{\Lambda}} \right), \\
V^\inst &= \frac{q^2 R}{2\pi} \sum_{n\neq 0} e^{i n q \theta_\elec} K_0(2 \pi R \abs{n q a}). \label{eq:V-inst}
\end{align}
Here $\Lambda$ is an ultraviolet cutoff related to the choice of
$\kappa_n$. \footnote{For example, if we choose $\kappa_n = (\vert
\tilde \Lambda\vert^2 + n^2)^{-1/2}$, then we can choose $\Lambda =
(qR)^{-1} \tilde\Lambda \exp[-2 \sum_{m=1}^\infty K_0(2\pi m \vert
\tilde \Lambda\vert )]$.}

To specify the metric fully we must also give $A(\vec{x})$ obeying
\eqref{eq:FdV}:
\begin{equation}
A = A^\sf + A^\inst,
\end{equation}
where
\begin{align}
A^\sf &= \frac{i q^2}{8 \pi^2} \left( \log \frac{a}{\Lambda} - \log \frac{\bar{a}}{\bar{\Lambda}} \right) d\theta_\elec, \label{eq:A0} \\
A^\inst &=  - \frac{q^2 R}{4 \pi} \left( \frac{da}{a} - \frac{d \bar a}{\bar a} \right) \sum_{n \neq 0} (\sgn n) e^{i n q \theta_\elec} \abs{a} K_1(2 \pi R \abs{n q a}).  \label{eq:A-inst}
\end{align}
At large $R$ the leading terms in $V$ and $A$ are   $V^\sf$ and
$A^\sf$. Keeping only these terms, $g$ becomes the semiflat metric
with
\begin{equation}
\tau = \frac{q^2}{2 \pi i} \log \frac{a}{\Lambda}.
\end{equation}
This is the running coupling which comes from integrating out the hypermultiplet in $d=4$.

The subleading terms $V^\inst$, $A^\inst$ yield corrections to the
semiflat metric.  They have the form of an instanton expansion as
we expected, because of the asymptotic behavior $K_\nu(x) \sim
\sqrt{\frac{\pi}{2x}} e^{-x}$ for $x \to +\infty$.   They also break
the translation invariance along $\theta_\elec$ as expected.  Finally,
they improve the singular behavior.  Recall that in $g^\sf$ there is a
singularity at $a=0$.  From
\eqref{eq:V-sum} we see that the only possible singularities of $g$
occur at $a = 0$, $q \theta_\elec = 2 \pi n$.  Studying
the metric near these points we find that there is an $A_{q-1}$
conical singularity at each one.  So the singularity
in $g^\sf$ is replaced by $q$ higher-codimension
singularities in $g$.
In the simplest case $q=1$, the singularity is completely smoothed.

\subsubsection*{Global issues}

There is a subtle issue regarding the global definition of
the coordinate $\theta_\magn$. We have chosen a gauge which is
convenient for discussing the periodicity in $\theta_\elec$. However, the
presence of the logarithm in $A^\sf$ signals that this gauge is
singular at $a = 0$.  Moreover $A^\sf$ shifts by $- \frac{q^2}{2
\pi} d \theta_\elec$  upon continuation around the origin $a \to e^{2 \pi
i} a$. This shift must be compensated by a gauge transformation
\begin{equation}
\theta_\magn \to \theta_\magn + q^2 \theta_\elec + C.
\end{equation}
To fix $C$ we make a gauge transformation to a new coordinate $\theta_\magn'$:
\begin{equation}
\theta_\magn' = \theta_\magn + \frac{i}{4 \pi} \left( \log \frac{a}{\Lambda} -
\log \frac{\bar a}{\bar\Lambda} \right) (q^2 \theta_\elec + C).
\end{equation}
The transformed $\theta_\magn'$ is single-valued as $a$ goes around the origin.
The gauge transformed $A^\sf$ is
\begin{equation}
(A')^\sf = - \frac{i}{4 \pi} \left( \frac{da}{a} - \frac{d \bar a}{\bar a} \right) (q^2 \theta_\elec + C).
\end{equation}
Now we focus on the behavior at $q \theta_\elec = \pi$.  Here we have
$A^\inst = 0$, so the exact gauge field is just given by $(A')^\sf$.
On the other hand, once the instanton corrections are included,
there is no singularity either of the metric or of the $U(1)$ bundle
at this point (recall that the only singularities occur at $a=0$, $q
\theta_\elec = 2 \pi n$.)  Since moreover $\theta_\magn'$ is single-valued, it
follows that $(A')^\sf$ cannot have a singularity here even if we
go to $a=0$ (or more
precisely the only allowed singularity is a quantized Dirac string),
which implies
\begin{equation}
C = - q \pi + 2 \pi  k
\end{equation}
for some integer $k$.
So we conclude that as we go around $a = 0$ the angular coordinates shift by
\begin{subequations}
\begin{align} \label{eq:angle-monodromy}
\theta_\elec &\to \theta_\elec, \\
\theta_\magn &\to \theta_\magn + q^2 \theta_\elec - q \pi. \label{angle-monodromy-varphi}
\end{align}
\end{subequations}
The shift by $q^2 \theta_\elec$ is as expected from the monodromy of the torus fibration.
The shift by $-q \pi$ is more surprising, but fits into our discussion in the end of Section
\ref{sec:reduction}, where we proposed that the Wilson lines are well defined only
after choosing a local quadratic refinement $\sigma$.  So far in this section we have
chosen the ``standard'' refinement $\sigma(\gamma_e,\gamma_m) = (-1)^{\gamma_e \gamma_m}$.
The monodromy shifts
$\gamma_e \to \gamma_e + q^2 \gamma_m$, and hence replaces $\sigma$ by
$\sigma'(\gamma_e,\gamma_m) = (-1)^{q^2 \gamma_m^2} \sigma(\gamma_e,\gamma_m)$.  This
change of refinement is compensated by the shift of $\theta_\magn$ by $-q \pi$.

\subsection{Hyperk\"ahler structure}

Next we want to describe $\CM$ as a \hk manifold.
The \hk structure of any Gibbons-Hawking metric is determined by the triplet of symplectic forms
\begin{equation}
\omega^\alpha = d x^\alpha \wedge \left(\frac{d \theta_\magn}{2\pi} +
A(\vec{x})\right) + \half \eps^{\alpha \beta \gamma} V dx^\beta
\wedge dx^\gamma.
\end{equation}
The holomorphic symplectic form \eqref{eq:Omegazeta} is then
\begin{equation}
\hsymp(\zeta) = \frac{1}{4 \pi^2 R} \xi_\magn \wedge \xi_\elec
\end{equation}
where
\begin{align}
\xi_\magn &= i d\theta_\magn + 2 \pi i A(\vec{x}) + \pi i V \left(\frac{1}{\zeta} da - \zeta d\bar{a}\right), \label{eq:xi-magn} \\
\xi_\elec &= i d\theta_\elec + \pi R \left(\frac{1}{\zeta} da + \zeta d\bar{a}\right). \label{eq:xi-elec}
\end{align}
In particular it follows that $\xi_\elec$ and $\xi_\magn$ are of type $(1,0)$.

Moreover, $\xi_\elec$ can be written as
\begin{equation}
\xi_\elec = \frac{d\CX_\elec}{\CX_\elec}
\end{equation}
where
\begin{equation} \label{eq:X1-def}
\CX_\elec = \exp \left[ \pi R \frac{a}{\zeta} + i \theta_\elec + \pi R \zeta \bar{a} \right].
\end{equation}
So $\CX_\elec$ is a holomorphic function on $\CM$ in complex structure $J^\pz$.  Notice that it coincides with
the semiflat coordinate $\CX^\sf_{\gamma}$ given in \eqref{eq:X-sf}, if we choose $\gamma$ to be the
unit electric charge, since in that case $Z_\gamma = a$ and $\theta_\gamma = \theta_\elec$.
In other words, the ``electric'' complex coordinate is unaffected
by the instanton corrections due to the electrically charged particle,
\begin{equation}
\CX_\elec = \CX^\sf_\elec.
\end{equation}

To finish describing the complex geometry of $\CM$ one should construct a second ``magnetic''
complex coordinate $\CX_\magn$, such that
\begin{equation} \label{eq:second-coord}
\hsymp(\zeta) =- \frac{1}{4 \pi^2 R}\frac{d \CX_\elec}{\CX_\elec} \wedge \frac{d \CX_\magn}{\CX_\magn}.
\end{equation}
Such a $\CX_\magn$ is necessarily of the form
\begin{equation} \label{eq:mag-form}
\CX_\magn = e^{i \theta_\magn + \Phi(a, \bar{a},\theta_\elec, \zeta)}.
\end{equation}

The most obvious way of constructing $\CX_\magn$ would be to write out the Cauchy-Riemann equations on $\CM$
and look for a particular solution of the form \eqref{eq:mag-form}.
In the next section we follow a different approach:  we give a particular solution for $\CX_\magn$ directly,
in a form which will be especially convenient for what follows, and then rather than checking the
Cauchy-Riemann equations we check \eqref{eq:second-coord} directly.

\subsection{The solution for $\CX_\magn$} \label{sec:solving-xmagn}

Now we specialize to our $\CM$.  In this case we have
\begin{equation}
\hsymp(\zeta) = \hsymp^\sf(\zeta) + \hsymp^\inst(\zeta)
\end{equation}
where
\begin{align}
\hsymp^\sf(\zeta) &= - \frac{1}{4 \pi^2 R} \xi_\elec \wedge \left[ i d \theta_\magn + 2 \pi i A^\sf + \pi i V^\sf \left(\frac{1}{ \zeta} da -  \zeta d\bar{a}\right) \right],  \\
\hsymp^\inst(\zeta) &= - \frac{1}{4 \pi^2 R} \xi_\elec \wedge \left[ 2 \pi i A^\inst +
\pi i V^\inst \left(\frac{1}{\zeta} da -  \zeta d\bar{a}\right)\right]. \label{eq:instanton-corrections-Omega}
\end{align}
If we neglect the instanton corrections, the desired magnetic coordinate is
\begin{equation} \CX_\magn^\sf(\zeta)= \exp\left[-i \frac{R q^2}{2 \zeta} \left(a
\log\frac{a}{\Lambda} -a\right) + i \theta_\magn + i \frac{\zeta R q^2}{2} \left(\bar a
\log\frac{\bar a}{\bar \Lambda} - \bar a\right)\right].
\end{equation}
This coincides with the expression
\eqref{eq:X-sf} for the holomorphic coordinate $\CX^\sf_\gamma$ in the semiflat geometry, if we choose $\gamma$ to be
the unit magnetic charge, with $Z_\gamma = \frac{q^2}{2 \pi i} (a \log \frac{a}{\Lambda} - a)$
and $\theta_\gamma = \theta_\magn$.  A direct computation verifies that
\begin{equation}
\frac{d \CX^\sf_\magn}{\CX^\sf_\magn} = \left[ i d \theta_\magn + 2 \pi i A^\sf + \pi i V^\sf \left(\frac{1}{\zeta} da -  \zeta d\bar{a}\right) \right] - \frac{i q^2}{4 \pi}\left( \log \frac{a}{\Lambda} - \log \frac{\bar{a}}{\bar{\Lambda}} \right) \frac{d \CX_\elec}{\CX_\elec},
\end{equation}
and hence in particular
\begin{equation} \label{eq:Omega-sf-repeat}
\hsymp^\sf(\zeta) = - \frac{1}{4 \pi^2 R} \frac{d\CX_\elec}{\CX_\elec} \wedge \frac{d\CX_\magn^\sf}{\CX_\magn^\sf},
\end{equation}
as expected.

Notice that $\CX_\magn^\sf$ has a nontrivial monodromy around $a=0$:
the monodromies of $\log a$ and $\log \bar a$ combine with the monodromy
of $e^{i \theta_\magn}$ given in \eqref{angle-monodromy-varphi} to give
\begin{equation}\label{eq:mono2}
\CX_\magn^\sf \to (-1)^q \CX_\elec^{q^2} \CX_\magn^\sf.
\end{equation}

Next we include the instanton corrections.  As we will demonstrate below, we can give the desired $\CX_\magn$
obeying \eqref{eq:second-coord} by the integral formula
\begin{equation}\label{eq:FullX2}
\begin{split}
\CX_\magn = \CX_\magn^\sf \exp\Biggl[ \frac{i q}{4\pi} & \int_{\ell_+} \frac{d \zeta'}{\zeta'}
\frac{\zeta' + \zeta}{\zeta' - \zeta} \log[1-\CX_\elec(\zeta')^q]  \\
- \frac{i q}{4\pi} & \int_{\ell_-}
 \frac{d \zeta'}{\zeta'} \frac{\zeta' + \zeta}{\zeta' - \zeta} \log[1-\CX_\elec(\zeta')^{-q}] \Biggr],
\end{split}
\end{equation}
where we choose the contours $\ell_\pm$ to be any paths connecting $0$ to $\infty$ which lie in the
two half-planes
\begin{equation}
\CU_\pm = \left\{ \zeta: \pm {\rm Re} \frac{a}{\zeta}<0 \right\}.
\end{equation}
The two integral contributions in \eqref{eq:FullX2} come respectively from instanton corrections of positive and negative winding around $S^1$.

In the rest of this section we verify that \eqref{eq:FullX2} is indeed correct.  This amounts to verifying
\begin{equation} \label{eq:tomatch}
- \frac{1}{4 \pi^2 R}\frac{d \CX_\elec}{\CX_\elec} \wedge \frac{d \CX_\magn}{\CX_\magn} = \hsymp(\zeta).
\end{equation}
From \eqref{eq:FullX2} we have
\begin{equation} \label{eq:dx2x2}
\frac{d \CX_\magn}{\CX_\magn} = \frac{d \CX^\sf_\magn}{\CX^\sf_\magn} + \CI_+ + \CI_-
\end{equation}
where
\begin{equation} \label{eq:ipm}
\CI_\pm = -\frac{i q^2}{4\pi} \int_{\ell_\pm} \frac{d \zeta'}{\zeta'}
\frac{\zeta' + \zeta}{\zeta' - \zeta} \left[ \frac{\CX_\elec(\zeta')^{\pm q}}{1-\CX_\elec(\zeta')^{\pm q}} \frac{d\CX_\elec(\zeta')}{\CX_\elec(\zeta')}\right].
\end{equation}
(Here we used the fact that the integrals in \eqref{eq:FullX2} depend on $a,\bar a,\theta_\magn,\theta_\elec$ only
through $\CX_\elec(\zeta')$, and are absolutely convergent, so we are free to bring the differential $d$ inside.)
Combining \eqref{eq:tomatch}, \eqref{eq:dx2x2}, and \eqref{eq:Omega-sf-repeat}, we see that
the integrals $\CI_\pm$ need to give the instanton part $\hsymp^\inst(\zeta)$ on the right side of
\eqref{eq:tomatch}, i.e. we need
\begin{equation} \label{eq:intermediate}
\frac{d \CX_\elec(\zeta)}{\CX_\elec(\zeta)} \wedge (\CI_+ + \CI_-) = \frac{d \CX_\elec(\zeta)}{\CX_\elec(\zeta)} \wedge \left[ 2 \pi i A^\inst + \pi i V^\inst \left(\frac{1}{\zeta} da - \zeta d\bar{a}\right)\right].
\end{equation}
To check this we first note that
\begin{equation}
\frac{d \CX_\elec(\zeta)}{\CX_\elec(\zeta)} \wedge \CI_\pm = -\frac{i q^2}{4 \pi} \int_{\ell_\pm} \frac{d \zeta'}{\zeta'}
 \left( \frac{\zeta' + \zeta}{\zeta' - \zeta} \frac{d \CX_\elec(\zeta)}{\CX_\elec(\zeta)} \wedge \frac{d\CX_\elec(\zeta')}{\CX_\elec(\zeta')} \right) \left[ \frac{\CX_\elec(\zeta')^{\pm q}}{1-\CX_\elec(\zeta')^{\pm q}} \right]
\end{equation}
and the two-form which appears here can be rewritten,
\begin{align}\label{eq:rearrange}
\frac{\zeta' + \zeta}{\zeta' - \zeta} \frac{d \CX_\elec(\zeta)}{\CX_\elec(\zeta)} \wedge \frac{d \CX_\elec(\zeta')}{\CX_\elec(\zeta')} &=
\frac{\zeta' + \zeta}{\zeta' - \zeta} \frac{d \CX_\elec(\zeta)}{\CX_\elec(\zeta)} \wedge \left[\frac{d \CX_\elec(\zeta')}{\CX_\elec(\zeta')}-\frac{d \CX_\elec(\zeta)}{\CX_\elec(\zeta)}\right]\\
&=  - \pi R \frac{d \CX_\elec(\zeta)}{\CX_\elec(\zeta)} \wedge \left[\left(\frac{1}{\zeta'} + \frac{1}{\zeta}\right) da - (\zeta'+\zeta)d\bar a\right],
\end{align}
using the explicit form \eqref{eq:X1-def} of $\CX_\elec$.
Hence the left side of \eqref{eq:intermediate} becomes
\begin{equation} \label{eq:intermediate-2}
\begin{split}
\frac{i q^2 R}{4} \frac{d \CX_\elec(\zeta)}{\CX_\elec(\zeta)} \wedge
\biggl( & \int_{\ell_+}  \frac{d \zeta'}{\zeta'}  \left[\left(\frac{1}{\zeta'}+\frac{1}{\zeta}\right)da - ( \zeta'+\zeta)d\bar a\right]\frac{\CX_\elec(\zeta')^{q}}{1-\CX_\elec(\zeta')^{q}}  \\
+ & \int_{\ell_-}  \frac{d \zeta'}{\zeta'}  \left[\left(\frac{1}{\zeta'}+\frac{1}{\zeta}\biggr)da - ( \zeta'+\zeta)d\bar a\right]\frac{\CX_\elec(\zeta')^{-q}}{1-\CX_\elec(\zeta')^{-q}}
\right).
\end{split}
\end{equation}

Now we are ready to evaluate the integrals.
It is convenient first to deform each of the contours $\ell_\pm$ to a canonical choice
lying exactly in the middle of $\CU_\pm$, i.e. to choose
\begin{equation}
\ell_\pm = \left\{ \zeta: \pm \frac{a}{\zeta} \in \IR_- \right\}.
\end{equation}
We first consider the terms which multiply
$\zeta$ or $\frac{1}{\zeta}$.  Expanding the geometric series we obtain:
\begin{align}
\int_{\ell_+}  \frac{d \zeta'}{\zeta'} \frac{\CX_\elec(\zeta')^{q}}{1-\CX_\elec(\zeta')^{q}} &=
\sum_{n>0} \int_{\ell_+} \frac{d \zeta'}{\zeta'}
\exp \left[ \pi R q n\frac{a}{\zeta'} + i q n\theta_\elec + \pi R q n \zeta' \bar{a} \right] \\
&= \sum_{n>0} 2 e^{i q n\theta_\elec}K_0(2 \pi R q|na|).
\end{align}
The integral over $\ell_-$ in \eqref{eq:intermediate-2} gives a similar sum over $n<0$.  Altogether
we find that the terms which multiply $\zeta$ or $\frac{1}{\zeta}$ in \eqref{eq:intermediate-2} equal
\begin{equation} \label{eq:final-part-1}
\frac{d\CX_\elec}{\CX_\elec} \wedge \left( \frac{i q^2 R}{2} \sum_{n \neq 0} e^{i q n\theta_\elec}K_0(2 \pi R q|na|) \right) \left(\frac{1}{\zeta} da - \zeta d\bar{a}\right) = \frac{d\CX_\elec}{\CX_\elec} \wedge i \pi V^\inst\left(\frac{1}{\zeta} da - \zeta d\bar{a}\right).
\end{equation}
For the remaining terms in \eqref{eq:intermediate-2} we use similarly
\begin{align}
\int_{\ell_+}  \frac{d \zeta'}{\zeta'} \zeta' \frac{\CX_\elec(\zeta')^{q}}{1-\CX_\elec(\zeta')^{q}} &=
\sum_{n>0} \int_{\ell_+} \frac{d \zeta'}{\zeta'}
\zeta' \exp \left[ \pi R q n\frac{a}{\zeta'} + i q n\theta_\elec + \pi R q n \zeta' \bar{a} \right] \\
&= - \sum_{n>0} 2 \frac{\abs{a}}{\bar{a}} e^{i q n\theta_\elec}K_1(2 \pi R q|na|)
\end{align}
and
\begin{align}
\int_{\ell_+}  \frac{d \zeta'}{\zeta'} \frac{1}{\zeta'} \frac{\CX_\elec(\zeta')^{q}}{1-\CX_\elec(\zeta')^{q}} &=
\sum_{n>0} \int_{\ell_+} \frac{d \zeta'}{\zeta'}
\frac{1}{\zeta'} \exp \left[ \pi R q n\frac{a}{\zeta'} + i q n\theta_\elec + \pi R q n \zeta' \bar{a} \right] \\
&= - \sum_{n>0} 2 \frac{\abs{a}}{a} e^{i q n\theta_\elec}K_{1}(2 \pi R q|na|).
\end{align}
Combining these with their counterparts from the integral over $\ell_-$ (which come with an extra minus sign),
we see that these terms in \eqref{eq:intermediate-2} equal
\begin{equation} \label{eq:final-part-2}
\frac{d\CX_\elec}{\CX_\elec} \wedge \left( - \frac{i q^2 R}{2} \sum_{n \neq 0} e^{i q n\theta_\elec} (\sgn n) \abs{a} K_1(2 \pi R q|na|) \right) \left( \frac{da}{a} - \frac{d \bar a}{\bar a} \right) = \frac{d\CX_\elec}{\CX_\elec} \wedge 2 \pi i A^\inst.
\end{equation}
So finally, summing \eqref{eq:final-part-1} and \eqref{eq:final-part-2}, we obtain \eqref{eq:intermediate}
as desired:  differentiating the contour integrals in $\CX_\magn$ has correctly produced the instanton corrections
$V^\inst$ and $A^\inst$.
This finishes the check that $\CX_\magn$ is the desired ``magnetic'' complex coordinate on $\CM$.

\bigskip
\textbf{Remark}:  $\CX_\magn$ is closely related to the so-called
``$Q$ function'' in the theory of quantum integrable
systems.\footnote{We thank S. Lukyanov for sharing his notes on
these functions with us.} We feel this is not a coincidence and
points to some deep relation to integrable field theories. This
feeling is reinforced by the fact that the crucial equation
\eqref{eq:X-integral-mult-explicit} below is a form of the
Thermodynamic Bethe Ansatz, as explained in Appendix \ref{app:TBA}.

\subsection{Analytic properties}

We now consider the analytic behavior of the pair $(\CX_\magn, \CX_\elec)$ on the $\zeta$-plane.

For $\CX_\elec$ the story is simple:  it is analytic for $\zeta \in
\IC^\times$, with essential singularities at $\zeta=0,\infty$.  For
$\CX_\magn^\sf$ the same is true, but for the full $\CX_\magn$ the
story is more intricate:  the integrals in \eqref{eq:FullX2} are
analytic in $\zeta$ only away from the contours $\ell_\pm$.  As
$\zeta$ crosses either of these contours, the pole in the integrand
crosses the path of integration. Therefore our expression for
$\CX_\magn$ defines a \ti{piecewise} analytic function, with the
discontinuity determined by the residue of the pole.
Introduce the notation $(\CX_\magn)_{\ell_+}^{+}$, $(\CX_\magn)_{\ell_+}^{-}$ for the limit of $\CX_\magn$
as $\zeta$ approaches $\ell_+$ in the clockwise or counterclockwise direction respectively, and similar
notation for $\ell_-$.  The discontinuity is then given by
\begin{subequations}  \label{eq:X2-disc}
\begin{align}
(\CX_\magn)_{\ell_+}^{+} &= (\CX_\magn)_{\ell_+}^{-} (1 - \CX_\elec^q)^{-q}, \label{eq:X2-disc-a} \\
(\CX_\magn)_{\ell_-}^{+} &= (\CX_\magn)_{\ell_-}^{-} (1 - \CX_\elec^{-q})^q.
\end{align}
\end{subequations}

These discontinuities will play a crucial role for us below:  indeed we will identify them with
Kontsevich-Soibelman symplectomorphisms, as follows.
We consider the pair of complex functions $(\CX_\magn, \CX_\elec)$
as giving a map
\begin{equation}
\CX: \CM_a \to T_a
\end{equation}
from the real 2-torus $\CM_a$ coordinatized by $(\theta_\magn, \theta_\elec)$
to a complexified 2-torus $T_a$ coordinatized by
$(X_\magn, X_\elec)$. The map $\CX$ varies as a function of $\zeta$
(and $a, \bar{a}, R$).  In Section \ref{sec:ReviewKS} we introduced
the Kontsevich-Soibelman factors $\CK_\gamma$ as symplectomorphisms of
$T_a$.  Our discontinuities \eqref{eq:X2-disc} say that at the ray
$\ell_\pm$, $\CX^{+}$ and $\CX^{-}$ differ by composition with
$\CK_{0,\pm q}$.

An interesting phenomenon has occurred here.  Consider the monodromy of $\CX_\magn$ in the $a$-plane around $a=0$.
This monodromy receives two contributions:  the monodromy of $\CX_\magn^\sf$ given in \eqref{eq:mono2}
and the contributions from
\eqref{eq:X2-disc}.  These two contributions actually cancel one another!  This fact is essentially related
to the fact that the singularity of the semiflat metric at $a=0$ has been smoothed out.
On the other hand, if we analytically continue $\CX_\magn$ around $\zeta = 0$ it does not
come back to itself.  This monodromy does not create any problems.
In particular, $\hsymp(\zeta)$ does behave well near $\zeta = 0$:  it just has a simple
pole, as one expects from the discussion in Section \ref{sec:hk-general}.

Now let us consider the asymptotics of $\CX_\elec, \CX_\magn$ as $\zeta \to 0,\infty$.
The asymptotics of $\CX_\elec$ can be trivially read off from \eqref{eq:X1-def},
\begin{equation} \label{eq:X1-asymptotics}
\CX_\elec \sim \begin{cases}
               \exp \left[ \pi R \frac{a}{\zeta} + i \theta_\elec \right] & \text{as } \zeta \to 0, \\
               \exp \left[ \pi R \zeta\bar{a} + i \theta_\elec \right] & \text{as } \zeta \to \infty.
               \end{cases}
\end{equation}
For $\CX_\magn$ the asymptotics are more interesting.
As $\zeta \to 0,\infty$ the integrand of
\eqref{eq:FullX2} simplifies:  the rational function just reduces to $\pm 1$.
Then expanding the logarithm and evaluating the integral gives
\begin{equation} \label{eq:X2-asymptotics}
\CX_\magn \sim \begin{cases}
               \exp \left[ - i \frac{R q^2}{2\zeta} (a \log (a/\Lambda) - a) + i \theta_\magn + \frac{q}{2\pi i} \sum_{s \neq 0} \frac{1}{s} e^{i s q \theta_\elec} K_0(2 \pi R q \abs{sa}) \right] & \text{as } \zeta \to 0, \\
               \exp \left[ i \frac{\zeta R q^2}{2} (\bar{a} \log (\bar{a}/\bar{\Lambda}) - \bar{a}) + i \theta_\magn - \frac{q}{2\pi i} \sum_{s \neq 0} \frac{1}{s} e^{i s q \theta_\elec} K_0(2 \pi R q \abs{sa}) \right]  & \text{as } \zeta \to \infty.
               \end{cases}
\end{equation}
These asymptotics hold for all phases of $\zeta$.  The discontinuities \eqref{eq:X2-disc} along $\ell_\pm$ do
not lead to discontinuities in the asymptotics, because the jump is exponentially close to $1$ as $\zeta \to 0,\infty$
along $\ell_\pm$:  along $\ell_+$ we have $\CX_\elec \to 0$ exponentially fast, and along $\ell_-$, $\CX_\elec^{-1} \to 0$
exponentially fast.

On the other hand, we could also have defined a different function $\CX'_\magn$, by
analytically continuing $\CX_\magn$ across $\ell_+$ clockwise.
It follows from \eqref{eq:X2-disc-a} that on the clockwise side of $\ell_+$ we have
\begin{equation} \label{eq:continued-vs-jumping}
\CX'_\magn = \CX_\magn (1 - \CX_\elec^q)^{q}.
\end{equation}
Suppose now that we analytically continue $\CX'_\magn$ further,
clockwise to the boundary of $\CU_+$ and then across into $\CU_-$.  In $\CU_-$,
$\CX_\elec$ is exponentially large as $\zeta \to 0$.  So from \eqref{eq:continued-vs-jumping} it
follows that the $\zeta \to 0$ asymptotics of $\CX'_\magn$ and $\CX_\magn$
are different; in particular, $\CX'_\magn$ does \ti{not} obey \eqref{eq:X2-asymptotics}.  \emph{Thus, the asymptotics of
the analytic continuation of the function $\CX_\magn$ is not the
analytic continuation of the asymptotics.} This is the hallmark of
Stokes' phenomenon.

Altogether, we have been led to consider a map $\CX: \CM_a \to T_a$,
which depends holomorphically on $\zeta$, and exhibits Stokes
phenomena at $\zeta \to 0, \infty$, with Stokes factors given by
composition with the Kontsevich-Soibelman symplectomorphisms acting
on $T_a$.  The crucial idea of this paper is that this picture is
valid for general gauge theories, not just the abelian theory we
considered here; indeed, as we will see in Section
\ref{sec:constructing-moduli}, it automatically incorporates
multi-instanton effects from mutually non-local particles, and gives
the exact metric on $\CM$.

\subsection{Differential equations}

Above we saw that the \hk geometry of $\CM$ is naturally described
in terms of a map $\CX$ which exhibits Stokes phenomena.  Stokes
phenomena typically arise in the theory of linear ordinary
differential equations with irregular singular points. Indeed, in
our case there is such a differential equation
\begin{equation} \label{eq:de-zeta}
\zeta \partial_\zeta \CX = \CA_\zeta \CX,
\end{equation}
with an irregular singularity.
In this section we identify this equation.  In fact, at the same
time we will find a companion equation, governing the dependence on the radius of $S^1$,
\begin{equation} \label{eq:de-R}
R \partial_R \CX = \CA_R \CX.
\end{equation}

Equations of the form \eqref{eq:de-zeta}, \eqref{eq:de-R} are commonly encountered for finite-dimensional matrices $\CA_\zeta$, $\CA_R$, $\CX$.
Then $\CA_\zeta$ and $\CA_R$ act on $\CX$ by matrix multiplication from the left, and the Stokes factors act from the right,
so in particular the two actions commute.
In our case the solution $\CX$ is a map $\CM_a \to T_a$.  The Stokes factors act as diffeomorphisms of $T_a$.
$\CA_\zeta$ and $\CA_R$ act as infinitesimal diffeomorphisms of $\CM_a$, i.e. as differential operators
in $(\theta_\magn, \theta_\elec)$.  These two actions commute with one another because they act on different spaces.

Now what is the origin of the desired equations?  They should be related to some symmetries of $(\CM, g)$.
At first glance $(\CM, g)$ would appear to have a $U(1)$ symmetry which just maps $a \mapsto e^{i \theta} a$.  Such a symmetry would have an obvious physical origin:  it would come from a $U(1)_R$ symmetry of the theory
in $d=4$.  However, we know that this symmetry is actually anomalous once we include the matter hypermultiplet.
Indeed, $A^\sf$ from \eqref{eq:A0} contains the factor $\log(a/\Lambda)$, which is invariant only under a
\ti{simultaneous} rotation of $a$ and $\Lambda$.  This simultaneous rotation hence leaves the metric invariant.
It does not preserve the \hk forms $\vec{\omega}$, but rather rotates $\omega_1$ and $\omega_2$ into one another;
hence it leaves $\hsymp(\zeta)$ invariant if combined with the action $\zeta \mapsto e^{i \theta} \zeta$.
By inspection, both $\CX_\elec$ and $\CX_\magn$ are invariant under this combined rotation of $a$,
$\Lambda$ and $\zeta$, which leads to a differential equation:
\begin{equation} \label{eq:R-symmetry}
\plog{\zeta} \CX = \left( -\plog{\Lambda} + \plog{\bar\Lambda} - \plog{a} + \plog{\bar a} \right) \CX.
\end{equation}
Similarly the anomalous scale invariance of the $d=4$ theory leads to a symmetry which rescales $R$, $a$
and $\Lambda$:
\begin{equation} \label{eq:conformal-symmetry}
\plog{R} \CX = \left( \plog{\Lambda} + \plog{\bar\Lambda} + \plog{a} + \plog{\bar a} \right) \CX.
\end{equation}
These equations are not yet of the desired form \eqref{eq:de-zeta}, \eqref{eq:de-R} since they still
involve derivatives with respect to the parameters $(\Lambda, \bar\Lambda, a, \bar a)$.
So let us consider the dependence on these parameters.

The dependence of $\CX$ on $(a, \bar{a})$ is completely
determined in terms of the dependence on $(\theta_\elec, \theta_\magn)$,
by the requirement that $(\CX_\elec, \CX_\magn)$ are holomorphic in complex structure $J^\pz$.
Indeed, using the basis \eqref{eq:xi-magn}, \eqref{eq:xi-elec} for $(T^*)^{1,0} \CM$, we see that the
Cauchy-Riemann equations on $\CM$ are simply
\begin{align}
\partial_a \CX &= \CA_a \CX, \label{eq:de-cr-1} \\
\partial_{\bar a} \CX &= \CA_{\bar a} \CX, \label{eq:de-cr-2}
\end{align}
where the connection form $\CA$ is defined by
\begin{align}
\CA_a &= \frac{1}{\zeta} \left[ - i \pi R \partial_{\theta_\elec} + \pi (V + 2 \pi i R A_{\theta_\elec}) \partial_{\theta_\magn} \right] + 2 \pi A_a \partial_{\theta_\magn}, \\
\CA_{\bar a} &= 2 \pi A_{\bar a} \partial_{\theta_\magn} - \zeta \left[ i \pi R \partial_{\theta_\elec} + \pi (V - 2 \pi i R A_{\theta_\elec}) \partial_{\theta_\magn} \right].
\end{align}
We can similarly dispose of the $(\Lambda, \bar\Lambda)$ dependence.  First note that $\CX_\elec$ is simply independent
of $(\Lambda, \bar\Lambda)$.  For $\CX_\magn$ we have $\Lambda \frac{\partial \CX_\magn}{\partial \Lambda} = \frac{i R q^2 a}{2  \zeta} \CX_\magn$, and similarly for $\bar\Lambda$.  So writing
\begin{equation}
\CA_\Lambda = \frac{q^2 R a}{2 \zeta} \partial_{\theta_\magn}, \quad \CA_{\bar\Lambda} = \frac{ \zeta q^2 R \bar{a}}{2} \partial_{\theta_\magn},
\end{equation}
we have
\begin{align}
\Lambda \partial_\Lambda \CX &= \CA_\Lambda \CX, \\
\bar\Lambda \partial_{\bar\Lambda} \CX &= \CA_{\bar\Lambda} \CX.
\end{align}
We can now recast the equations
\eqref{eq:R-symmetry}, \eqref{eq:conformal-symmetry} in the desired form \eqref{eq:de-zeta}, \eqref{eq:de-R}, with
\begin{align}
\CA_\zeta &= -a \CA_a + \bar{a} \CA_{\bar a} - \Lambda \CA_{\Lambda} + \bar\Lambda \CA_{\bar\Lambda}, \label{eq:zeta-connection-abelian} \\
\CA_R &= a \CA_a + \bar{a} \CA_{\bar a} + \Lambda \CA_{\Lambda} + \bar\Lambda \CA_{\bar\Lambda}.
\end{align}
Now we come to the crucial point:  $\CA_\zeta$ as given in \eqref{eq:zeta-connection-abelian} depends on
$\zeta$ in a very simple way --- it has only simple poles at $\zeta = 0, \infty$:
\begin{equation}
\CA_\zeta = \frac{1}{\zeta} \CA^{(-1)}_\zeta + \CA^{(0)}_\zeta + \zeta \CA^{(1)}_\zeta.
\end{equation}
The equation
\eqref{eq:de-zeta} thus defines a meromorphic connection on $\IC\IP^1$, with two irregular
singularities of rank 1.  This motivates the appearance of Stokes phenomena, which we saw explicitly
in the previous section.

A family of differential equations very
similar to \eqref{eq:de-zeta}, \eqref{eq:de-R}, \eqref{eq:de-cr-1}, \eqref{eq:de-cr-2},
defining the ``$tt^*$ connection,'' appeared in \cite{Cecotti:1993rm,MR1213301} in the context
of the analysis and classification of $\CN=(2,2)$ field theories in $d=2$.  The similarity is
not just formal.  In particular, the interpretation of their equations for the $\zeta$ and $R$ dependence
was also in terms of $U(1)_R$ symmetry and scale transformations of the underlying field theory.
A crucial point of their analysis is a direct relation between the large $R$ asymptotics
of the connection $\CA$, the explicit form of the Stokes factors, and the degeneracies of BPS states in the $d=2$
theory.  There is a similar relation in our problem as well.  Indeed this relation is the key to understanding the Kontsevich-Soibelman wall-crossing formula.

A look at \cite{Cecotti:1993rm,MR1213301} also suggests a very useful technical tool for making further progress:
we should convert the differential equations into a Riemann-Hilbert problem for $\CX$, defined directly in terms of
the Stokes data and asymptotics as $\zeta \to 0, \infty$.  Using this tool we can immediately write down the
generalization to multiple mutually non-local BPS instantons.  We move to that problem in Section
\ref{sec:constructing-moduli}.

\subsection{Higher spin multiplets}

So far we have considered in some detail the corrections to $g$ which come from integrating out
a single electrically charged hypermultiplet.  One can ask similarly about the corrections due to
a single electrically charged higher spin multiplet --- for example the vector multiplet containing the massive $W$ boson.

In principle these corrections could be determined by a careful one-loop computation
in three dimensions.  Instead we exploit a trick:  we consider the massive vector multiplet of an
$\CN=4$ supersymmetric theory.  Decomposing under $\CN=2$ supersymmetry this multiplet contains
two hypermultiplets and one vector multiplet.
On the other hand, because of the higher supersymmetry in the $\CN=4$ theory, one expects that the metric on
$\CM$ will not receive any instanton corrections.  The reason is that
according to standard nonrenormalization
theorems the Higgs branch is uncorrected \cite{Argyres:1996eh}, but the nonanomalous
$R$-symmetry mixes the Higgs and Coulomb branches.  It follows that the
corrections from the $\CN=2$ vector multiplet must precisely cancel those
from the two $\CN=2$ hypermultiplets.  In other words, at least as far as these two $\CN=2$ multiplets are
concerned, the corrections are weighted by the helicity supertrace $\degen(\gamma;u)$.

More generally we may consider integrating out $\CN=2$ multiplets with arbitrary spin.
Let $a_j$ denote the weight multiplying the instanton correction from the $j$-th $\CN=2$ multiplet
($j=0$ for the hypermultiplet, $j=1$ for the
vector, \dots), normalized
to $a_0 = 1$.  We saw above that $a_1 = -2$.  Moreover, from the fact that the contribution from \ti{any}
multiplet of $\CN=4$ supersymmetry should vanish,
we get $a_{j+2} + 2 a_{j+1} + a_j = 0$.  This determines $a_j = (-1)^j (j+1)$, so indeed the instanton
corrections are weighted by the second helicity supertrace.

\subsection{Higher rank generalization} \label{sec:higher-rank}

All of our discussion can be
easily generalized to the case of a rank $r$ abelian gauge theory
coupled to a set of electrically charged hypermultiplets.
Let the charges be $q_I^{(s)}$, where $I=1, \dots, r$ runs over the electric
gauge fields, and $s$ labels the set of hypermultiplets.

There is a $4r$-dimensional generalization of the Gibbons-Hawking
ansatz, with base $(\IR^3)^r$ and a fiber $(S^1)^r$. We use
coordinates $(x^{\alpha I}) = \vec x^I$ for the base, $\theta_{\magn,I}$ for the fiber, and write
\begin{equation}
g = [V(\vec{x})^{-1}]^{IJ} \left(\frac{d \theta_{\magn,I}}{2\pi} +
A_I(\vec{x})\right)\left(\frac{d \theta_{\magn,J}}{2\pi} +
A_J(\vec{x})\right) + V(\vec{x})_{IJ} d\vec{x}^I d\vec{x}^J,
\end{equation}
where $A$ and $V$ are related by differential equations stating that
\begin{equation}
\omega^{\alpha} = dx^{\alpha I} \wedge \left( \frac{d\theta_{\magn,I}}{2\pi} + A_I
\right) + \half V_{IJ} \epsilon^{\alpha\beta\gamma} dx^{\beta I } \wedge
dx^{\gamma J}
\end{equation}
is a closed 2-form for $\alpha = 1,2,3$.

The 1-loop integral gives the natural result
\begin{equation}
\label{eq:V-sum-r} V_{IJ} =  \mathrm{Im} \tau^0_{IJ} + \sum_{s}
\frac{q^{(s)}_I q^{(s)}_J R}{4 \pi} \sum_{n=-\infty}^\infty \left(
\frac{1}{\sqrt{| q^{(s)}_K R a^K|^2 + (q^{(s)}_K \frac{\theta_\elec^K}{2
\pi} + n)^2}} - \kappa_n \right).
\end{equation}
Poisson resummation of \eqref{eq:V-sum} shows that
\begin{equation}
V_{IJ} = V_{IJ}^\sf + V_{IJ}^\inst,
\end{equation}
with
\begin{align}
V_{IJ}^\sf &= \mathrm{Im} \tau^0_{IJ} - \sum_s \frac{q^{(s)}_I
q^{(s)}_J R}{4\pi} \left(\log  \frac{q^{(s)}_K a^K}{\Lambda} + \log \frac{q^{(s)}_K \bar{a}^K}{\bar{\Lambda}} \right), \\
V_{IJ}^\inst &= \sum_s \frac{q^{(s)}_I q^{(s)}_J R}{2\pi}
\sum_{n\neq 0} e^{i n q^{(s)}_K \theta_\elec^K} K_0(2 \pi R \abs{n
q^{(s)}_I a^I}). \label{eq:V-inst-r}
\end{align}
Also
\begin{equation}
A_I = A_I^\sf + A_I^\inst,
\end{equation}
where
\begin{align}
A_I^\sf &= \mathrm{Re} \tau^0_{IJ} \frac{d \theta_\elec^J}{2 \pi} + \sum_s
\frac{i q^{(s)}_I
q^{(s)}_J}{8 \pi^2} \left( \log \frac{q^{(s)}_K a^K}{\Lambda} - \log \frac{q^{(s)}_K \bar{a}^K}{\bar{\Lambda}} \right) d\theta_\elec^J, \label{eq:A0-rankr} \\
A_I^\inst &=  - \frac{q^{(s)}_I q^{(s)}_J R}{4 \pi} \left(
\frac{da^J}{q^{(s)}_K a^K} - \frac{d\bar a^J}{q^{(s)}_K \bar a^K}
\right) \sum_{n \neq 0} (\sgn n) e^{i n q^{(s)}_I \theta_\elec^I}
\abs{q^{(s)}_K a^K} K_1(2 \pi R \abs{n q^{(s)}_K a^K}).
\label{eq:A-inst-rankr}
\end{align}
At large $R$ the leading terms in $V$ and $A$ are   $V^\sf$ and
$A^\sf$. Keeping only these terms, $g$ becomes the semiflat metric
with
\begin{equation}
\tau_{IJ} =\tau^0_{IJ} + \sum_s \frac{q^{(s)}_I q^{(s)}_J}{2 \pi i}
\log \frac{q^{(s)}_K a^K}{\Lambda}.
\end{equation}
This is the coupling which comes from integrating out the
hypermultiplets in $d=4$.

The holomorphic symplectic form is
\begin{equation}
\hsymp(\zeta) = - \frac{1}{4 \pi^2 R} \xi^I_e \wedge \xi_{m,I}
\end{equation}
where
\begin{align}
\xi_e^I &= i d \theta_\elec^I + \pi R \left(\frac{da^I}{\zeta} +
\zeta d \bar a^I \right), \\
\xi_{m,I} &= i d \theta_{\magn,I} + 2 \pi i A_I + i \pi V_{IJ} \left(
\frac{da^J}{\zeta} - \zeta d \bar a^J \right).
\end{align}
As before, the electric coordinates agree with their semiflat
approximation,
\begin{equation}
\CX^I_e = \exp \left[ \pi R \frac{a^I}{\zeta} + i \theta_\elec^I +
\pi R \zeta \bar a^I \right].
\end{equation}
The semiflat approximation to the magnetic ones is
\begin{multline}
\CX^\sf_{m,I} = \exp \Biggl[ \frac{\pi R}{\zeta} \left(\tau^0_{IJ}
a^J  + \sum_s \frac{q^{(s)}_I}{2 \pi i} q^{(s)}_K a^K \log
\frac{q^{(s)}_K a^K}{e \Lambda}\right) + i \theta_{\magn,I}\,+ \\
\pi R \zeta
\left( \tau^0_{IJ} \bar a^J + \sum_s \frac{q^{(s)}_I}{2 \pi
i} q^{(s)}_K \bar a^K \log \frac{q^{(s)}_K \bar a^K}{e \bar
\Lambda}\right) \Biggr].
\end{multline}
The full magnetic coordinates are given by the integral
formula
\begin{equation}\label{eq:FullX2-r}
\begin{split}
\CX_{\magn,I} = \CX_{\magn,I}^\sf \exp\Biggl[ \sum_s \frac{i
q^{(s)}_I}{4\pi} & \int_{\ell^s_+} \frac{d \zeta'}{\zeta'}
\frac{\zeta' + \zeta}{\zeta' - \zeta} \log[1-\prod_J \CX^J_\elec(\zeta')^{q^{(s)}_J}]  \\
- \frac{i q^{(s)}_I}{4\pi} & \int_{\ell^s_-}
 \frac{d \zeta'}{\zeta'} \frac{\zeta' + \zeta}{\zeta' - \zeta} \log[1-\prod_J \CX^J_\elec(\zeta')^{-q^{(s)}_J}] \Biggr],
\end{split}
\end{equation}
where $\ell^s_\pm$ are any paths connecting $0$ to $\infty$ which lie in the
two half-planes
\begin{equation}
\CU^s_\pm = \left\{ \zeta: \pm {\rm Re} \frac{a^K q^{(s)}_K}{\zeta}<0 \right\}.
\end{equation}

Notice that $\CX_{m,I}$ have discontinuities for each
hypermultiplet, which as before are given by the Kontsevich-Soibelman
symplectomorphisms $\CK_{0,q_I^{(s)}}$.

\section{Mutually non-local corrections} \label{sec:constructing-moduli}

As we have just seen in the simplest nontrivial case,
the exact \hk metric $g$ is not equal to the semiflat metric $g^\sf$,
because of the quantum corrections from instantons corresponding to $d=4$ BPS states.

In general we expect such a quantum correction for each charge $\gamma$
supporting a BPS state.
These corrections should be weighted by the BPS multiplicities $\degen(\gamma;u)$. However, we
know that $\degen(\gamma;u)$ can jump as $u$ crosses a wall of
marginal stability!  So there seems to be a puzzle:  will not the
quantum corrections to $g$ also jump discontinuously?  How is
this consistent with the field theory expectation that $g$
should be smooth?

In this section we will give an explicit
construction of the exact \hk metric $g$ for large $R$.  We will see that it
is indeed smooth, provided that the Kontsevich-Soibelman wall-crossing formula
is satisfied.  This is our physical interpretation of the WCF.

Expanding $g$ around $R \to \infty$, we find the resolution of our puzzle:
in addition to the contributions from single BPS particles, there are also multi-particle
contributions.  The discontinuity in the 1-particle contributions is compensated
by a discontinuity in the multi-particle sector.
See \cite{GarciaEtxebarria:2007zv} for a related discussion
in the $\CN=1$ context.

\subsection{Defining the Riemann-Hilbert problem} \label{sec:defining-rh}

We take our inspiration from the abelian theory studied in Section \ref{sec:periodicnut} and construct the
metric by solving a Riemann-Hilbert problem in the $\zeta$-plane.  We work initially at fixed
$u \in \vm$, away from the walls of marginal stability.  We also choose a fixed quadratic refinement
$\sigma$ at $u$.

The Riemann-Hilbert problem is formulated in
terms of maps $\CX$ from the real torus $\widetilde\CM_u$ to the complexified symplectic $2r$-torus
$\tilde T_u$ which we introduced in our review of the Kontsevich-Soibelman formula (Section \ref{sec:ReviewKS}).
Given any such $\CX$, we pull back the coordinate functions $X_\gamma$ on $\tilde T_u$ to give functions
$\CX_\gamma$ on $\widetilde\CM_u$, defined by $\CX_\gamma(\theta) = X_\gamma(\CX(\theta))$.
In particular, the $\CX^\sf_\gamma$ given in Section \ref{sec:semiflat-twistor} come from
a reference map $\CX^\sf$; it is the zeroth approximation to the $\CX$ we construct below.

To formulate the Riemann-Hilbert problem we need to fix the \ti{asymptotic behavior} of $\CX$ as $\zeta \to 0, \infty$
and its \ti{discontinuities} in the $\zeta$-plane.

We begin with the asymptotics.  Introduce
\begin{equation} \label{eq:def-upsilon}
\Upsilon := \CX (\CX^\sf)^{-1}.
\end{equation}
In this section we are using an unconventional notation for composition of maps: $(fg)(x)$
means $g(f(x))$.\footnote{One virtue of this notation can be seen by observing that the
diagram $A \stackrel{f}{\to} B \stackrel{g}{\to} C$ composes to $A \stackrel{fg}{\to} C$.  A second
virtue will become apparent in Section \ref{sec:differential-problem}.}
Thus $\Upsilon$ is a map from $\widetilde\CM_u$ to itself (or more precisely to its complexification).
Concretely $\Upsilon$ maps
\begin{equation} \label{eq:upsilon-concrete}
e^{i \theta_i} \mapsto \CX^i(\theta) \exp\biggl[  - \pi R
\frac{Z_i}{\zeta} - \pi R  \zeta \bar Z_i \biggr].
\end{equation}
We require that the limit of $\Upsilon$ as $\zeta \to 0$ and $\zeta \to \infty$ exists,
\begin{equation} \label{eq:RH-asymptotics}
\lim_{\zeta \to 0} \Upsilon = \Upsilon_0, \quad \lim_{\zeta \to \infty} \Upsilon = \Upsilon_\infty,
\end{equation}
and moreover obeys
\begin{equation} \label{eq:upsilon-reality}
\Upsilon_0 = \bar \Upsilon_\infty.
\end{equation}

Next we need to specify the discontinuities of $\CX$, considered as
a piecewise-analytic function of $\zeta$. Assume temporarily that
$u$ does not lie on any wall of marginal stability. The
discontinuities will be given in terms of the Kontsevich-Soibelman
symplectomorphisms $\CK_\gamma: \tilde T_u \to \tilde T_u$ associated to the BPS states.
To each ray $\ell$ through the origin in the $\zeta$-plane, we
associate a subset of $\clat_u$,
\begin{equation} \label{eq:clat-ell}
(\clat_u)_\ell := \{ \gamma: Z_\gamma(u)/\zeta \in \IR_- \text{ for } \zeta \in \ell \},
\end{equation}
and a corresponding product over BPS states:
\begin{equation}
S_\ell := \prod_{\gamma \in (\clat_u)_\ell} \CK_{\gamma}^{\degen(\gamma;u)}.
\end{equation}
(Since $u$ does not lie on a wall, $(\clat_u)_\ell$ is at most one-dimensional, and the $\CK_\gamma$ for $\gamma \in (\clat_u)_\ell$ all commute; hence we do not have to specify the ordering in this product.)
Since the charge lattice $\clat_u$ is countable, for all but a countable set of rays $\ell$ we have $(\clat_u)_\ell = \emptyset$ and thus $S_\ell = 1$. We refer to rays for which $(\clat_u)_\ell \neq \emptyset$
as ``BPS rays''.

The most canonical choice of discontinuities is to require that
\begin{equation} \label{eq:RH-jumps}
\CX^{+} = \CX^{-} S_\ell
\end{equation}
where $\CX^{+}$, $\CX^{-}$ are the limit of $\CX$ as $\zeta$
approaches $\ell$ clockwise, counterclockwise respectively. This is
the most straightforward generalization of what we found in Section
\ref{sec:periodicnut}:  there we found a map $\CX = (\CX_\magn,
\CX_\elec)$ which was sectionally analytic in $\zeta$, with two BPS
rays $\ell_\pm$ across which $\CX_\elec$ was continuous and $\CX_\magn$ jumped
according to \eqref{eq:X2-disc}.  These two BPS rays corresponded to
the single hypermultiplets of charge $(0,\pm q)$, and the discontinuity was
exactly \eqref{eq:RH-jumps}, with $S_{\pm \ell} = \CK_{0,\pm q}$.  We are now generalizing to
include many BPS particles, just by requiring jumps along many BPS
rays. In this more general situation there will be no $\CX_\gamma$
that is continuous everywhere.

We have now formulated our Riemann-Hilbert problem.
Its solution is not unique:  rather it is determined only up to a transformation
$\CX \to b \CX$, with $b$ an arbitrary diffeomorphism of $\widetilde\CM_u$.  We will fix
this ambiguity in a convenient way when we solve the problem in the next section.

This Riemann-Hilbert problem might appear a bit unconventional since
it is formulated in terms of $\CX$ and $S_\ell$, which are not linear maps, but more general maps of manifolds.
The concerned reader should feel free
to ``linearize'' the problem by considering, instead of $\CX$, the
operation $\CX^*$ of pullback $C^\infty(\tilde T_u) \to C^\infty(\widetilde\CM_u)$. The
price of doing so is that not every map $Q: C^\infty(\tilde T_u) \to
C^\infty(\widetilde\CM_u)$ can be obtained as $\CX^*$ for some map $\CX$; so
if we find a solution $Q$ to the linear version of the
Riemann-Hilbert problem, we face the extra difficulty of checking
that $Q = \CX^*$ for some $\CX$.  Fortunately the ``functoriality''
of the Riemann-Hilbert problem comes to the rescue.  $Q$ will be
$\CX^*$ for some $\CX$ if and only if it preserves multiplication,
$Q(fg) = Q(f) Q(g)$. Since all the data defining the linear problem
is compatible with this structure, the solution is as well.

Finally let us discuss a reality property of our problem, which will be crucial for our construction of
the \hk metric.
Thanks to the relations $\degen(\gamma;u) = \degen(-\gamma;u)$ our discontinuity conditions
enjoy a discrete symmetry:  given any solution
$\CX$, we can obtain another solution which we call $\bar\CX$ by
\begin{equation}
\bar\CX_\gamma(\zeta) = \overline{\CX_{-\gamma}(-1/\bar\zeta)}.
\end{equation}
We claim that in fact our solution is invariant under this transformation,
\begin{equation} \label{eq:reality-X}
\CX = \bar \CX.
\end{equation}
To see this, consider the map $Y = \bar \CX \CX^{-1}$.  Because both $\bar \CX$ and $\CX$ have the same
discontinuities, $Y$ is actually \ti{analytic} in $\zeta$.  On the other hand, because of our asymptotic
condition \eqref{eq:upsilon-reality}, $Y \to 1$ as $\zeta \to 0,\infty$.  Therefore by Liouville's theorem we get $Y=1$.

\subsection{The role of the KS formula} \label{sec:continuity-rh}

Now we come to an important point, which was the main reason for writing this paper.
Suppose we find a $\CX$ which solves our Riemann-Hilbert problem for any fixed $u$
away from the walls of marginal stability.  Then its behavior as a function of $\zeta$ is
completely determined:  it is continuous except at the BPS rays, where it jumps according to
\eqref{eq:RH-jumps}.  But what can we say about its behavior as a function of $u$?

The $u$ dependence in our
Riemann-Hilbert problem comes from two places.  One is in the
asymptotic boundary conditions \eqref{eq:RH-asymptotics}; this dependence is certainly
continuous.  The other is in the discontinuity prescription \eqref{eq:RH-jumps}.  Here
too the dependence is continuous as long as $u$ stays away from the walls of marginal stability.
But what happens as we cross the wall?
Let $u_w$ denote a generic point on a wall.  As $u \to u_w$ from one side of the wall,
BPS rays corresponding to charges $\gamma = n \gamma_1 + m \gamma_2$ collide with one
another, coalescing into a single ray $\ell$.
Let $A$ denote the total discontinuity of $\CX$ across this group of rays,
\begin{equation} \label{eq:disc-product}
A = \prod^\ccwarrow_{\substack{\gamma = n \gamma_1 + m \gamma_2 \\ m>0,\,n>0}}
\CK_\gamma^{\degen(\gamma;u)}.
\end{equation}
Assuming that $\lim_{u \to u_w} \CX$ from this side exists, it is the solution to a
Riemann-Hilbert problem in which the
discontinuity across $\ell$ is $A$ (while the discontinuities along all other BPS rays are specified
as before).

On the other hand, we could also consider
$\lim_{u \to u_w} \CX$ from the other side of the wall.
For the two limits to agree, it is necessary and sufficient that they are solutions of the \ti{same}
Riemann-Hilbert problem:  so this requires that $A$ computed
by \eqref{eq:disc-product} is the same on both sides of
the wall.  As we reviewed in Section \ref{sec:ReviewKS}, this is precisely the content of the
KS wall-crossing formula!\footnote{The fact that the product in \eqref{eq:disc-product} is
counterclockwise, while it was clockwise in Section \ref{sec:ReviewKS}, comes from our unusual convention
on composition of maps in Section \ref{sec:constructing-moduli}.}

We conclude that, \ti{assuming the BPS degeneracies obey the KS formula}, a solution $\CX$ of the
Riemann-Hilbert problem is continuous as a function of $u$ and $\zeta$, except at the BPS
rays. Moreover, the discontinuity across the BPS ray is given by a symplectomorphism.

\subsection{Solving the Riemann-Hilbert problem} \label{sec:solving-rh}

Having formulated the Riemann-Hilbert problem, we would like to see that it has a
solution, and understand its large-$R$ behavior.  Unlike the simple
cases we considered in Section \ref{sec:periodicnut}, for which all of the
$S_\ell$ commute with one another, here we cannot write an explicit integral
formula for the desired $\CX$; we have to proceed more indirectly.
We exploit the fact that the problem has a structure very similar to
one considered in \cite{MR1213301,Cecotti:1993rm}. Indeed our
problem is an infinite-dimensional version of the one
considered there.

In \cite{Cecotti:1993rm} the Riemann-Hilbert
problem is re-expressed as an integral equation for an analog $\Phi$ of
$\Upsilon(\zeta)$. For large enough $R$, this equation describes
$\Phi$ as a small correction of the identity matrix.  It can therefore
be solved iteratively, which proves the
existence of a solution for large enough $R$, and also gives an explicit
formula for the leading corrections to the zeroth-order approximation $\Phi=1$.
These leading corrections are expressed directly in terms of the discontinuity factors.

This is exactly the sort of information we would like to find about
our map $\CX$.  One direct approach would be to write down an
infinite dimensional analogue of the integral equation in
\cite{Cecotti:1993rm}.  This approach is directly applicable only to a
linear Riemann-Hilbert problem, so one would have
to pass to the linear problem mentioned at the end of the previous
subsection.  The solution of the integral equation
 would then give a linear map between the function spaces; as
we have described, this linear map would be $\CX^*$ for some map
$\CX: \widetilde\CM_u \to \tilde T_u$.

One minor issue is that if we follow precisely the prescription of \cite{Cecotti:1993rm}
we will get a solution obeying the boundary condition
$\Upsilon_0 = 1$.  For our construction we need a different choice of boundary condition,
namely \eqref{eq:upsilon-reality}, which has the advantage of being compatible with the
reality condition $\CX_\gamma(\zeta) = \overline{\CX_{-\gamma}(-1/\bar \zeta)}$.\footnote{In
this section we write $\CX = \CX(\zeta)$ explicitly,
thinking of $\CX$ as a \ti{map} which varies with $\zeta$, and hence suppress the dependence
on the coordinates $\theta$ of $\widetilde\CM_u$.}
Fortunately, it is straightforward to write a variant of the integral equation
which takes into account this different choice of boundary condition,
by a slight modification of the integral kernel.

This strategy seems good enough to prove the existence of a solution, but
it has an important drawback: the intermediate steps of
the iterative solution need not be of the form $\CX^*$ for any $\CX$.
It is useful to have a realization of the problem where each step in
the approximation scheme is itself a
map $\widetilde\CM_u \to \tilde T_u$. This is possible if we write the following
integral equation, using the abelian group structure on $\tilde T_u$:
\begin{equation} \label{eq:X-integral-mult}
\CX_\gamma(\zeta) = \CX^\sf_\gamma(\zeta) \exp \left[ \frac{1}{4 \pi
i}\sum_{\ell} \int_{\ell } \frac{d\zeta'}{\zeta'} \frac{\zeta' +
\zeta}{\zeta' - \zeta} \log \frac {\CX_\gamma(\zeta')}{(\CX S_\ell)_\gamma(\zeta')}\right]
\end{equation}
Here the sum runs over BPS rays $\ell$.
Any solution of \eqref{eq:X-integral-mult} obeys the discontinuity
conditions \eqref{eq:RH-jumps}.  Moreover, our choice of
integral kernel ensures that the solution will also obey the reality condition
\eqref{eq:upsilon-reality}.  Hence a solution of \eqref{eq:X-integral-mult}
is a solution of the Riemann-Hilbert problem.\footnote{Note
that although the Riemann-Hilbert problem is invariant under diffeomorphisms
of $\widetilde\CM_u$ the equation \eqref{eq:X-integral-mult} is not; its solution is unique,
not unique up to diffeomorphism.}

Using the explicit form of the Kontsevich-Soibelman factors from \eqref{eq:ks-symplect}, we have
\begin{equation}
(\CX S_\ell)_\gamma = \CX_\gamma \prod_{\gamma' \in (\clat_u)_\ell} (1-
\sigma(\gamma') \CX_{\gamma'})^{\degen(\gamma';u)\langle
\gamma,\gamma' \rangle}
\end{equation}
(with $(\clat_u)_\ell$ defined in \eqref{eq:clat-ell}).  Plug this into
\eqref{eq:X-integral-mult} to get the final integral equation for $\CX$:
\begin{equation} \label{eq:X-integral-mult-explicit}
\CX_\gamma(\zeta) = \CX^\sf_\gamma(\zeta) \exp \left[ -\frac{1}{4
\pi i} \sum_{\gamma'} \degen(\gamma';u) \langle \gamma,\gamma'
\rangle \int_{\ell_{\gamma'}} \frac{d\zeta'}{\zeta'} \frac{\zeta' +
\zeta}{\zeta' - \zeta} \log (1- \sigma(\gamma')
\CX_{\gamma'}(\zeta'))\right].
\end{equation}
As we have mentioned, equation \eqref{eq:X-integral-mult-explicit}
is a form of the Thermodynamic Bethe Ansatz. See Appendix
\ref{app:TBA}.

In Appendix \ref{app:integral-asymptotics} we argue that \eqref{eq:X-integral-mult-explicit}
has a solution for sufficiently large $R$, and describe its expansion as $R \to \infty$ for $u$
away from the walls.  The first nontrivial approximation is
\begin{equation} \label{eq:X-first-approx}
\CX_\gamma(\zeta) \sim \CX^\sf_\gamma(\zeta) \exp \left[ -\frac{1}{4
\pi i}  \sum_{\gamma'} \degen(\gamma';u) \langle \gamma,\gamma'
\rangle \int_{\ell_{\gamma'}} \frac{d\zeta'}{\zeta'} \frac{\zeta' +
\zeta}{\zeta' - \zeta} \log (1- \sigma(\gamma')
\CX^\sf_{\gamma'}(\zeta'))\right],
\end{equation}
and is essentially a linear superposition of the 1-instanton corrections
that we found in the abelian theory.  Higher-order corrections involve
multilinears in the $\degen(\gamma';u)$, and have an $R$ dependence
which identifies them as multi-instanton contributions.

Our arguments in Appendix \ref{app:integral-asymptotics} are closely related to ones
given in \cite{Cecotti:1993rm} in the finite-dimensional $tt^*$ context.  In fact, our approach leads to a
simplification of the asymptotic analysis even in the finite-dimensional case; hence in
Appendix \ref{app:integral-asymptotics} we re-analyze that case as well.

\subsubsection*{Global issues}

By solving the Riemann-Hilbert problem, we have obtained a map $\CX: \widetilde\CM_u \to \tilde T_u$
depending on the choice of the local
quadratic refinement $\sigma(\gamma)$.  This choice affects the Riemann-Hilbert problem through the
definition of the discontinuities $\CK_\gamma$.
However, the solution $\CX$ depends on $\sigma$ in a simple way.  Recall that for any two refinements
$\sigma, \sigma'$ there is some $c(\sigma, \sigma') \in \clat_u^* / 2 \clat_u^*$ such that $\sigma(\gamma) \sigma'(\gamma) = (-1)^{\gamma \cdot c(\sigma, \sigma')}$.  Given a solution
$\CX^{[\sigma]}$ of \eqref{eq:X-integral-mult} with refinement
$\sigma$, there is a corresponding solution $\CX^{[\sigma']}$ with refinement $\sigma'$,
\begin{equation}
\CX_\gamma^{[\sigma']}(u,\theta;\zeta) = (-1)^{\gamma \cdot c(\sigma,\sigma')} \CX_\gamma^{[\sigma]}(u,\theta + c \pi;\zeta).
\end{equation}
It follows that if we use the refinement to identify $\widetilde\CM_u \simeq \CM_u$ and also
$\tilde T_u \simeq T_u$, we obtain $\CX: \CM_u \to T_u$ which is \ti{independent} of the choice of refinement.

\subsection{Constructing the symplectic form} \label{sec:symplectic-form}

So far, we have solved the Riemann-Hilbert problem to give a map
$\CX: \CM_u \to T_u$, obeying the asymptotic conditions
\eqref{eq:RH-asymptotics}, the jump conditions \eqref{eq:RH-jumps},
and the reality condition \eqref{eq:reality-X}.  Now letting $u$
vary we obtain a map $\CX: \CM \to T$.   We then
construct a complex 2-form $\hsymp(\zeta)$ on $\CM$ by
pullback of the canonical fiberwise symplectic form on $T$,
\begin{equation} \label{eq:omega-pullback}
\hsymp(\zeta) = \frac{1}{4 \pi^2 R} \CX^* \hsymp^T = \frac{1}{8 \pi^2 R} \eps_{ij} \frac{d\CX_{\gamma^i}}{\CX_{\gamma^i}} \wedge \frac{d\CX_{\gamma^j}}{\CX_{\gamma^j}}.
\end{equation}
A few properties of $\hsymp(\zeta)$ follow directly from \eqref{eq:omega-pullback}:
\begin{itemize}
\item Although $\CX$ is only piecewise analytic in $\zeta$, $\hsymp(\zeta)$ is honestly analytic
(because the discontinuities $S_\gamma$ are symplectomorphisms, i.e. they preserve $\hsymp^T$.)

\item Using \eqref{eq:reality-X}, we have $\hsymp(- 1 / \bar\zeta) = \overline {\hsymp(\zeta)}$.

\item As $\zeta \to 0,\infty$ we can determine the behavior of $\hsymp(\zeta)$
using the asymptotics \eqref{eq:RH-asymptotics} of $\CX$ and the explicit form
\eqref{eq:Omega-sf-2} of $\hsymp^\sf(\zeta)$.  We find that $\hsymp(\zeta)$ has a simple pole in
each case, with residue
\begin{equation} \label{eq:omega-residues}
\Res_{\zeta = 0} \hsymp(\zeta) = \frac{i}{8 \pi} \Upsilon_0^*\inprod{dZ, d\theta}, \quad \Res_{\zeta = \infty} \hsymp(\zeta) = -\frac{i}{8 \pi} \Upsilon_\infty^*\inprod{d\bar Z, d\theta}.
\end{equation}

\item Using $\lim_{R \to \infty} \CX = \CX^\sf$, it follows that $\hsymp(\zeta)$ is nondegenerate (in the holomorphic sense) for large enough $R$.
\end{itemize}
These properties will be important in our construction of the \hk metric.

\subsection{Differential equations} \label{sec:differential-problem}

Our Riemann-Hilbert problem has been formulated in terms of discontinuity
factors which are universal
(locally independent of all parameters of the gauge theory),
together with asymptotics given by the functions $\CX^\sf_\gamma$, which
depend on the parameters only in a very simple way.
In this section, following a standard recipe, we show that
this implies that the solution $\CX$ obeys
a family of differential equations.

As we will see, the physical meaning of these equations
is rather transparent.  One group
expresses the fact that that the functions $\CX(\zeta)$ which solve the Riemann-Hilbert
problem are holomorphic on $\CM$ in complex structure $J^\pz$.  These equations
are essential for the construction of the \hk metric.
Another pair describe the renormalization group flow and a
$U(1)_R$-symmetry action.  These are important for relating the metric to the KS
wall-crossing formula.

A very similar family of equations were crucial in the story of ``$tt^*$
geometry'' which appeared in the context of massive $\CN=(2,2)$
$2$-dimensional theories
\cite{Cecotti:1993rm,Cecotti:1992qh,Cecotti:1991me,Cecotti:1993vy}.

We begin by
recalling that the solution $\CX$ of our Riemann-Hilbert problem over $\IC\IP^1$
is only sectionally analytic; it has jumps of the form
$\CX \to \CX S_\ell$ along various rays $\ell \subset \IC\IP^1$.
So consider instead\footnote{This is the standard notation, but in our context it is
somewhat mnemonic, so here is a longer description.  The infinitesimal variation of
the map $\CX$ by applying $\zeta \partial_\zeta$ gives a vector field on $T$, which we
call $\zeta \partial_\zeta \CX$.  We then pull this back using $\CX$ to get the vector
field $\CA_\zeta$ on $\CM_u$.  We write this pullback operation as $\CX^{-1}$, and because of
our non-standard convention for composition, this $\CX^{-1}$ appears on the right rather
than the left; this makes our equation agree with the usual form for Riemann-Hilbert problems,
and in fact this agreement is the reason we use the non-standard convention in Section \ref{sec:constructing-moduli}.  In local coordinates one would write
\begin{equation}
\CA_\zeta =  \frac{\partial \CX^i}{\partial \zeta}
\left[ \left(\frac{\partial \CX}{\partial \theta}\right)^{-1}
\right]^i_j \frac{\partial}{\partial \theta^j}.
\end{equation}
}
\begin{equation}
\CA_\zeta := \zeta \partial_\zeta \CX \CX^{-1}.
\end{equation}
The discontinuities of $\CX$ along the BPS rays cancel out in $\CA_\zeta$, which is therefore honestly
analytic in $\zeta$, except possibly for $\zeta = 0, \infty$ where $\CX$ becomes singular.
So we can think of $\CX$ as a solution of an ordinary differential equation in $\zeta$,
\begin{equation} \label{eq:zeta-eq-general}
\zeta \partial_\zeta \CX = \CA_\zeta \CX.
\end{equation}
We can describe this equation rather concretely, using our asymptotic information
about $\CX$.  Note first that $\CX^\sf$ obeys an equation of the same form.
To write it we first introduce two vector fields on $\CM_u$,
\begin{equation}
\CA_\zeta^{(-1),\sf} := i \pi Z \cdot \partial_\theta, \quad \CA_\zeta^{(1),\sf} := i \pi \bar{Z} \cdot \partial_\theta.
\end{equation}
Then we have
\begin{equation} \label{eq:zeta-eq-sf}
\zeta \partial_\zeta \CX^\sf = \CA^\sf_\zeta \CX^\sf,
\end{equation}
where
\begin{equation}
\CA^\sf_\zeta = \frac{1}{\zeta} \CA_\zeta^{(-1),\sf} + \zeta \CA_\zeta^{(1),\sf}.
\end{equation}
The important point is that the $\zeta$ dependence of $\CA^\sf_\zeta$ is very simple:  just a simple pole at each of
$\zeta = 0,\infty$.  We can convert this information to information about $\CA_\zeta$, since we know from
\eqref{eq:RH-asymptotics} that $\Upsilon = \CX (\CX^\sf)^{-1}$ remains finite at both $\zeta = 0, \infty$.
This shows that $\CA_\zeta$ also has only a simple pole at $\zeta = 0, \infty$, and even determines the residue,
\begin{equation}
\CA_\zeta = \frac{1}{\zeta}\CA_\zeta^{(-1)} + \CA_\zeta^{(0)} +  \zeta \CA_\zeta^{(1)},
\end{equation}
where
\begin{equation}
\CA_\zeta^{(-1)} = \Upsilon_0 \CA_\zeta^{(-1),\sf} \Upsilon_0^{-1}, \quad \CA_\zeta^{(1)} = \Upsilon_\infty \CA_\zeta^{(1),\sf} \Upsilon_\infty^{-1}.
\end{equation}
So we see that \eqref{eq:zeta-eq-general} defines a flat connection $\zeta \partial_\zeta - \CA_\zeta$
over $\IC\IP^1$, valued in
the infinite-dimensional algebra of vector fields on $\CM_u$, with rank-1 irregular singularities
at $\zeta = 0, \infty$.  $\CX$ is a flat section for this connection.

So our solution to the Riemann-Hilbert problem leads directly to the construction of a flat connection over $\IC\IP^1$.
In fact, this is a standard maneuver in the theory of ordinary differential equations.
The connection we obtained has irregular singularities at $\zeta = 0$ and $\zeta = \infty$, and
hence it exhibits Stokes' phenomenon.  One of the virtues of the Riemann-Hilbert construction
is that it is easy to determine the Stokes factors:
they are simply the discontinuities $S_\ell$ which entered the Riemann-Hilbert problem.

The above discussion has an important extension.  We have not just a single Riemann-Hilbert problem
but a whole family of them, varying with additional parameters.  These parameters include the
coordinates $u^i$ on $\vm$, as well as the scale $\Lambda$, the radius $R$ of $S^1$,
and perhaps some bare gauge couplings $\tau^0$.  (For the moment we do \ti{not} introduce mass parameters;
but see Section \ref{sec:masses} below.)  We introduce the generic notation $t^n$ to encompass
all of these parameters.

Importantly, the discontinuities $S_\ell$ which define the Riemann-Hilbert problem
do not depend on any of the $t^n$.  Hence just as we did above for the $\zeta$ dependence,
we consider
\begin{equation}
\CA_n := \partial_{t^n} \CX \CX^{-1}.
\end{equation}
As before,
the discontinuities of $\CX$ cancel out, so $\CA_n$ is analytic in $\zeta$
away from $\zeta = 0, \infty$.  Also as before, we can control the behavior
near these singularities by first checking the behavior of $\CA^\sf_n := \partial_{t^n} \CX^\sf (\CX^\sf)^{-1}$.
For all of our $t^n$ we have
\begin{equation}
\CA^\sf_n = \frac{1}{\zeta} \CA^{(-1),\sf}_n + \zeta \CA^{(1),\sf}_n
\end{equation}
for some simple vector fields $\CA^{(\pm 1),\sf}_n$;
then using the fact that $\Upsilon$ is finite as $\zeta \to 0,\infty$ as before, we obtain
\begin{equation}
\CA_n = \frac{1}{\zeta} \CA_n^{(-1)} + \CA_n^{(0)} + \zeta \CA_n^{(1)},
\end{equation}
where
\begin{equation}
\CA_n^{(-1)} = \Upsilon_0 \CA_n^{(-1),\sf} \Upsilon_0^{-1}, \quad \CA_n^{(1)} = \Upsilon_\infty \CA_n^{(1),\sf} \Upsilon_\infty^{-1}.
\end{equation}
Also including \eqref{eq:zeta-eq-general}, the full set of equations we obtain is
\begin{align}
\partial_{u^j} \CX &= \left( \frac{1}{\zeta} {\CA^{(-1)}_{u^j}} + \CA^{(0)}_{u^j} \right) \CX, \label{eq:conn-holomorphy-1} \\
\partial_{\bar{u}^{\bar j}} \CX &= \left( {\CA}^{(0)}_{\bar{u}^{\bar j}} + \zeta {\CA}^{(1)}_{\bar{u}^{\bar j}} \right) \CX, \label{eq:conn-holomorphy-2} \\
\Lambda \partial_\Lambda \CX &= \left( \frac{1}{\zeta} {\CA_\Lambda^{(-1)}} + \CA_\Lambda^{(0)} \right) \CX, \label{eq:lambda-equation} \\
\bar\Lambda \partial_{\bar\Lambda} \CX &= \left( \CA_{\bar \Lambda}^{(0)} + \zeta \CA_{\bar \Lambda}^{(1)} \right) \CX, \label{eq:lambda-bar-equation} \\
R \partial_R \CX &= \left( \frac{1}{\zeta} {\CA_R^{(-1)}} + \CA_R^{(0)} + \zeta \CA_R^{(1)} \right) \CX, \label{eq:R-equation} \\
\zeta \partial_\zeta \CX &= \left( \frac{1}{\zeta} {\CA_\zeta^{(-1)}} + \CA_\zeta^{(0)} + \zeta \CA_\zeta^{(1)} \right) \CX. \label{eq:zeta-equation}
\end{align}
One also gets the extra relations
\begin{equation} \label{eq:M-relations}
\CA_R^{(-1)} = - \CA_\zeta^{(-1)}, \quad \CA_R^{(1)} = \CA_\zeta^{(1)},
\end{equation}
from the fact that $\CX^\sf$ is annihilated by $\zeta \partial_\zeta + R \partial_R$ as $\zeta \to 0$,
and by $\zeta \partial_\zeta - R \partial_R$ as $\zeta \to \infty$.

We have finished constructing our equations.  In Appendix \ref{app:diff-asymptotics} we
discuss how to write them more concretely given the asymptotic expansion of $\CX$
around $\zeta = 0$.
We conclude this section with a few remarks:

\begin{itemize}
\item
Since the symplectic form $\hsymp(\zeta)$ was constructed from $\CX$,
\eqref{eq:zeta-equation}, \eqref{eq:R-equation} trivially imply equations
for the $\zeta$ and $R$ dependence of $\hsymp(\zeta)$, of the form
\begin{align}
\plog{\zeta} \hsymp &= \left( \frac{1}{\zeta} \CL_{\CA_\zeta^{(-1)}} +
\CL_{\CA_\zeta^{(0)}} + \zeta \CL_{\CA_\zeta^{(1)}} \right)  \hsymp,
\\ \plog{R} (R\hsymp) &= \left( \frac{1}{\zeta} \CL_{\CA_R^{(-1)}} +
\CL_{\CA_R^{(0)}} + \zeta \CL_{\CA_R^{(1)}} \right) (R \hsymp). \label{eq:omega-R}
\end{align}
Recalling from \eqref{eq:Omegazeta} that
$\hsymp(\zeta) = - \frac{i}{2 \zeta}\omega^+ + \omega_3 - \frac{i}{2} \zeta
\omega_-$, these equations can be expanded in powers of $\zeta$ to
derive some interesting differential equations for the hyperkahler
forms $\vec\omega$.

\item
Recall that the solution $\CX$ of the Riemann-Hilbert problem was ambiguous up to a transformation
$\CX \to b \CX$, with $b$ any diffeomorphism of $\CM_u$.  This ambiguity leads
to $\zeta$-independent gauge transformations of the connection $\CA$.  There are several particularly
convenient gauges.
One is a gauge in which $\CA_R^{(0)} = 0$.
It follows from \eqref{eq:omega-R} that in this gauge the restriction of $R \omega_3$ to each $\CM_u$
is independent of $R$ (and hence equals its $R \to \infty$ limit, namely
$- \frac{1}{8 \pi^2} \inprod{d\theta, d\theta}$.)
It would be interesting to know whether this gauge is the one chosen by our integral equation
\eqref{eq:X-integral-mult}.
If we allow $b$ to be a \ti{complexified} diffeomorphism, then at least formally we can also pick a gauge in which
$\Upsilon_0 = 1$, so $\CA^{(-1)} = \CA^{(-1),\sf}$; this
is an analogue of the ``topological gauge'' of \cite{Cecotti:1993rm} (dually $\Upsilon_\infty = 1$ would
be an ``antitopological gauge'').

\item
Two linear combinations of our equations have a simple physical meaning:  they express the
invariance under overall changes of scale and R-symmetry transformations.  To see this first note that
\begin{equation}
\left(a^I \partial_{a^I} + \Lambda \partial_\Lambda\right) Z_\gamma = Z_\gamma
\end{equation}
for all $\gamma \in \clat$.  It follows that
\begin{align}
\left(R \partial_R - a^I \partial_{a^I} - \bar{a}^I \partial_{\bar{a}^I} - \Lambda \partial_\Lambda - \bar{\Lambda} \partial_{\bar{\Lambda}} \right) \CX^\sf &= 0, \\
\left(\zeta \partial_\zeta  + a^I \partial_{a^I} - \bar{a}^I \partial_{\bar{a}^I} + \Lambda \partial_\Lambda - \bar{\Lambda} \partial_{\bar{\Lambda}} \right) \CX^\sf &= 0.
\end{align}
These equations can be interpreted as the (anomalous) scale and R-symmetry invariance of the semiflat
geometry.  They imply relations among the $\CA^\sf_n$ (just by replacing $\partial \to \CA$) which
in turn give relations among $\CA_n^{(\pm 1)}$:  we find that
\begin{align}
\left(R \partial_R - a^I \partial_{a^I} - \bar{a}^I \partial_{\bar{a}^I} - \Lambda \partial_\Lambda - \bar{\Lambda} \partial_{\bar{\Lambda}} \right) \CX &= \Delta \CX, \label{eq:rescaling-1} \\
\left(\zeta \partial_\zeta  + a^I \partial_{a^I} - \bar{a}^I \partial_{\bar{a}^I} + \Lambda \partial_\Lambda - \bar{\Lambda} \partial_{\bar{\Lambda}} \right) \CX &= \Delta' \CX, \label{eq:rescaling-2}
\end{align}
where $\Delta$, $\Delta'$ are $\zeta$-independent vector fields on $\CM_u$.
We can set $\Delta = 0$, $\Delta' = 0$ by a gauge transformation.  Indeed, our integral equation
automatically picks the appropriate gauge:
the recursive solution we give in Appendix \ref{app:integral-asymptotics}
for large $R$ satisfies \eqref{eq:rescaling-1}, \eqref{eq:rescaling-2}
term-by-term with $\Delta = \Delta' = 0$.  So there is a sense in which the scale and R-symmetry
invariance survive the instanton corrections.

\item
The compatibility between \eqref{eq:R-equation} and \eqref{eq:zeta-equation}, together with
the relations \eqref{eq:M-relations}, implies a set of nonlinear
differential equations for the $R$ dependence of the quadruple
$(\CA_\zeta^{(\pm 1)}, \CA_\zeta^{(0)}, \CA_R^{(0)})$.  These equations are a
deformation of the Nahm equations, as we explain in Appendix \ref{app:diff-asymptotics};
the large $R$ expansion of this
quadruple can be produced directly by solving them iteratively.
They are a possible tool for studying the behavior of our
construction at \ti{small} $R$.  A generic solution of the Nahm
equations would become singular at a finite value of $R$, and we
expect that the same is true for our problem. Nevertheless, we
expect that the particular solutions which we have described here,
determined by the BPS degeneracies in $\CN=2$, $d=4$ field theories,
actually \ti{are} regular for all values of $R$.  It is possible
that this gives an interesting constraint on the possible BPS
spectra and IR prepotentials of $\CN=2$ theories.  A very similar
strategy was employed in \cite{Cecotti:1993rm} to constrain the
properties of $d=2$ theories.

\item
Our discussion in this section gives a new perspective on the
role of the wall-crossing formula.  The collection of equations
\eqref{eq:conn-holomorphy-1}-\eqref{eq:zeta-equation} describe a flat
connection over $\IC\IP^1 \times \CP$, where $\CP$ is the parameter space
coordinatized by the $t^n$.  This flat connection can be viewed equivalently as
an \ti{isomonodromic family} of connections over $\IC\IP^1$, with irregular
singularities of rank $1$ at $\zeta=0, \infty$.  At each $t \in \CP$ the
Stokes data of the connection on $\IC\IP^1$ are given by the Kontsevich-Soibelman
factors.
Using the parallel transport along $\CP$, one shows that the Stokes data
at the irregular singularities are ``invariant'' in an appropriate sense.
To be precise:  choosing any convex sector $\CV$ in the $\zeta$-plane, the
product
\begin{equation}
A_\CV = \prod_{\ell \subset \CV}^\ccwarrow S_\ell
\end{equation}
is invariant, under any variation of $t \in \CP$ for which no Stokes line $\ell$ enters
or leaves $\CV$.  Applying this statement to variations of $u$, we recover
the wall-crossing formula.

\end{itemize}

\subsection{Constructing the metric and its large $R$ asymptotics}

So far we have constructed a family of functions $\CX_\gamma(\zeta)$ on $\CM$,
and the corresponding holomorphic symplectic form $\hsymp(\zeta)$.
As we have discussed in Section \ref{sec:twistor-general},
given $\hsymp(\zeta)$ with the properties listed in Section \ref{sec:symplectic-form},
and $\CX_\gamma$ obeying ``Cauchy-Riemann'' equations of the form \eqref{eq:conn-holomorphy-1},
\eqref{eq:conn-holomorphy-2},
there exists a corresponding \hk metric $g$ on $\CM$.  This is our construction of $g$.

Given the exact functions $\CX_\gamma(\zeta)$ solving
\eqref{eq:X-integral-mult-explicit}, we can write $g$ in
closed form as follows.  Use the expansion of the kernel in \eqref{eq:X-integral-mult-explicit}
for $\vert \zeta'/\zeta\vert < 1$ to obtain an asymptotic expansion for $\zeta \to 0$,
\begin{equation}
\log \CX_{\gamma} = \frac{1}{\zeta} F_{-1}^\gamma + F_0^\gamma +
\zeta F_1^\gamma + \CO(\zeta^2),
\end{equation}
where
\begin{equation}
\begin{split}
F_{-1}^{\gamma} & = \pi R Z_\gamma, \\
F_0^\gamma & = i \theta_\gamma -  \frac{1}{4 \pi i} \sum_{\gamma'}
\Omega(\gamma';u) \langle \gamma,\gamma' \rangle
\int_{\ell_{\gamma'}} \frac{d\zeta'}{\zeta'}   \log (1-
\sigma(\gamma') \CX_{\gamma'}(\zeta')),\\
F_1^\gamma & = \pi R \bar Z_\gamma -  \frac{1}{2 \pi i}
\sum_{\gamma'} \Omega(\gamma';u) \langle \gamma,\gamma' \rangle
\int_{\ell_{\gamma'}} \frac{d\zeta'}{\zeta'^2}   \log (1-
\sigma(\gamma') \CX_{\gamma'}(\zeta')).
\end{split}
\end{equation}
Then, substituting into $\varpi$ we extract
\begin{equation}
\begin{split}
\omega_+ & = \frac{i}{2\pi^2 R} \epsilon_{ij} dF_{-1}^{\gamma^i}
\wedge d F_0^{\gamma^j}, \\
\omega_3 & = \frac{1}{8\pi^2 R} \epsilon_{ij} \left( 2
dF_1^{\gamma^i}\wedge d F_{-1}^{\gamma^j} + d F_0^{\gamma^i} \wedge
d
F_0^{\gamma^j} \right).\\
\end{split}
\end{equation}
From these symplectic forms it is straightforward to obtain $g$.

Now let us consider the behavior of $g$ for large $R$.  In Section \ref{sec:solving-rh}
we have discussed the large $R$ asymptotics of the $\CX_\gamma$,
including the first BPS instanton correction, given in \eqref{eq:X-first-approx}.  Now we translate
this into the correction to $\hsymp(\zeta)$.  We begin by computing the correction to $\frac{d\CX_\gamma}{\CX_\gamma}$:
\begin{equation} \label{eq:dX-correction}
\frac{d\CX_\gamma}{\CX_\gamma} = \frac{d\CX^\sf_\gamma}{\CX^\sf_\gamma} + \CI_\gamma + \cdots,
\end{equation}
where
\begin{equation} \label{eq:correction-integral}
\CI_\gamma =  \frac{1}{4\pi i} \sum_{\gamma'} \degen(\gamma';u)
\langle \gamma,\gamma' \rangle \int_{\ell_{\gamma'}}
\frac{d\zeta'}{\zeta'} \frac{\zeta' + \zeta}{\zeta' - \zeta}
\frac{d\CX^\sf_{\gamma'}(\zeta')}{\CX^\sf_{\gamma'}(\zeta')}
\frac{\sigma(\gamma') \CX^\sf_{\gamma'}(\zeta')}{1- \sigma(\gamma')
\CX^\sf_{\gamma'}(\zeta')}.
\end{equation}
Note that $\CI_\gamma$ is exponentially suppressed as $R \to \infty$ as promised, since on $\ell$ we have $\CX_{\gamma'}^\sf \to 0$ exponentially as $R \to \infty$.  The ellipsis in \eqref{eq:dX-correction} indicates the
multi-instanton corrections, which are even more suppressed.
The leading correction to $\hsymp(\zeta)$ therefore arises from the wedge product between $\frac{d\CX^\sf_\gamma}{\CX^\sf_\gamma}$ and $\CI_\gamma$.

To describe the correction more explicitly, it is convenient to consider each $\gamma'$ separately, and adopt a symplectic
basis $\{\gamma^1, \dots, \gamma^{2r}\}$ in which $\gamma' = q_{\gamma'} \gamma^1$.  Then the integral in \eqref{eq:correction-integral} becomes essentially identical to the integral \eqref{eq:ipm},
which gave the instanton corrections to $\CX_\magn$ in Section \ref{sec:solving-xmagn}.
Evaluating the corresponding correction to $\hsymp(\zeta)$ just as we did there, we obtain
\begin{equation}
\hsymp(\zeta) = \hsymp^\sf(\zeta) + \sum_{\gamma' \in \clat} \hsymp^\inst_{\gamma'}(\zeta) + \cdots,
\end{equation}
where
\begin{equation} \label{eq:Omega-inst-general}
\hsymp^\inst_{\gamma'}(\zeta) = - \degen(\gamma';u) \frac{1}{4 \pi^2 R}\frac{d \CX^\sf_{\gamma'}(\zeta)}{\CX^\sf_{\gamma'}(\zeta)} \left[ A^\inst_{\gamma'} +
\frac{1}{2} V^\inst_{\gamma'} \left(\frac{1}{\zeta} da_{\gamma'} - \zeta d\bar{a}_{\gamma'}\right)\right],
\end{equation}
with (cf. \eqref{eq:V-inst}, \eqref{eq:A-inst})
\begin{align} \label{eq:A-inst-general}
V^\inst_{\gamma'} &= \frac{R q_{\gamma'}^2}{2\pi} \sum_{n>0}
\sigma(n \gamma') e^{i n \theta_{\gamma'}} K_0(2 \pi R \abs{n
Z_{\gamma'}}), \\
A^\inst_{\gamma'} &= - \frac{R q_{\gamma'}^2}{4 \pi} \left( \frac{dZ_{\gamma'}}{Z_{\gamma'}} - \frac{d \bar Z_{\gamma'}}{\bar Z_{\gamma'}} \right) \sum_{n>0} \sigma(n \gamma') e^{i n \theta_{\gamma'}} \abs{Z_{\gamma'}} K_1(2 \pi R \abs{n Z_{\gamma'}}).
\end{align}
From here one may expand in $\zeta$ to extract the leading corrections to $\omega_+$, $\omega_3$ and hence obtain the leading
correction to $g$.

\subsection{Comparison to the physical metric} \label{sec:comparison}

Having constructed a \hk metric $g$ on $\CM$ for large enough $R$, we now summarize
some of its properties:

\begin{enumerate}
\item $g$ is continuous,
\item $g$ approaches the semiflat metric $g^\sf$ if all BPS particles have $\abs{Z} \to \infty$,
\item $g$ is smooth except for specific physically expected singularities, located over the singular loci in $\vm$,
\item $g$ has $\vol(\CM_u) = \left(\frac{1}{R}\right)^{r}$,
\item ($\CM$, $g$) in complex structure $J_3$ can be identified with the Seiberg-Witten torus
fibration in its standard complex structure, and after this identification, the holomorphic symplectic
form is $\omega_+ = - \frac{1}{4 \pi} \inprod{dZ, d\theta}$.
\end{enumerate}

All of these properties agree with what is expected for the physical metric on $\IR^3 \times S^1$
as described in \cite{Seiberg:1996nz}.
The simplest consistent picture is therefore that the metric we have constructed is indeed the
physical one.  (In the rank $1$ case it was suggested in \cite{Seiberg:1996nz} that these
properties indeed \ti{determine} the metric, by a non-compact analogue of Yau's theorem.  It is
plausible that there could be a similar theorem more generally.)

In the rest of this section we establish these properties from
our construction:
\begin{enumerate}
\item The continuity of $g$ follows from the wall-crossing formula, as we have
explained.

\item We need only look at the form of the corrections
\eqref{eq:Omega-inst-general}:  they are all exponentially
suppressed in $R \abs{Z_{\gamma'}}$, and hence vanish exponentially
fast if all $\abs{Z_{\gamma'}} \to \infty$.

\item In any limit where $R
\abs{Z_\gamma} \to \infty$ for all $\gamma$, the instanton
contributions are exponentially suppressed and $g$ approaches
$g^\sf$. This is enough to establish the smoothness of $g$ at large
$R$, \ti{except} near a singular locus where some BPS particles with charges $\gamma_i$
become massless ($Z_{\gamma_i} = 0$ and $\degen(\gamma_i;u) \neq 0$). To
understand the behavior near these points, we consider a scaling
limit where $R \to \infty$ holding $R Z_{\gamma_i}$ finite.  One can
approximate the Riemann-Hilbert problem in this limit by one in
which we keep only the BPS rays $\ell_{\gamma_i}$, dropping all the
others.  Indeed all other discontinuities involve factors of the form
$(1-\sigma(\gamma')\CX_{\gamma'})$, which become exponentially close to $1$ in this
scaling limit.

In the simplest case where only a single $Z_\gamma = 0$,
we can always choose a duality frame such that $\gamma$ is an electric charge.
By shifting some of the angles $\theta$ by $\pi$, we can also arrange that the refinement
$\sigma$ is of the standard form $\sigma = (-1)^{\gamma_e \cdot \gamma_m}$ for this frame.
Then we are in the situation we studied in Section \ref{sec:periodicnut}, where we found a \hk metric
which is smooth except for a periodic array of $q$ $A_{q-1}$ singularities.  This agrees with
the expectation from effective field theory in $d=3$: a singularity occurs at the point where one
of the Kaluza-Klein tower of charge-$q$ hypermultiplets becomes massless.

In addition to the physical singularities we have examined, where a
set of mutually local BPS particles become massless, there can also
be superconformal points, where mutually nonlocal particles
simultaneously become massless \cite{Argyres:1995jj,Argyres:1995xn}.
We have not analyzed these singularities, although we expect them to
be interesting, and we expect the quadratic refinement to play
an important role in their analysis.

\item Since $\CM_u$ is a complex torus with respect to $J_3$,
its volume is just $\frac{1}{r!} \int_{\CM_u} \omega_3^r$.  On the other hand,
using \eqref{eq:Omegazeta} and the fact that
$\omega_{\pm}$ restrict to zero on $\CM_u$ by \eqref{eq:omega-residues}, this is
\begin{equation}
\vol(\CM_u) = \frac{1}{r!} \int_{\CM_u} \hsymp^r(\zeta) = \frac{1}{(4 \pi^2 R)^r r!} \int_{\CX(\CM_u)} (\hsymp^T)^r = \left(\frac{1}{R}\right)^r,
\end{equation}
as desired.

\item Complex structure $J_3$ can be determined
from $\omega_1$ and $\omega_2$, just by $J_3 = \omega_1^{-1} \omega_2$.  But this information in turn
is given by the residue of $\hsymp(\zeta)$ at $\zeta = 0$; recall from \eqref{eq:Omegazeta} that
$\omega_+ = \omega_1 + i \omega_2$ is given by
\begin{equation}
\omega_+ = 2i\,\Res_{\zeta = 0} \, \hsymp(\zeta).
\end{equation}
Our asymptotic condition \eqref{eq:RH-asymptotics} on $\hsymp(\zeta)$ precisely ensures that this is related to the residue
of $\hsymp^\sf(\zeta)$:  indeed we just have
\begin{equation}
\omega_+ = \Upsilon_0^* \omega_+^\sf.
\end{equation}
It follows that $(\CM, J_3)$ can be identified with $(\CM, J_3^\sf)$ just by acting with the fiberwise
diffeomorphism $\Upsilon_0$.  As we explained in Section \ref{sec:semiflat-twistor}, the complex structure $J_3^\sf$
on $\CM$ is just that of the Seiberg-Witten torus fibration.  Moreover, under this identification $\omega_+$
is identified with $\omega_+^\sf$ given in \eqref{eq:omegaplus-sf}.
\end{enumerate}

\section{Adding masses} \label{sec:masses}

In this section we briefly indicate how the results of the previous sections should be modified
to include nontrivial mass parameters.

\subsection{Single-particle corrections with masses}

There is a simple variant of the $U(1)$ theory considered in Section
\ref{sec:periodicnut}:  we can consider the $U(1)$ theory with
several electrically charged hypermultiplets, of charges $q_i$.  A
theory with more than one species of particle will involve flavor
charges, and depend non-trivially on mass parameters.  The mass
parameters in four dimensions are complex numbers $m_i$.  Upon
compactification to three dimensions an extra real periodic mass
parameter $m^3_i$ appears, which is essentially a Wilson line for
the flavor symmetry.  We write $\psi_i := 2 \pi R m^3_i$, with
period $2\pi$.

The mass parameters enter the corrected metric in a very simple fashion: each
particle gives an additive contribution to $V$ and $A$ similar to
the one we met before,
\begin{equation} \label{eq:V-sum-masses}
V = \sum_i \frac{q_i^2 R}{4 \pi} \sum_{n=-\infty}^\infty \left(
\frac{1}{\sqrt{R^2 \abs{q_i a + m}^2 + (q \frac{\theta_\elec}{2 \pi} +
\frac{\psi_i}{2 \pi} + n)^2}} - \kappa_n \right)
\end{equation}

The coordinate $\CX_\elec$ is unchanged:
\begin{equation}
\CX_\elec = \exp \left[ \pi R \frac{a}{\zeta} + i \theta_\elec +
\pi R \zeta \bar{a} \right].
\end{equation}
It is also useful to introduce a similar combination of the mass
parameters:
\begin{equation}
\mu_i := \exp \left[ \pi R \frac{m_i}{\zeta} + i \psi_i + \pi R
\zeta \bar{m_i} \right].
\end{equation}
The semiflat $\CX_\magn$ receives contributions from integrating out all
of the particles in $d=4$:
\begin{multline} \CX_\magn^\sf(\zeta)= e^{i \theta_\magn} \times  \\
\prod_i \exp\left[-i\frac{R q_i}{2
\zeta} \left((q_i a+m_i) \log\frac{q_i a+ m_i}{e \Lambda} \right) +
i \frac{\zeta R q_i}{2} \left((q_i \bar a + \bar
m_i)\log\frac{q_i \bar a+ \bar m_i}{e\bar \Lambda} \right)\right].
\end{multline}
The monodromy of $\CX_\magn^\sf(\zeta)$ around $q_i a+m_i =0$ is
\begin{equation}\label{eq:mono2-masses}
\CX_\magn^\sf \to (-\mu_i)^{q_i} \CX_\elec^{q_i^2}
\CX_\magn^\sf.
\end{equation}
The full coordinate similarly receives instanton contributions from all
of the particles,
\begin{equation}\label{eq:FullX2-masses}
\begin{split}
\CX_\magn = \CX_\magn^\sf \prod_i \exp\Biggl[ & \frac{i q_i}{4\pi}
\int_{\ell^i_+} \frac{d \zeta'}{\zeta'}
\frac{\zeta' + \zeta}{\zeta' - \zeta} \log[1-\mu_i \CX_\elec(\zeta')^{q_i}] �\\
& - \frac{i q_i}{4\pi}\int_{\ell^i_-}
\frac{d \zeta'}{\zeta'} \frac{\zeta' + \zeta}{\zeta' - \zeta}
\log[1-\mu_i^{-1} \CX_\elec(\zeta')^{-q_i}] \Biggr],
\end{split}
\end{equation}
where we choose the contours $\ell^i_\pm$ to be any paths in the $\zeta$-plane connecting
$0$ to $\infty$ which lie in the two half-planes
\begin{equation}
\CU^i_\pm = \left\{ \zeta: \pm {\rm Re} \frac{q_i a + m_i}{\zeta} < 0 \right\}.
\end{equation}

The discontinuities depend now on the masses:
\begin{subequations} \label{eq:X2-disc-masses}
\begin{align}
(\CX_\magn)_{\ell^i_+}^{+} &= (\CX_\magn)_{\ell^i_+}^{-} (1 - \mu_i \CX_\elec^q)^{-q}, \\
(\CX_\magn)_{\ell^i_-}^{+} &= (\CX_\magn)_{\ell^i_-}^{-} (1 - \mu_i^{-1}
\CX_\elec^{-q})^q.
\end{align}
\end{subequations}

All of the formulas of this section can also be extended to higher rank
along the lines of Section \ref{sec:higher-rank}.

\subsection{Multiple-particle corrections with masses}

Now we are ready to understand the role of the mass parameters in the general
Riemann-Hilbert and differential problems.  Consider a gauge theory with
$n_f$ flavor symmetries.  Denote the flavor charges as $\gamma^f$, and
build $\mu_{\gamma^f}$ in the obvious way from the masses and
flavor Wilson lines,
\begin{equation}
\mu_{\gamma^f} := \exp \left[ \pi R \frac{m_{\gamma^f}}{\zeta} + i \psi_{\gamma^f} + \pi R
\zeta \bar{m}_{\gamma^f} \right].
\end{equation}
The discontinuities of the abelian problem
suggest generalized Kontsevich-Soibelman factors,
of the form
\begin{equation}
\CK_{\gamma,\gamma^f} := \CX_{\gamma'} \to \CX_{\gamma'} (1-
\sigma(\gamma) \mu_{\gamma^f} \CX_\gamma)^{\langle \gamma,\gamma'
\rangle}.
\end{equation}
We can then define a Riemann-Hilbert problem similar to that of
Section \ref{sec:defining-rh}, which associates the discontinuity
$\CK_{\gamma,\gamma^f}$ to each particle of charge $\gamma$ and flavor
charge $\gamma^f$. Assuming that the wall-crossing formula still
gives the correct BPS degeneracies when generalized to use these
modified symplectomorphisms, we can use this Riemann-Hilbert problem
to construct a \hk metric on $\CM$, which we propose is the correct
one.

A standard trick in supersymmetric field theory is to regard the
mass parameters as vacuum expectation values of vector multiplet
scalars of an enhanced theory in which the flavor symmetry has been
weakly gauged \cite{Seiberg:1993vc}.  Because of the weak gauging, the
particles with magnetic flavor charge are very heavy and can be
neglected in the limit in which the flavor gauge coupling goes to
zero; the only flavor gauge charges that remain are electric.  Using
this trick, the generalized wall-crossing formula with masses can be
interpreted as a zero-coupling limit of the standard wall-crossing formula.

Finally we would like to extend the differential formulation of Section
\ref{sec:differential-problem} to deal with the mass parameters.
In that section we relied on the fact that the factors $S_\ell$ were
independent of the parameters.  In our modified problem the $S_\ell$
depend explicitly on $\mu_{\gamma^f}$, hence on $m_i$, $R$, $\zeta$.
However, it is true that all $S_\ell$ are annihilated by
$\partial_{m_i} + i \frac{\pi R}{\zeta} \partial_{\psi_i}$ and by $\partial_{\bar m_i} +
i \pi R \zeta \partial_{\psi_i}$.
Then a slight modification of the
arguments of that section shows that the solutions $\CX$ of the Riemann-Hilbert
problem obey differential equations of the form
\begin{align}
\partial_{m_i} \CX &= \left(\frac{1}{\zeta} \CA^{(-1)}_{m_i} + \CA^{(0)}_{m_i} \right) \CX,
\label{eq:mass-equation} \\
\partial_{\bar{m}_i} \CX &= \left(
\CA^{(0)}_{\bar{m}_i} + \zeta \CA^{(1)}_{\bar{m}_i} \right) \CX, \label{eq:mass-bar-equation}
\end{align}
where $\CA^{(-1)}_{m_i}$ is not just a vector field on $\CM_u$ but also includes
the operator $-i \frac{\pi R}{\zeta} \partial_{\psi_i}$, and similarly for $\CA^{(1)}_{\bar{m}_i}$.

The $S_\ell$ are also annihilated by the R-symmetry and scale invariance operators,
so we obtain analogues of \eqref{eq:rescaling-m1}, \eqref{eq:rescaling-m2} (after passing to an appropriate gauge),
\begin{align}
\left(R \partial_R - a^I \partial_{a^I} - \bar{a}^I \partial_{\bar{a}^I} - \Lambda \partial_\Lambda - \bar{\Lambda} \partial_{\bar{\Lambda}} - m^i \partial_{m^i} - \bar{m}^i \partial_{\bar{m}^i} \right) \CX &= 0, \label{eq:rescaling-m1} \\
\left(\zeta \partial_\zeta + a^I \partial_{a^I} - \bar{a}^I \partial_{\bar{a}^I} + \Lambda \partial_\Lambda - \bar{\Lambda} \partial_{\bar{\Lambda}} + m^i \partial_{m^i} - \bar{m}^i \partial_{\bar{m}^i} \right) \CX &= 0. \label{eq:rescaling-m2}
\end{align}
Using these equations we can obtain our standard form \eqref{eq:R-equation}, \eqref{eq:zeta-equation} for
the $R$ and $\zeta$ dependence of $\CX$, again with the modification that $\CA_R$ and $\CA_\zeta$
now involve derivatives with respect to the $\psi_i$.

\section{A proof of the wall-crossing formula} \label{sec:proof}

In this paper we have given a construction of a \hk\ metric $g$ on $\CM$
and argued that it matches the physical metric on the moduli space
of the gauge theory on $\IR^3 \times S^1$. The Kontsevich-Soibelman
wall-crossing formula arose as a consistency condition:  without it
our construction would not have given a smooth metric.  We view this
as strong circumstantial evidence that the wall-crossing formula is
indeed correct.

However, these constructions do not quite give a \ti{proof} of the wall-crossing formula.
To give a proof we need to work directly from the physics of the gauge theory, rather than making
any assumptions about what form the metric should take.  In this approach we do not
have the power of the Riemann-Hilbert construction available to us (at least initially).
We use instead the alternative perspective which we described in Section \ref{sec:differential-problem}.
Let $\CM$ be the moduli space of the gauge theory on $\IR^3 \times S^1$, and
consider maps $\CX(\zeta): \CM \to T$ (for $\zeta \neq 0,\infty$).
We aim to construct an integrable set of equations for such
$\CX(\zeta)$, of the form \eqref{eq:conn-holomorphy-1}-\eqref{eq:zeta-equation},
such that the connection \eqref{eq:zeta-equation} over $\IC\IP^1$
has Stokes rays $\ell$ carrying Stokes factors
\begin{equation} \label{eq:sf-redux}
S_\ell = \prod_{\gamma \in (\clat_u)_\ell} \CK_\gamma^{\degen(\gamma;u)}.
\end{equation}
Having constructed such differential equations, the WCF would be the statement of isomonodromic deformation
for the connection \eqref{eq:zeta-equation}.

We now describe how to derive these differential equations
directly from gauge theory.
As we show in Appendix \ref{app:holomorphy}, \eqref{eq:conn-holomorphy-1}, \eqref{eq:conn-holomorphy-2}
have a simple geometric meaning:  they are just the Cauchy-Riemann equations, expressing the holomorphy
of $\CX(\zeta)$ in the complex structure $J^\pz$.  In particular, these equations can be understood purely
in terms of the $\CN=4$ supersymmetry of the reduced theory.  Next note that
\eqref{eq:lambda-equation}, \eqref{eq:lambda-bar-equation} are of exactly the same form
as \eqref{eq:conn-holomorphy-1}, \eqref{eq:conn-holomorphy-2}.  Indeed they would become
identical if we consider $\Lambda$ as the scalar component of a ``background''
vector multiplet.  This is a standard technique for proving
non-renormalization theorems, see e.g. \cite{Seiberg:1993vc}; applying it here
should lead to the desired \eqref{eq:lambda-equation}, \eqref{eq:lambda-bar-equation}.
If the theory involves mass parameters we can prove \eqref{eq:mass-equation}, \eqref{eq:mass-bar-equation}
similarly, by weakly gauging the flavor symmetry.

Finally we need to establish the key equations
\eqref{eq:R-equation}, \eqref{eq:zeta-equation} giving the $R$ and
$\zeta$ dependence of $\CX$.  These follow from the anomalous
$U(1)_R$ symmetry and scale invariance of the $d=4$ theory, as
expressed by \eqref{eq:rescaling-1}, \eqref{eq:rescaling-2} or
\eqref{eq:rescaling-m1}, \eqref{eq:rescaling-m2}, together with the
equations we have already established above. The functions
$\CX_{\gamma}$ have a physical interpretation which we hope to
describe elsewhere. They can be viewed as elements of a chiral ring
of a three-dimensional topological field theory, or as certain line
operator expectation values in the four-dimensional theory. Viewed
in these terms the equations \eqref{eq:R-equation},
\eqref{eq:zeta-equation} are anomalous Ward identities.

To finish the proof we have to show that the Stokes factors for the connection on
$\IC\IP^1$ are indeed given by \eqref{eq:sf-redux}.  For this we use the fact that
the Stokes factors are invariant under variation of $R$, thanks to \eqref{eq:R-equation}.
We can therefore go to very large $R$, where (away from the walls) the corrections to the metric
should be well approximated by a linear superposition of the 1-instanton corrections we
know from the abelian theory.  Passing from the connection on $\IC\IP^1$ to the corresponding
Riemann-Hilbert problem, and running the same arguments we used in Section \ref{sec:constructing-moduli},
we can show that these corrections correspond directly to the Stokes factors.
This completes the proof of the wall-crossing formula, at least at a physical level of
rigor.

\section*{Acknowledgements}

We would like to give special thanks to  F. Denef, N. Nekrasov, and
X. Yin for collaboration in the early stages of this project, and to
B. Pioline for important discussions on the semiflat metric.  We
also thank Wu-yen Chuang, E. Diaconescu, D. Jafferis, D. Joyce, M.
Kontsevich, S. Lukyanov, N. Seiberg, Y. Soibelman, V. Toledano
Laredo, C. Vafa, E. Witten, and E. Zaslow for valuable discussions. The work of
GM is supported by the DOE under grant DE-FG02-96ER40949. GM also
thanks the Aspen Center for Physics and the KITP at UCSB (supported
in  part by the National Science Foundation under Grant No.
PHY05-51164) for hospitality during the completion of this work. The
work of AN is supported in part by the Martin A.\ and Helen
Chooljian Membership at the Institute for Advanced Study, and by the
NSF under grant number PHY-0503584.  The work of DG is supported in part
by DOE Grant DE-FG02-90ER40542.

\appendix

\section{Verifying the KS identity for some $SU(2)$ gauge theories} \label{app:ks-product}
There is an instructive way to prove the simple formula involving
$\CK_{1,0}$ and $\CK_{0,1}$. Consider a sequence of numbers $x_n$
satisfying the recursion
\begin{equation} x_{n+1} x_{n-1} = 1 - x_n. \end{equation}
Surprisingly, the recursion is periodic with period five:
\begin{equation} x_2 = \frac{1-x_1}{x_0}, \quad x_3 = \frac{x_0+
x_1-1}{x_0 x_1}, \quad x_4 = \frac{1-x_0}{x_1}, \quad x_5 = x_0, \quad
x_6=x_1. \end{equation} Now,
set   $X_{1,0}=X_{1,0}^{(1)}= x_1^{-1}$ and $ X_{0,1}
=X_{0,1}^{(1)}= x_0$. Our strategy will be to define successive
transformations $(X_{1,0}^{(n+1)},X_{0,1}^{(n+1)}) = \CK_n
(X_{1,0}^{(n)},X_{0,1}^{(n)})$ for an appropriate sequence of KS
transformations $\CK_n$ until we obtain the identity transformation as
$(X_{1,0}^{(1)},X_{0,1}^{(1)})\to (X_{1,0}^{(N)},X_{0,1}^{(N)})$
(where $N=6$ in our first example but will be  infinite in the
remaining examples). In order to avoid cluttering the notation we do
not indicate the superscript ${}^{(n)}$ in what follows.

If we apply $\CK_{1,0}$ it does not change the value of  $X_{1,0} =
x^{-1}_1$, but modifies $X_{0,1}$ to $X_{0,1}=x_0(1-x^{-1}_1)^{-1}
=- x_1 x^{-1}_2$. Notice that $X_{1,1} = X_{1,0} X_{0,1} =
-x_2^{-1}$.

We can then apply $\CK_{1,1}$: this leaves $X_{1,1} = -x^{-1}_2$ and
changes $X_{1,0}$. As a result, now $X_{0,1}=- x_1 x^{-1}_2
(1-x^{-1}_2)^{-1}= x_3^{-1}$.

If we apply $\CK_{0,1}$, $X_{0,1} = x_3^{-1}$ and $X_{1,0} = - x_3
x^{-1}_2 (1-x^{-1}_3)=x_4$. If we apply $\CK_{1,0}^{-1}$ $X_{1,0} =
x_4$, $X_{0,1} =x^{-1}_3 (1-x_4) = x_5$. Finally if we apply
$\CK_{0,1}^{-1}$ we get $X_{0,1}=x_5=x_0$, $X_{1,0} = x_4 (1-x_5)^{-1}
= x^{-1}_6=x^{-1}_1$. Hence we derive the desired
\begin{equation} \CK_{0,1}^{-1}\CK_{1,0}^{-1}\CK_{0,1}\CK_{1,1}\CK_{1,0} = 1.
\end{equation}

This was a useful warm-up exercise for more interesting formulae.
Consider now a different recursion relation:
\begin{equation} x_{n+1} x_{n-1} = (1- x_n)^2. \end{equation}
This recursion is not in general periodic: it has general solution
\begin{equation} x_n = -\frac{\cosh^2 (a n +b)}{\sinh^2 a}.
\end{equation} We can again relate the recursion to a product of $\CK$
factors.

We start again with $X_{0,1}=x_0$ and $X_{1,0} = x^{-1}_1$. If we
apply $\CK_{1,0}^2$ the result is $X_{1,0}=x^{-1}_1$, $X_{2,1} =
x_1^{-2} x_0 (1- x^{-1}_1)^{-2}=x^{-1}_2$. If we apply $\CK_{2,1}^2$
the result is $X_{2,1}=x^{-1}_2$, $X_{3,2}=x_2^{-2} x_1
(1-x^{-1}_2)^{-2} = x^{-1}_3$. We can keep acting with $\CK_{n+1,n}^2$
for all $n$, following the recursion to arbitrarily large $n$. We
can compute the infinite product by the infinite $n$ limit of the
relations $X_{n+1,n} = x^{-1}_{n+1}$ and $X_{n,n-1} = x^{-1}_{n}$.
If we pick the real part of $a,b$ positive, $X_{1,1} = e^{-2 a}$ and
$X_{1,0} =- e^{-2 b} (1-e^{-2 a})^2$.

On the other hand we can follow the recursion in the opposite
direction: $X_{0,1}=x_0$ and $X_{1,0} =x^{-1}_1$ under $\CK_{0,1}^2$
goes to $X_{0,1}=x_0$ and $X_{1,0} = x^{-1}_1 (1-x_0)^2 = x_{-1}$.
$\CK_{1,0}^2$ sends this to $X_{1,0} = x_{-1}$ and $X_{0,1} = x_0
(1-x_{-1})^{-2} = x^{-1}_{-2}$.

The latter relation is the image under $\CK_{0,1}^2$ of $X_{0,1} =
x^{-1}_{-2}$ and $X_{1,2} = x_{-2}^{-2} x_{-1} (1-x^{-1}_{-2})^{-2}
= x^{-1}_{-3}$. We can now keep acting with the inverse of
$\CK_{n,n+1}$ for all $n$, computing again an infinite product. The
large $n$ limit of $X_{n,n+1} = x^{-1}_{-n-2}$ and $X_{n-1,n} =
x^{-1}_{-n-1}$ is $X_{1,1} = e^{-2a}$ and $X_{1,0} =- e^{-2
b}(1-e^{-2 a})^{-2}$.

Hence by following the whole recursion from $n= - \infty$ to $n=\infty$
we can derive an expression for the infinite product
\begin{equation}
\cdots \CK_{4,3}^2 \CK_{3,2}^2 \CK_{2,1}^2 \CK_{1,0}^2 \CK_{0,1}^{-2}
\CK_{1,0}^{-2} \CK_{0,1}^2 \CK_{1,2}^2 \CK_{2,3}^2 \cdots
\end{equation}
The map between the limiting values of the recursion is
$(X_{1,1},X_{1,0}) \to (X_{1,1}, X_{1,0}(1-e^{-2 a})^4 =
X_{1,0}(1-e^{-4 a})^4 (1+e^{-2 a})^{-4}) $, which is the expected
$\CK_{2,2}^2 \CK_{1,1}^{-4}$!

In the main text we related this formula to the wall-crossing
behavior of a $SU(2)$ Seiberg-Witten theory with two flavors
($SO(4)=SU(2)_A \times SU(2)_B$ flavor symmetry). We argued that a
similar relation should hold, which carries information about flavor
charges. The relation should give the reordering of a product
$\CK_{1,0;1,0} \CK_{1,0;-1,0} \CK_{0,1;0,1} \CK_{0,1;0,-1}$, i.e. the
wall-crossing formula for a theory with a $SU(2)_A$ doublet of
particles of charge $(1,0)$ (flavor charge $(\pm 1,0)$ under the
Cartan generators of $SU(2)_A$ and $SU(2)_B$) and a $SU(2)_B$
doublet of particles of charge $(0,1)$ (flavor charge $(0,\pm 1)$
under the Cartan generators of $SU(2)_A$ and $SU(2)_B$).

The basic transformations are
\begin{equation}
\CK_{1,0;1,0} \CK_{1,0;-1,0}: (X_{1,0},X_{0,1}) \to (X_{1,0},X_{0,1}(1-
\mu_A X_{1,0})^{-1}(1- \mu_A^{-1} X_{1,0})^{-1})
\end{equation}
and
\begin{equation}
\CK_{0,1;0,1} \CK_{0,1;0,-1}: (X_{1,0},X_{0,1}) \to (X_{1,0}(1- \mu_B
X_{0,1})(1- \mu_B^{-1} X_{0,1}),X_{0,1}).
\end{equation}

For this problem we need to alternate the factors from particles in
doublets of $SU(2)_A$ or $SU(2)_B$. Let's take
\begin{multline}
x_n = -\frac{1}{2}\frac{\cosh u \cosh v}{\sinh^2 a} + (-1)^n
\frac{1}{2}\frac{\sinh u \sinh v}{\sinh^2 a}  \\
- \frac{\sqrt{(\cosh 2
a + \cosh 2 u)(\cosh 2 a + \cosh 2 v)}}{\sinh^2 2 a} \cosh (2 a n +
2 b)
\end{multline}
with $u,v$ to be determined in terms of $\mu_A, \mu_B$ below.

This satisfies the recursion
\begin{equation}
x_{n+1} x_{n-1} = (1 - e^{u + (-1)^n v} x_n)(1 - e^{-u - (-1)^n v}
x_n). \end{equation}

We can again initialize the recursion as $X_{0,1}=x_0$ and $X_{1,0}
= x^{-1}_1$. If we apply $\CK_{1,0;1,0} \CK_{1,0;-1,0}$ the result is
again $X_{1,0}=x^{-1}_1$, $X_{2,1} =x^{-1}_2$, as long as we
identify $\mu_A = e^{u-v}$. If we apply then $\CK_{2,1;0,1}
\CK_{2,1;0,-1}$ the result is $X_{2,1}=^{-1}x_2$, $X_{3,2}= x^{-1}_3$,
as long as we identify $\mu_B = e^{u+v}$. We can keep acting
alternatingly with the $\CK_{2n+1,2n;1,0} \CK_{2n+1,2n;-1,0}$ and the
$\CK_{2n+2,2n+1;0,1} \CK_{2n+2,2n+1;0,-1}$ for all $n$, following the
recursion to arbitrarily large $n$. We can compute the infinite
product by the infinite $n$ limit of the relations $X_{n+1,n} =
x^{-1}_{n+1}$ and $X_{n,n-1} = x^{-1}_{n}$. If we pick the real part
of $a,b$ positive, $X_{1,1} = e^{-2 a}$ and
\begin{equation}X_{1,0} =- e^{-2 b} (1-e^{-4 a})^2(1+e^{-2a
-2u})^{-1/2}(1+e^{-2a +2u})^{-1/2}(1+e^{-2a -2v})^{-1/2}(1+e^{-2a
+2v})^{-1/2}.\end{equation}

On the other hand we can follow the recursion in the opposite
direction. The large $n$ limit of $X_{n,n+1} = x^{-1}_{-n-2}$ and
$X_{n-1,n} = x^{-1}_{-n-1}$ is $X_{1,1} = e^{-2a}$ and
\begin{equation}X_{1,0} =- e^{-2 b} (1-e^{-4 a})^{-2}(1+e^{-2a
-2u})^{1/2}(1+e^{-2a +2u})^{1/2}(1+e^{-2a -2v})^{1/2}(1+e^{-2a
+2v})^{1/2}.\end{equation}

The total map is
\begin{multline} (X_{1,1},X_{1,0}) \to  \\
\left(X_{1,1},X_{1,0} \frac{(1-X_{1,1}^2)^{4}}{(1+ \mu_A \mu_B X_{1,1})(1+ \mu_A^{-1} \mu_B X_{1,1})
(1+ \mu_A \mu_B^{-1}X_{1,1})(1+ \mu_A^{-1} \mu_B^{-1}X_{1,1})}\right).\end{multline}

We recognize the expected answer: a vector multiplet of charge
$(2,2)$ and no flavor charges, and a hypermultiplet of charge
$(1,1)$ in the $(2_A)\otimes (2_B)$ representation of the flavor
symmetry (the vector of $SO(4)$).

\section{Cauchy-Riemann equations on $\CM$} \label{app:holomorphy}

In this appendix we explain how the Cauchy-Riemann equations on $(\CM,g)$ in complex
structure $J^\pz$ may be
recast as flatness equations for a connection over $\vm$, with a very simple $\zeta$ dependence.
We do not assume that $g$ arises from the construction we described in
Section \ref{sec:constructing-moduli}; rather, we use only general facts that
follow from identifying $(\CM, g)$ as the moduli space of the gauge theory
on $\IR^3 \times S^1$.

For each $\zeta \in \IC^\times$ we now consider the Cauchy-Riemann equations
\begin{equation} \label{eq:cauchy-riemann}
\bar\partial f = 0
\end{equation}
with respect to complex structure $J^\pz$ on $\CM$.  We will rewrite these equations in the form
\begin{align}
\partial_{u^i} f &= \CA_{u^i} f, \label{eq:cr-connection-1} \\
\partial_{\bar{u}^{\bar i}} f &= \CA_{{\bar u}^{\bar i}} f, \label{eq:cr-connection-2}
\end{align}
where $\CA_{u^i}$ and $\CA_{{\bar u}^{\bar i}}$ are first-order
differential operators acting along the torus fibers (so in coordinates $(u,\bar
u,\theta)$ for $\CM$ they just involve derivatives with respect to
$\theta$), and moreover they depend on $\zeta$ in a simple way,
\begin{align}
\CA_{u^i} &= \frac{1}{\zeta} \CA^{(-1)}_{u^i} + \CA^{(0)}_{u^i}, \\
\CA_{\bar u^{\bar i}} &= \CA_{\bar u^{\bar i}}^{(0)} + \zeta \CA_{\bar u^{\bar i}}^{(1)},
\end{align}
with the $\CA^{(-1)}_{u^i}$ linearly independent at every point,
and similarly $\CA^{(1)}_{\bar{u}^{\bar{i}}}$.

We begin by rewriting \eqref{eq:cauchy-riemann} as
\begin{equation} \label{eq:holomorphy}
(1 - i J^{(\zeta,\bar\zeta)}) df = 0.
\end{equation}
If we treat $\zeta$ and $\bar\zeta$ as independent complex variables, then
this equation is actually independent of $\bar\zeta$.  To see this, it is enough to work at
a single fixed $\zeta$, say $\zeta = 0$.
Specialize the general complex structure \eqref{eq:Jzeta} to $\zeta = 0$,
\begin{equation}
J^{(\zeta=0,\bar\zeta)} = J_3 + i \bar\zeta J_+
\end{equation}
where we introduced $J_+ = J_1 + i J_2$.
Next note that $J_+ J_3 = i J_+$, so $J_+ (J_3 - i) = 0$, so $J_+$ annihilates the $-i$ eigenspace of $J_3$.
So we have shown that $J^{(\zeta=0,\bar\zeta)}$ and $J_3$ share an $n$-dimensional eigenspace
with eigenvalue $-i$.  To finish the argument we would like to know that $J^{(\zeta=0,\bar\zeta)}$ does
not have any $\ti{other}$ eigenvectors with eigenvalue $-i$.
To see this we run a similar argument where we fix $\bar\zeta$
and let $\zeta$ vary; this produces $n$ eigenvectors of $J^{(\zeta=0,\bar\zeta)}$ with eigenvalue $+i$.
Then by dimension counting there is no room for any more.
So finally we see that the $-i$ eigenspace of $J^{(\zeta,\bar\zeta)}$ is independent of $\bar\zeta$ as
desired.

Thus we are free to choose any convenient $\bar\zeta$ in studying the Cauchy-Riemann equations
\eqref{eq:holomorphy}.
Since we want to understand how \eqref{eq:holomorphy} looks in terms of the Seiberg-Witten
fibration over $\vm$, it is natural to choose $\bar\zeta = 0$; substituting this in \eqref{eq:holomorphy}
gives
\begin{equation} \label{eq:hol-specialized}
(1 - i J_3 - \zeta J_-) df = 0.
\end{equation}
We assume given an identification of the complex symplectic manifold $(\CM, J_3, \omega_+)$
with the Seiberg-Witten torus fibration $(\CM, J_3^\sf, \omega_+^\sf = - \frac{1}{4 \pi} da^I \wedge dz_I)$.
(It was argued in \cite{Seiberg:1996nz} that such an identification should exist at least for $J_3$,
using a weak coupling of the gauge theory to gravity; a similar argument shows the identification also for
$\omega_+$.)  Then contracting \eqref{eq:hol-specialized} with a vector field tangent to the torus fiber,
$\frac{\partial}{\partial \bar{z}_I}$, gives
\begin{equation}
2 \partial_{\bar{z}_I} f - \zeta \left( \partial_{\bar{z}_I} \cdot J_- df \right) = 0.
\end{equation}
To deal with the second term, we use
\begin{equation}
J_- = g^{-1} \omega_- =  \frac{1}{4 \pi} g^{-1}(d\bar{a}^{\bar I}
\wedge d\bar{z}_{\bar I})
\end{equation}
and multiply by $4 \pi / \zeta$ to get
\begin{equation}
g^{-1}(df, d \bar{a}^I) =  \frac{8 \pi}{\zeta} \partial_{\bar{z}_I} f.
\end{equation}
This is almost of the form \eqref{eq:cr-connection-1} which we want, but not quite:
$g^{-1}(df, d \bar{a}^I)$ is a
mixture of derivative operators acting on $f$.  We want to make a change of basis to
extract an equation for $\frac{\partial f}{\partial a^I}$.  To do this we consider the restriction
of $g$ to a horizontal subspace orthogonal to $\CM_u$; write this as $g = h_{\bar{I} J} d\bar{a}^{\bar I} da^J$.
Then multiplying by $h_{\bar{I} J}$ we get
\begin{equation}
h_{\bar{I} J} g^{-1}(df, d \bar{a}^I) =  \frac{8 \pi}{\zeta}
h_{\bar{I} J} \partial_{\bar{z}_I} f.
\end{equation}
Now consider the special case where $f$ depends only on the base coordinates $(a, \bar{a})$.  Erecting an orthonormal
basis at a point we see that $g^{-1}(df, d \bar{a}^I) = (h^{-1})^{\bar{I} J} \partial_{a^J} f$.
This implies that for general $f$ the left side can be written as
\begin{equation}
\left( \partial_{a_J} - \CA^J \right) f
\end{equation}
where $\CA^J$ is a differential operator acting only in the fiber direction.  (More intrinsically
the full connection operator $\partial_{a_J} - \CA^J$ is the derivative of $f$ along the horizontal lift of the vector field $\partial_{a_J}$ from $\vm$ to $\CM$.)

Altogether then we have obtained
\begin{equation}
\frac{\partial}{\partial a_J} f = \frac{8 \pi}{\zeta} h_{\bar{I} J}
\frac{\partial f}{\partial \bar{z}_I} + \CA^J f,
\end{equation}
which is of the desired form \eqref{eq:cr-connection-1}.
An identical argument (starting with $\bar\zeta = \infty$ instead of $\bar\zeta = 0$) shows
the conjugate equation \eqref{eq:cr-connection-2}.

The structure we have discovered here is very similar to the ``improved connection'' introduced
in \cite{Cecotti:1991me}.
To see the similarity most clearly, introduce an infinite-dimensional bundle $V$ over
$\vm$, such that the fiber of $V$ over $u \in \vm$ is simply the space of real-analytic functions on the torus
$\CM_u$,
\begin{equation}
V_u = C^\omega(\CM_u).
\end{equation}
So a real-analytic complex-valued function $f$ on the whole $\CM$ is equivalently a real-analytic section
of $V$ over $\vm$.  Then what we have found above is that the Cauchy-Riemann equations can be
thought of as flatness equations for a 1-parameter family of connections in $V$, of a specific form.
The flatness of these connections is a consequence of the integrability of the complex structures on $\CM$.
In \cite{Cecotti:1991me} one also has a moduli space $\vm$ (parameterizing
$\CN=(2,2)$ supersymmetric field theories in $d=2$) and a vector bundle $V$ over $\vm$
(the bundle of Ramond ground states.)  One finds a family of flat connections in $V$ parameterized by
$\zeta \in \IC^\times$, of the form
\begin{align}
\nabla_i &= \frac{1}{\zeta} C_i + D_i, \\
\nabla_{\bar i} &= \zeta \bar{C}_{\bar i} + \bar{D}_{\bar i},
\end{align}
where $D_i$ is the standard connection provided by adiabatic variation of the couplings,
and $C_i$ are the ``chiral ring'' operators.  The flatness of these connections is a consequence
of the famous $tt^*$ equations.

Throughout this paper, particularly in Section \ref{sec:constructing-moduli}, many of the constructions ---
as well as their physical interpretations --- are parallel to those which appeared in the $tt^*$ story.

\section{Asymptotics of integral equations} \label{app:integral-asymptotics}

In this appendix we will first show how to modify the asymptotic analysis of
\cite{Cecotti:1993rm} in a situation with several BPS rays, and
then adapt this analysis to our problem.

\subsection*{Finite-dimensional case}

In \cite{Cecotti:1993rm} one studies a Riemann-Hilbert problem on the
complex $x$-plane for an $m \times m$ matrix $\Psi(x)$,
with a discontinuity along the real axis:
\begin{align} \label{eq:cv-discont}
\Psi(y e^{- i \epsilon}) &= \Psi(y e^{ i \epsilon}) S \qquad \text{for } y \in
\IR^+, \notag \\
\Psi(y e^{- i \epsilon}) &= \Psi(y e^{ i \epsilon}) S^t \qquad \text{for } y \in
\IR^-.
\end{align}
(We have now returned to the standard conventions for compositions
of operators.)  The analogues of the central charges $Z_\gamma$ here
are complex numbers $\Delta_{ij}$, $i,j = 1, \dots, m$, which obey
$\Delta_{ij} = w_i-w_j$ for some $w_i$. The matrix $S$ is
triangular, with $1$ on the diagonal and $S_{ij}=0$ if $\re
\Delta_{ij} <0$.

The asymptotic behavior of $\Psi$ is determined by the constants $w_i$. If one
defines
\begin{equation} \Phi_{ij}(x) = \Psi_{ij}(x) e^{- \beta x w_j - \frac{\beta}{x} \bar w_j},
\end{equation}
then $\Phi(x)$ tends to the identity matrix at $x\to \infty$ and to a certain ``metric'' $g_{ij}$ at $x\to 0$.

The matrix $S$ is the ``Stokes multiplier'' of the problem.  It is
convenient to re-express it as a product of more elementary ``Stokes factors.''
Indeed, assuming no three $w_i$ are collinear in the complex plane,
there are unique decompositions
\begin{equation}
S = \prod^\cwarrow_{(ij): \re \Delta_{ij} > 0} s_{(ij)}, \quad S^t = \prod^\cwarrow_{(ij): \re \Delta_{ij} < 0} s_{(ji)}
\end{equation}
where the products are taken in the order of increasing $\arg \Delta_{ij}$,
and each $s_{(ij)}$ has ones on the diagonal, and a single non-zero
off-diagonal element at the location $(ji)$, with value $- \mu_{ij} = -\mu_{ji}$.

Using this decomposition, we can introduce our ``multi-ray'' version
of the Riemann-Hilbert problem.
Namely, introduce a set of rays through the origin in the $x$-plane,
\begin{equation}
\ell_{(ij)} = \{ x: x \Delta_{ij} \in \IR_+ \},
\end{equation}
and require
\begin{equation}
\Psi(y e^{- i \epsilon}) = \Psi(y e^{ i \epsilon}) s_{(ij)} \qquad \text{for } y \in
\ell_{(ij)}.
\end{equation}
Importantly, one can show that $g_{ij}$ --- which was the main object of interest
in \cite{Cecotti:1993rm} --- is the same whether we use
the single-ray or multi-ray problem.

The integral equation $(4.17)$ of \cite{Cecotti:1993rm} for $\Phi(x)$ reads (with
some slight modifications to the $\epsilon$ conventions, for later
convenience):
\begin{align}
\Phi_{ij}(x) = \delta_{ij} &+ \frac{1}{2 \pi i} \int_0^\infty
\frac{dy}{y-x} \sum_k \Phi_{ik}(e^{i \epsilon} y)(1-S)_{kj}e^{-
\beta y \Delta_{kj} - \beta/y \bar \Delta_{kj}}  \nonumber  \\
&+ \frac{1}{2 \pi i}
\int_{-\infty}^0 \frac{dy}{y-x} \sum_k \Phi_{ik}(e^{-i \epsilon}
y)(1-S^t)_{kj}e^{- \beta y \Delta_{kj} - \beta/y \bar \Delta_{kj}}.  \label{eq:cv-integral}
\end{align}
A solution to this equation gives a solution to the single-ray Riemann-Hilbert problem.
Now we formulate a new integral equation which is equivalent to the multi-ray problem:
\begin{equation} \label{eq:cv-multi-line}
\Phi(x)_{ij} = \delta_{ij} + \sum_{k} \frac{1}{2 \pi i}
\int_{\ell_{(kj)}} \frac{dy}{y-x}\Phi_{ik}(y)\mu_{kj}e^{- \beta y
\Delta_{kj} - \frac{\beta}{y} \bar \Delta_{kj}}
\end{equation}
The integration here is understood to be the principal part integration.
Note that when no three $w_i$ are collinear the matrix
elements $\Phi_{ik}$ are continuous across all the rays $\ell_{(kj)}$.

The recursive solution of \eqref{eq:cv-multi-line} takes a simple form:
\begin{equation}
\Phi(x)_{ij} = \delta_{ij} + \Phi(x)_{ij}^{(1)} + \Phi(x)_{ij}^{(2)} + \cdots,
\end{equation}
where
\begin{align}
\Phi(x)_{ij}^{(1)} &= \frac{1}{2 \pi i} \mu_{ij}
\int_{\ell_{(ij)}} \frac{dy}{y-x} e^{ -\beta y
\Delta_{ij} - \frac{\beta}{y} \bar \Delta_{ij} }, \\ \notag
\Phi(x)_{ij}^{(2)} &= \frac{1}{(2 \pi i)^2} \sum_{i_2} \mu_{i i_2} \mu_{i_2 j}
\int_{\ell_{(i i_2)}} \int_{\ell_{(i_2 j)}} \frac{dy_1}{y_1-y_2}
\frac{dy_2}{y_2-x}e^{ -\beta
y_1 \Delta_{i i_2} - \frac{\beta}{y_1} \bar \Delta_{i i_2} }e^{ -\beta
y_2 \Delta_{i_2 j} - \frac{\beta}{y_2} \bar \Delta_{i_2 j} },
\end{align}
and in general $\Phi(x)^{(n)}$ involves integrals over all chains of $n$ rays $\ell_{(i_k i_{k+1})}$,
where $i_1 = i$ and $i_{n+1} = j$.
Each integral along $\ell_{(i_k i_{k+1})}$ contains the factor
$e^{ -\beta y \Delta_{i_k i_{k+1}} - \frac{\beta}{y} \bar \Delta_{i_k i_{k+1}} }$.  On the ray
this exponent is real and negative, with a single peak at $y = \exp (-i \arg \Delta_{i_k i_{k+1}})$.
As $\beta$ is taken to be large, the integral is thus well approximated by
the saddle point method, replacing the rest of the integrand by its value at the peak.
As a result, the large $\beta$ asymptotics at fixed $x$ are simply:\footnote{These asymptotics are
valid except when $x$ lies exactly on the saddle point for the integral over $y_n$,
i.e. $x = \exp [-i \arg \Delta_{i_n j}]$.  At this point we find similar large-$\beta$ asymptotics
except that one of the $\sqrt{\beta}$ suppression factors is absent.}
\begin{equation}
\Phi(x)^{(1)}_{ij}  \sim \frac{1}{2 \sqrt{\pi \beta |\Delta_{ij}|} i}
\mu_{ij}\frac{1}{\exp[-i \arg \Delta_{ij}]-x} e^{- 2 \beta
|\Delta_{ij}|},
\end{equation}
and
\begin{multline}
\Phi(x)^{(2)}_{ij} \sim \sum_{i_2} \mu_{i i_2} \mu_{i_2 j} \frac{1}{2 \sqrt{\pi \beta
|\Delta_{i i_2}|} i}\frac{1}{2 \sqrt{\pi \beta |\Delta_{i_2 j}|} i} \times \\
\frac{1}{\exp[-i \arg \Delta_{i_2 j}]-x}\frac{1}{\exp[-i \arg
\Delta_{i i_2}]-\exp[-i \arg \Delta_{i_2 j}]}e^{- 2 \beta |\Delta_{i
i_2}|}e^{- 2 \beta |\Delta_{i_2 j}|},
\end{multline}
with similar estimates for the higher $\Phi^{(n)}$.
By the triangle inequality we see that in the large $\beta$ limit $\Phi^{(2)}$
is exponentially suppressed relative to $\Phi^{(1)}$, and similarly
$\Phi^{(n+1)}$ is suppressed relative to $\Phi^{(n)}$.  $\Phi^{(n)}$ has exactly the
exponential suppression expected for an $n$-instanton correction.

We note that this asymptotic analysis is much simpler than the corresponding
analysis of \eqref{eq:cv-integral}; in that case one has to deform the
integration contour to pass through the appropriate saddle,
and one encounters cuts and poles along the way, whose contributions have to be carefully
tracked.

To finish this section we briefly discuss the analytic properties of this expansion.
As we saw above, the $n$-th correction to $\Phi$ can be expressed in terms of certain
iterated integrals:
\begin{multline} \CF^{(n)}[x,\beta;\Delta_{i_1 i_2},\dots, \Delta_{i_{n} i_{n+1}}]=  \\
\prod_{k=1}^n \left[ \int_{\ell_{(i_k i_{k+1})}} \frac{dy_k}{2 \pi i} e^{ -\beta y_k \Delta_{i_k i_{k+1}} -
\frac{\beta}{y_k} \bar \Delta_{i_k i_{k+1}} } \right] \prod_{k=1}^n
\frac{1}{y_k - y_{k+1}} \Bigg\vert_{y_{n+1} = x}.
\end{multline}
This $\CF^{(n)}$ has an obvious discontinuity on the ray  $\arg x = -\arg \Delta_{i_n i_{n+1}}$, which equals
\begin{equation}
\CF^{(n-1)}[x,\beta;\Delta_{i_1 i_2}, \dots, \Delta_{i_{n-1}
i_{n}}]e^{ -\beta x \Delta_{i_{n} i_{n+1}} - \frac{\beta}{x} \bar
\Delta_{i_{n} i_{n+1}} }.
\end{equation}
Note that $\CF^{(n)}$ makes sense for generic values of the arguments $d_k =
\Delta_{i_k i_{k+1}}$, without the restriction $\Delta_{ij} =
w_i-w_j$.  (In fact, one could even make $\Delta_{ij}$ and $\bar
\Delta_{ij}$ into independent complex parameters, and put the rays
at $x^2 \Delta_{ij}/\bar \Delta_{ij} \in \IR^+$.)  It has cuts
whenever the phases of two consecutive arguments $d_k, d_{k+1}$
align.  The discontinuity is $\CF^{(n-1)}$ with the same arguments,
except for the substitution $d_k, d_{k+1} \to d_k + d_{k+1}$.

Essentially the same functions $\CF^{(n)}$ appeared in the asymptotic
analysis of \cite{Cecotti:1993rm}; ours differ from those only in the placement of branch cuts.

\subsection*{Infinite-dimensional case}

Now we give a similar analysis for the multiplicative
integral equation \eqref{eq:X-integral-mult-explicit}.
First, for any vector $\gamma \in \clat$, we define a vector $f^\gamma \in \clat_\IQ$, by
the power series expansion
\begin{equation}
- \sum_{\gamma' \in \clat} \left( \degen(\gamma';u)
\log (1- \sigma(\gamma')
\CX_{\gamma'} ) \right) \gamma' = \sum_{\gamma' \in \clat} f^{\gamma'} \CX_{\gamma'}
\end{equation}
or more explicitly
\begin{equation}
f^\gamma = \sum_{n \ge 1 \text{ s.t. } \gamma = n \gamma'} \frac{\sigma(\gamma')^n}{n} \degen(\gamma';u) \gamma'.
\end{equation}
The point of this definition is that then \eqref{eq:X-integral-mult-explicit} takes the form
\begin{equation} \label{eq:X-integral-mult-std}
\CX_\gamma(\zeta) = \CX^\sf_\gamma(\zeta) \exp \left\langle \gamma,
\frac{1}{4 \pi i} \sum_{\gamma'} f^{\gamma'} \int_{\ell_{\gamma'}} \frac{d\zeta'}{\zeta'} \frac{\zeta' +
\zeta}{\zeta' - \zeta} \CX_{\gamma'}(\zeta') \right\rangle.
\end{equation}

We aim to construct a solution $\CX$ to \eqref{eq:X-integral-mult-std} as a limit of successive approximations
$\CX^{(n)}$, or the corresponding approximations $\Upsilon^{(n)}$ to $\Upsilon$ defined in \eqref{eq:def-upsilon}.
$\Upsilon$ is a map from $\CM_u$ to its complexification; we write its components as functions,
$\Upsilon_\gamma(\theta) := \gamma \cdot \Upsilon(\theta)$.
We begin by choosing $\Upsilon_\gamma^{(0)} = \theta_\gamma$.
Recalling that $\Upsilon$ is defined so that
\begin{equation}
\CX^\sf(\Upsilon^{(n)}) = \CX^{(n)}(\theta),
\end{equation}
we can write the iteration step as
\begin{equation} \label{eq:recursion}
e^{i \Upsilon^{(n+1)}_\gamma} = e^{i \theta_\gamma} \exp \left\langle \gamma,
\frac{1}{4 \pi i} \sum_{\gamma'} f^{\gamma'} \int_{\ell_{\gamma'}}
\frac{d\zeta'}{\zeta'} \frac{\zeta' + \zeta}{\zeta' - \zeta}
\CX^\sf_{\gamma'}(\Upsilon^{(n)},\zeta')\right\rangle
\end{equation}
A fixed point of this iteration, $\Upsilon^{(n+1)} = \Upsilon^{(n)}$, would give a solution of \eqref{eq:X-integral-mult}.  So to see
that a solution exists we should verify that the iteration is a contraction, i.e. that
\begin{equation}
\max_{\zeta,\theta} \norm{\Upsilon^{(n+1)} - \Upsilon^{(n)}} < C \max_{\zeta,\theta} \norm{\Upsilon^{(n)} - \Upsilon^{(n-1)}}
\end{equation}
for some constant $C < 1$.  More precisely, we will verify that this iteration is a contraction
when acting on $\Upsilon^{(n)}$ which have $\max_{\zeta,\theta} \norm{\Upsilon^{(n)} - \theta} < \infty$,
and which obey a side condition expressing the fact that they are not too far from the real torus:
we require
$\max_{\zeta,\theta} |e^{i \Upsilon^{(n)}_\gamma}|<e^{\eps \norm{\gamma}}$, for a constant $\eps>0$ to be determined shortly.

First we need to see that the iteration preserves our side condition.
Taking the absolute value of \eqref{eq:recursion} and making the saddle point analysis,
for large enough $R$ we get the estimate\footnote{As in the previous section, this analysis has to
be supplemented by a separate discussion when $\zeta$ hits the saddle point, but that only
reduces the suppression by a factor $\sqrt{R}$, and still allows us to establish \eqref{eq:estimate}.}
\begin{equation} \label{eq:estimate}
|e^{i \Upsilon^{(n+1)}_\gamma}| <  \exp \left[ \sum_{\gamma'} \abs{\inprod{f^{\gamma'},\gamma}}
e^{- 2 \pi R |Z_{\gamma'}| + \eps \norm{\gamma'}} \right].
\end{equation}
Now we assume that $\sum_{\gamma'} f^{\gamma'} \norm{\gamma'}
e^{- 2 \pi R |Z_{\gamma'}|}$ converges for large enough $R$.  (This amounts
to a requirement that the $\degen(\gamma';u)$ do not grow too quickly with $\gamma'$;
it appears very reasonable for field theory but would almost certainly be violated
in the gravitational case.)  We also use the Support Property recalled in Section
\ref{sec:ReviewKS}, to bound $\norm{\gamma'}$ by $K \abs{Z_{\gamma'}}$.
Then for large enough $R$ we can pick $\eps$ so that the right side
is smaller than $e^{\eps \norm{\gamma}}$.  Doing this for $\gamma$ running over a basis of $\clat$,
we obtain our desired $\eps$.  (Indeed, we can take $\epsilon \to 0$ exponentially
fast for large $R$.)

Now we want to estimate
\begin{align}
\norm{\Upsilon^{(n+1)}-\Upsilon^{(n)}} &=
\frac{1}{4 \pi} \bigg\lVert \sum_{\gamma' \in \clat} \int_{\ell_{\gamma'}}
\frac{d\zeta'}{\zeta'} \frac{\zeta' + \zeta}{\zeta' - \zeta}
f^{\gamma'}\left(\CX^\sf_{\gamma'}(\Upsilon^{(n)},\zeta')-\CX^\sf_{\gamma'}(\Upsilon^{(n-1)},\zeta')\right) \bigg\rVert \\
&\le \frac{1}{4 \pi}\bigg\lVert \sum_{\gamma' \in \clat} \int_{\ell_{\gamma'}}
\frac{d\zeta'}{\zeta'} \frac{\zeta' + \zeta}{\zeta' - \zeta}
f^{\gamma'} \abs{\CX^\sf_{\gamma'}(\theta, \zeta')} \left(e^{i \Upsilon^{(n)}_{
\gamma'}(\zeta')} - e^{i \Upsilon^{(n-1)}_{\gamma'}(\zeta')}\right) \bigg\rVert \\
&\le \frac{1}{4 \pi}\bigg\lVert \sum_{\gamma' \in \clat} \int_{\ell_{\gamma'}}
\frac{d\zeta'}{\zeta'} \frac{\zeta' + \zeta}{\zeta' - \zeta}
f^{\gamma'} \abs{\CX^\sf_{\gamma'}(\theta, \zeta')}  \bigg\rVert e^{\eps \norm{\gamma'}} \norm{\gamma'} \max_{\zeta}  \norm{\Upsilon^{(n)} -
\Upsilon^{(n-1)}}.
\end{align}
The large-$R$ saddle point analysis then gives
\begin{equation}
\norm{\Upsilon^{(n+1)}-\Upsilon^{(n)}} \le
\frac{1}{4\pi}\max_{\zeta} \norm{\Upsilon^{(n)}(\zeta) -
\Upsilon^{(n-1)}(\zeta)} \bigg\lVert \sum_{\gamma' \in \clat}
f^{\gamma'} \norm{\gamma'} e^{- 2 \pi R |Z_{\gamma'}| + \eps
\norm{\gamma'}} \bigg\rVert.
\end{equation}
For large enough $R$, with our convergence assumptions,
this establishes the contraction property; indeed
the iteration converges very quickly, with a speed
determined by the largest $e^{- 2 \pi R |Z_{\gamma'}|}$.

One can give an explicit expression for $\CX_\gamma(\zeta)$
in terms of functions like the $\CF$ of the finite-dimensional case.
The presence of the exponential in the recursion relation makes
things a bit more intricate:  instead of summing over chains one
now gets a sum over decorated rooted trees.
Let $\CT$ denote a rooted tree, with
edges labeled by pairs $(i,j)$ (where $i$ is the node closer to the root),
and each node decorated by a choice of $\gamma_i \in \clat$.
Also call the decoration at the root node $\gamma_\CT$.
Then define the weight of the tree to be
an element of $\clat_\IQ$, determined by the $\degen(\gamma;u)$,
\begin{equation}
\CW_\CT = \frac{f^{\gamma_\CT}}{\abs{\mathrm{Aut}(\CT)}} \prod_{(i,j) \in {\mathrm {Edges}}(\CT)} \inprod{\gamma_i, f^{\gamma_j}}.
\end{equation}
The iterative solution for $\CX_\gamma(\zeta)$ then takes the form
\begin{equation} \label{eq:iterative-X}
\CX_\gamma(\zeta) = \CX^\sf_\gamma (\zeta)\exp \left\langle \gamma, \sum_\CT
\CW_\CT \CG_\CT(\zeta) \right\rangle,
\end{equation}
for some functions $\CG_\CT(\zeta)$.  The integral equation
\eqref{eq:X-integral-mult-std} for $\CX$ becomes a formula expressing each
$\CG_\CT(\zeta)$ in terms of the ones for smaller trees.  Namely, deleting the root
from $\CT$ leaves behind a set of rooted trees $\CT_a$, and \eqref{eq:X-integral-mult-std} will
be satisfied if
\begin{equation}
\CG_\CT(\zeta) = \frac{1}{4 \pi i} \int_{\ell_{\gamma_\CT}} \frac{d \zeta'}{\zeta'} \frac{\zeta' + \zeta}{\zeta' - \zeta} \CX^\sf_{\gamma_\CT}(\zeta') \prod_a \CG_{\CT_a}(\zeta').
\end{equation}
It follows that, as for the $\CF^{(n)}$ of the finite-dimensional case, the discontinuity of $\CG_\CT$
along $\ell_{\gamma_\CT}$ is determined by the product of the lower $\CG_{\CT_a}$.

From \eqref{eq:iterative-X} we can also obtain $\Upsilon$ directly:
\begin{equation}
e^{i \Upsilon_\gamma} = \exp \left\langle \gamma, \sum_\CT
\CW_\CT \CG_\CT(\zeta) \right\rangle.
\end{equation}
In particular this allows us to evaluate $\Upsilon(\zeta = 0)$.
The expansion of the symplectic form $\hsymp(\zeta)$ can similarly be
analyzed in this fashion, and organized as a sum over trees.

\section{Asymptotics of differential equations} \label{app:diff-asymptotics}
In this appendix we would like to understand how to compute the
coefficients of the differential equations satisfied by $\CX$.
Consider the asymptotic expansion of $\CX$ around $\zeta=0$:
\begin{equation} \label{eq:as-X}
\CX(\theta,\zeta)_\gamma \sim
\CX^\sf_\gamma(\Upsilon_0(\theta),\zeta) \exp \sum_{n>0} \zeta^n
\gamma \cdot g_n(\theta).
\end{equation}
We consider the differential operators defined in Section \ref{sec:differential-problem},
\begin{equation}
\frac{1}{\CX_\gamma}\frac{\partial \CX_\gamma}{\partial t^n}  =
\frac{1}{\CX_\gamma} \CA_n \CX_\gamma.
\end{equation}
Plug in the expansion \eqref{eq:as-X} and keep only the first few terms:
\begin{equation}
\frac{\partial \log \CX^\sf_\gamma(\Upsilon_0(\theta))}{\partial
t^n} = \left(\frac{1}{\zeta}\CA^{(-1)}_n + \CA^{(0)}_n\right)
\Upsilon_0(\theta)_\gamma+ \CA_n^{(-1)} \gamma \cdot g_1(\theta).
\end{equation}
The leading part in $\zeta$ is a statement we already understood:
\begin{equation}
\pi R \frac{\partial Z_\gamma}{\partial t^n} = \CA^{(-1)}_n
\Upsilon_0(\theta)_\gamma.
\end{equation}
This means that $\CA^{(-1)}$ is the pull-back by $\Upsilon_0$ of
$\CA^{(-1)}_\sf$.

The next term in the expansion is
\begin{equation}
i \frac{\partial \Upsilon_0(\theta)_\gamma}{\partial t^n} =
\CA^{(0)}_n \Upsilon_0(\theta)_\gamma+ \CA_n^{(-1)} \gamma \cdot
g_1(\theta),
\end{equation}
which determines $\CA^{(0)}_n$, given a knowledge of $\Upsilon_0(\theta)$.

There is an alternative point of view, which is quite useful:
consider the compatibility conditions between the various
differential equations.
For example, consider the equation $[R\partial_R - \CA_R, \zeta \partial_\zeta - \CA_\zeta] = 0$, i.e.
\begin{equation} \left[ \plog{R} + \frac{1}{\zeta} {\CA_{\zeta}^{(-1)}} - \CA_R^{(0)} -
\zeta \CA^{(1)}_{\zeta} \, , \, \plog{\zeta} -
\frac{1}{\zeta} {\CA_{\zeta}^{(-1)}} - \CA_{\zeta}^{(0)} - \zeta
\CA^{(1)}_{\zeta} \right] = 0,
\end{equation}
and expand it in powers of $\zeta$. This gives three equations:
\begin{align}
\plog{R} \CA_{\zeta}^{(-1)} - [\CA_R^{(0)},\CA_{\zeta}^{(-1)}] &=
[\CA_{\zeta}^{(0)},\CA_{\zeta}^{(-1)}] + \CA_{\zeta}^{(-1)},\\
\plog{R} \CA_{\zeta}^{(0)} - [\CA_R^{(0)},\CA_{\zeta}^{(0)}] &=
2 [\CA_{\zeta}^{(1)},\CA_{\zeta}^{(-1)}],\\
\plog{R} \CA_{\zeta}^{(1)} - [\CA_R^{(0)},\CA_{\zeta}^{(1)}] &=
[\CA_{\zeta}^{(1)},\CA_{\zeta}^{(0)}] + \CA_{\zeta}^{(1)}.
\end{align}
These equations are strongly reminiscent of the Nahm equations, differing
from them only by the two extra linear pieces on the right hand side.
These extra pieces are dominant at large radius. An alternative strategy
to derive the large $R$ asymptotics is again an iterative solution
of these three equations around the semiflat solution.

Another interesting set of ``isomonodromic'' equations can be derived by similarly
expanding $[\partial_u - \CA_u, \zeta\partial_\zeta - \CA_\zeta] = 0$:
\begin{align}
0 &=[\CA^{(-1)}_u, \CA^{(-1)}_\zeta], \\
\frac{\partial}{\partial u} \CA_{\zeta}^{(-1)} -
[\CA_u^{(0)},\CA_{\zeta}^{(-1)}] &=
[\CA_u^{(-1)},\CA_{\zeta}^{(0)}] - \CA_{u}^{(-1)},\\
\frac{\partial}{\partial u} \CA_{\zeta}^{(0)} -
[\CA_u^{(0)},\CA_{\zeta}^{(0)}]  &=
[\CA_u^{(-1)},\CA_{\zeta}^{(1)}], \\
\frac{\partial}{\partial u} \CA_{\zeta}^{(1)} -
[\CA_u^{(0)},\CA_{\zeta}^{(1)}] &=
[\CA_u^{(1)},\CA_{\zeta}^{(0)}] + \CA_{u}^{(-1)}, \\
0 &= [\CA^{(1)}_u, \CA^{(1)}_\zeta].
\end{align}

\section{A relation to the Thermodynamic Bethe Ansatz} \label{app:TBA}

{\it Note added Nov. 20, 2009}:

It was pointed out to us some time
ago by A. Zamolodchikov that one of the central results of this
paper, equation \eqref{eq:X-integral-mult-explicit}, is in fact a
version of the Thermodynamic Bethe Ansatz
\cite{Zamolodchikov:1989cf}.  In this appendix we explain that
remark.  Another relation between four-dimensional super Yang-Mills
theory and the TBA has recently been discussed by Nekrasov and
Shatashvili \cite{Nekrasov:2009ui}.

The TBA equations for an integrable system of particles $a$ with
masses $m_a$, at inverse temperature $\beta$, with integrable
scattering matrix $S_{ab}(\theta-\theta')$, where $\theta$ is the
rapidity, are
\begin{equation}\label{eq:gen-TBA}
\epsilon_a(\theta) = m_a \beta \cosh \theta - \sum_b
\int_{-\infty}^{+\infty} \frac{d \theta'}{2\pi}
\phi_{ab}(\theta-\theta') \log (1 + e^{\beta \mu_b -
\epsilon_b(\theta')})
\end{equation}
where $\phi_{ab}(\theta) = - i\frac{\p}{\p \theta} \log
S_{ab}(\theta)$.  Here the scattering matrix is diagonal, that is,
the soliton creation operators obey $\Phi_a(\theta) \Phi_b(\theta') =
S_{ab}(\theta-\theta') \Phi_b(\theta') \Phi_a(\theta)$.

We can put the logarithm of \eqref{eq:X-integral-mult-explicit} in
the form of \eqref{eq:gen-TBA} as follows.  Clearly the particle
labels $a, b, \dots$ correspond to $\gamma, \gamma', \dots$. Now let $Z_\gamma =
e^{i \alpha_\gamma} \vert Z_\gamma \vert$, where $\alpha_\gamma$ is
real and only defined modulo $2\pi$.  For any $\gamma$ we can make the
change of variables $\zeta = - e^{i \alpha_\gamma+ \theta}$, so that
the BPS ray $\ell_\gamma$ is mapped out by $-\infty < \theta <
\infty$.  Under this change of
variables the semiflat coordinate \eqref{eq:X-sf} becomes
\begin{equation}
\log \CX_{\gamma}^{\sf} = - 2 \pi R \lvert Z_\gamma \rvert \cosh \theta
+ i \varphi_\gamma.
\end{equation}
(Note that to avoid confusion with the rapidity $\theta$
we have changed the notation for the angular coordinate along the torus
from $\theta_\gamma$, used in the rest of this paper,
to $\varphi_\gamma$.) Now we set
\begin{equation}\label{eq:defmugamma}
\beta \mu_\gamma := i \varphi_\gamma + \log(-\sigma(\gamma))  \quad \mod \ 2
\pi i.
\end{equation}
Note that $\beta \mu_\gamma$ is $i \varphi_\gamma$ or differs by
$\pm i \pi$.  In particular, it is pure imaginary.  Define
``quasiparticle energies'' $\epsilon_{\gamma}(\theta)$ by
\begin{equation} \label{eq:qparticle}
\CX_{\gamma}(\zeta = - e^{i \alpha_\gamma+ \theta}) :=
-\sigma(\gamma) e^{\beta \mu_\gamma - \epsilon_{\gamma}(\theta)}=
e^{i \varphi_\gamma - \epsilon_\gamma(\theta)}.
\end{equation}
More precisely, this defines $\eps_\gamma$ on the BPS ray $\ell_\gamma$,
where $\theta$ is real.  For other $\theta$ we define $\eps_\gamma$ by analytic continuation ---
in contrast to $\CX_\gamma$, which has discontinuities along certain
lines of constant $\Im\,\theta$ (the BPS rays).

We have chosen \eqref{eq:defmugamma} so that the logarithm of
\eqref{eq:X-integral-mult-explicit} reads as
\begin{equation}\label{eq:GMN-2}
\epsilon_{\gamma}(\theta) = 2\pi R \vert Z_\gamma \vert \cosh\theta
+ \sum_{\gamma'} \Omega(\gamma') \int_{-\infty}^{+\infty}
\frac{d\theta'}{2\pi} K_{\gamma,\gamma'}(\theta-\theta') \log( 1 +
e^{\beta \mu_{\gamma'} - \epsilon_{\gamma'}(\theta')} ),
\end{equation}
with
\begin{equation}
K_{\gamma,\gamma'}(\theta-\theta') = \frac{i}{2} \langle \gamma,
\gamma'\rangle \frac{ e^{ \theta - \theta' + i \alpha_{\gamma} - i
\alpha_{\gamma'}} +1 }{e^{ \theta - \theta' + i \alpha_{\gamma} - i
\alpha_{\gamma'}} -1 }.
\end{equation}
This kernel can also be written as
\begin{equation}
\begin{split}
K_{\gamma,\gamma'}(\theta-\theta') & = i \langle \gamma, \gamma'
\rangle \frac{\p}{\p \theta} \log \bigl[\sinh\left(\half (\theta -
\theta' + i \alpha_{\gamma} - i \alpha_{\gamma'})\right)\bigr]\\
& =\frac{i}{2}\langle \gamma, \gamma' \rangle  \coth\left( \frac{
\theta-\theta' + i \alpha_{\gamma} - i \alpha_{\gamma'}}{2} \right).
\end{split}
\end{equation}
 The argument of the logarithm is not a pure phase, so
$K_{\gamma,\gamma'}(\theta-\theta')$ does not correspond to a
unitary scattering matrix, in general.

Let us comment briefly on the reality properties of the
``quasiparticle energies.''  The twistor coordinates satisfy the
reality property
\begin{equation}
\overline{\CX_{\gamma}(\zeta)} = \CX_{-\gamma}(-1/\bar \zeta).
\end{equation}
Since $Z_{-\gamma} = - Z_\gamma$, we have
\begin{equation}
e^{i \alpha_{-\gamma}} = - e^{i \alpha_\gamma}.
\end{equation}
Hence, if $\zeta = - e^{i \alpha_{\gamma} + \theta}$ and $\theta$ is
real, then $-1/\bar \zeta = - e^{i \alpha_{-\gamma} - \theta}$. Now,
using $\sigma(-\gamma) = \sigma(\gamma)$ and $\varphi_{-\gamma} = -
\varphi_{\gamma}$, we get the reality condition on the
``quasiparticle energies''
\begin{equation}\label{eq:reality}
\overline{\epsilon_{\gamma}(\theta)} = \epsilon_{-\gamma}(-\theta).
\end{equation}

The integral equation \eqref{eq:X-integral-mult-explicit} is
consistent with the reality condition (\ref{eq:reality}) since for
$\theta, \theta'$ both real
\begin{equation}
\overline{K_{\gamma, \gamma'}(\theta-\theta')} = K_{-\gamma, -
\gamma'}(-\theta + \theta').
\end{equation}

\bibliography{wcf-paper}

\end{document}